\documentclass[final,3p,times]{elsarticle}

\usepackage{amssymb}
\usepackage{amsthm}
\usepackage{amsmath}
\usepackage{lineno}
\usepackage{algorithm} 
\usepackage{algorithmic} 
\usepackage{multirow} 
\usepackage{lineno}
\usepackage{xcolor}
\usepackage{graphicx}
\usepackage{caption}
\usepackage{booktabs}
\usepackage{enumerate}
\usepackage{picinpar}
\usepackage{amsfonts}
\usepackage{stfloats}
\usepackage{url}
\usepackage{flushend}
\usepackage{colortbl}
\usepackage{soul}
\usepackage{multirow}
\usepackage{pifont}
\usepackage{color}
\usepackage{alltt}
\usepackage{subfigure}
\usepackage{setspace}
\usepackage{url}
\usepackage{pdfpages}
\usepackage{hyperref}
\usepackage{appendix}
\usepackage{pdflscape}
\usepackage{enumerate}
\usepackage{siunitx}
\usepackage{breakurl}
\usepackage{epstopdf}
\usepackage{pbox}
\usepackage{color}
\definecolor{b}{rgb}{0.00,0.00,1.00}    
\definecolor{r}{rgb}{1.00,0.00,0.00}    
\usepackage{appendix}
\usepackage{verbatim}

\newcommand\figref[1]{Fig.~\ref{#1}}
\newcommand\eqrefA[1]{Eq.~\ref{#1}}

\journal{XXXX}
\biboptions{numbers,sort&compress} 
\begin{document}
	
\captionsetup[figure]{labelfont={bf},labelformat={default},labelsep=period,name={Fig.}}	
		
\begin{frontmatter}

\title{Deep Learning Algorithms for Rotating Machinery Intelligent Diagnosis: An Open Source Benchmark Study}

\author[add1]{Zhibin Zhao}%
\ead{zhibinzhao1993@gmail.com}%
\author[add1]{Tianfu Li}%
\author[add1]{Jingyao Wu}%
\author[add1]{Chuang Sun}%
\author[add1]{Shibin Wang}%
\author[add1]{Ruqiang Yan\corref{cor1}}%
\ead{yanruqiang@xjtu.edu.cn}%
\author[add1]{Xuefeng Chen}%

\cortext[cor1]{Corresponding author}%

\address[add1]{School of Mechanical Engineering, Xi'an Jiaotong University, Xi'an, China}%

\begin{abstract}
Rotating machinery intelligent diagnosis based on deep learning (DL) has gone through tremendous progress, which can help reduce costly breakdowns.
However, different datasets and hyper-parameters are recommended to be used, and few open source codes are publicly available, resulting in unfair comparisons and ineffective improvement.
To address these issues, we perform a comprehensive evaluation of four models, including multi-layer perception (MLP), auto-encoder (AE), convolutional neural network (CNN), and recurrent neural network (RNN), with seven datasets to provide a benchmark study.
We first gather nine publicly available datasets and give a comprehensive benchmark study of DL-based models with two data split strategies, five input formats, three normalization methods, and four augmentation methods.
Second, we integrate the whole evaluation codes into a code library and release it to the public for better comparisons.
Third, we use specific-designed cases to point out the existing issues, including class imbalance, generalization ability, interpretability, few-shot learning, and model selection.
Finally, we release a unified code framework for comparing and testing models fairly and quickly, emphasize the importance of open source codes, provide the baseline accuracy (a lower bound), and discuss existing issues in this field.
The code library is available at: \url{https://github.com/ZhaoZhibin/DL-based-Intelligent-Diagnosis-Benchmark}.
\end{abstract}

\end{frontmatter}


\section{Introduction}
\label{S:1}
Rotating machinery, as key mechanical equipment in the modern industry, is chronically running in a complex environment with elevated temperature, fatigue, and heavy load.
Generated faults might cause severe accidents, resulting in enormous economic loss and casualties.
Intelligent diagnosis, as a key ingredient of prognostics health management (PHM), which is one of the most essential systems in a wide range of rotating machinery, such as helicopter, aero-engine, wind turbine, and high-speed train, is designed to detect faults effectively.
Traditional intelligent diagnosis methods mainly consist of feature extraction using signal processing methods \cite{zhao2018enhanced,wang2016matching} and fault classification \cite{sun2018sparse} using machine learning approaches, which have made considerable progress.
However, facing with heterogeneous massive data, feature extraction methods and mapping abilities from signals to conditions that are designed and chosen by experts, to a great extent depending on prior knowledge, are time-consuming and empirical.
Thus, how to perform diagnosis more precisely and efficiently is still a challenging problem.

DL, as a booming data mining technique, has swept many fields including computer vision (CV) \cite{krizhevsky2012imagenet,farabet2012learning}, natural language processing (NLP) \cite{hirschberg2015advances,sun2017review,young2018recent}, and other fields \cite{feng2019resilience,li2020development}.
In 2006, the concept of DL was first introduced via proposing deep belief network (DBN) \cite{hinton2006reducing}.
In 2013, MIT Technology Review ranked the DL technology as the top ten breakthrough technologies \cite{MIT2013}.
In 2015, a review \cite{lecun2015deep} published in nature stated that DL allows computational models composed of multiple processing layers to learn data representations with multiple levels of the abstraction.
Due to its strong representation learning ability, DL is well-suited to data analysis and classification.
Therefore, in the field of intelligent diagnosis, many researchers have applied DL-based techniques, such as multi-layer perception (MLP), auto-encoder (AE), convolutional neural network (CNN), DBN, and recurrent neural network (RNN) to boost the performance.
However, different researchers often recommended to use different inputs (such as time domain input, frequency domain input, time-frequency domain input, etc.) and set different hyper-parameters (such as the learning rate, the batch size, the network architecture, etc.).
Unfortunately, a few authors made their codes available for comparisons, resulting in unfair comparisons and ineffective improvement.
To address this problem, it is crucial to evaluate and compare different DL-based intelligent diagnosis algorithms to provide the benchmark study and open souce codes, thereby helping further studies to propose more persuasive and appropriate algorithms.

For comprehensive performance comparisons, it is necessary to collect different datasets in a library and evaluate the performance of algorithms for different datasets on a unified platform.
In addition, one common issue in intelligent diagnosis is data split, and researchers often use the random split strategy.
This strategy is dangerous since if the preparation process exists any overlap, the evaluation of classification algorithms will have test leakage \cite{riley2019three}.
As for industrial data, they are rarely random and are always sequential (they might contain trends in the time domain), and it is more appropriate to split data according to time sequences (we simply call it order split).
Conversely, if we randomly split the data, it might be possible for diagnosis algorithms to record the future patterns, which might cause another pitfall with test leakage.

To address these problems, in this paper, we collect nine publicly available datasets and discuss whether it is suitable for intelligent diagnosis.
After that, we evaluate DL-based intelligent diagnosis algorithms from different perspectives including the data preparation for all datasets and the whole evaluation framework with different input formats, normalization methods, data split ways, augmentation methods, and DL-based models.
Based on the benchmark study, we highlight some evaluation results which are very important for comparing or testing new models.
First, not all datasets are suitable for comparing the classification effectiveness of the proposed methods since basic models can achieve very high accuracy on these datasets, like CWRU and XJTU-SY.
Second, the frequency domain input can achieve the highest accuracy in all datasets, so researchers should first try to use the frequency domain as the input.
Third, it is not necessary for CNN models to get the best results in all cases, and we also should consider the overfitting problem.
Fourth, when the accuracy of datasets is not very high, data augmentation methods improve the performance of models, especially for the time domain input.
Thus, more effective data augmentation methods need to be investigated.
Finally, in some cases, it may be more suitable for splitting the datasets according to time sequences (order split) since the random split may provide a virtually high accuracy.
We also release a code library to evaluate DL-based intelligent diagnosis algorithms and provide the benchmark accuracy (a lower bound) to avoid useless improvement.
Meanwhile, we use specific-designed cases to discuss existing issues, including class imbalance, generalization ability, interpretability, few-shot learning, and model selection.
Through these works, we aim to allow comparisons fairer and quicker, emphasize the importance of open source codes, and provide deep discussions of existing issues.
To the best of our knowledge, this is the first work to comprehensively perform the benchmark study and release the code library to the public.
The code library is available at: \url{https://github.com/ZhaoZhibin/DL-based-Intelligent-Diagnosis-Benchmark}.

The main contributions of this paper can be summarized as three aspects:
\begin{enumerate}[1)]
	\item \textbf{\textit{Various datasets and data preparing}}. We gather nine publicly available datasets and give a detailed discussion about its adaptability.
	For data preparation, we first discuss different kinds of input formats and normalization methods, and then we perform different data augmentation approaches to clarify that they have not been fully investigated.
	We also discuss the way of data split and state that it may be more appropriate to split data according to time sequences.
	\item \textbf{\textit{Benchmark accuracy and existing issues}}. We evaluate various DL-based intelligent diagnosis algorithms for seven datasets and provide the benchmark accuracy to make future studies more comparable.
	Based on the benchmark study, we highlight some evaluation results which are very important for comparing or testing new models.
	We also discuss the existing issues, including class imbalance, generalization ability, interpretability, few-shot learning, and model selection.
	\item \textbf{\textit{Open source codes}}. To emphasize the reproducibility of DL-based intelligent diagnosis algorithms, we release the code library for the better comparisons.
	Meanwhile, it is a unified intelligent diagnosis library, which retains an extended interface for everyone to load their own datasets and models to carry out new studies.
\end{enumerate}

The outlines of this paper are listed as follows:
In Section \ref{S:2}, we give a brief review of the recent development of DL-based intelligent diagnosis algorithms.
Then, Sections \ref{S:3} to \ref{S:9} discuss the evaluation algorithms, datasets, data preprocessing, data augmentation, data split, evaluation methodologies, and evaluation results, respectively. After that, Section \ref{S:10} makes some further discussions and the results, followed by conclusions in Section \ref{S:11}.

\section{Brief Review}
\label{S:2}
Recently, DL has become a promising method in a large scope of fields, and a huge amount of papers related to DL have been published since 2012.
This paper mainly focuses on a benchmark study of intelligent diagnosis, rather than providing a comprehensive review on DL for other fields.
Some famous DL researchers have published more professional references, and readers could refer to \cite{lecun2015deep,goodfellow2016deep}.

\begin{figure}[!t]
	\centering
	\subfigure{\includegraphics[scale = 0.55]{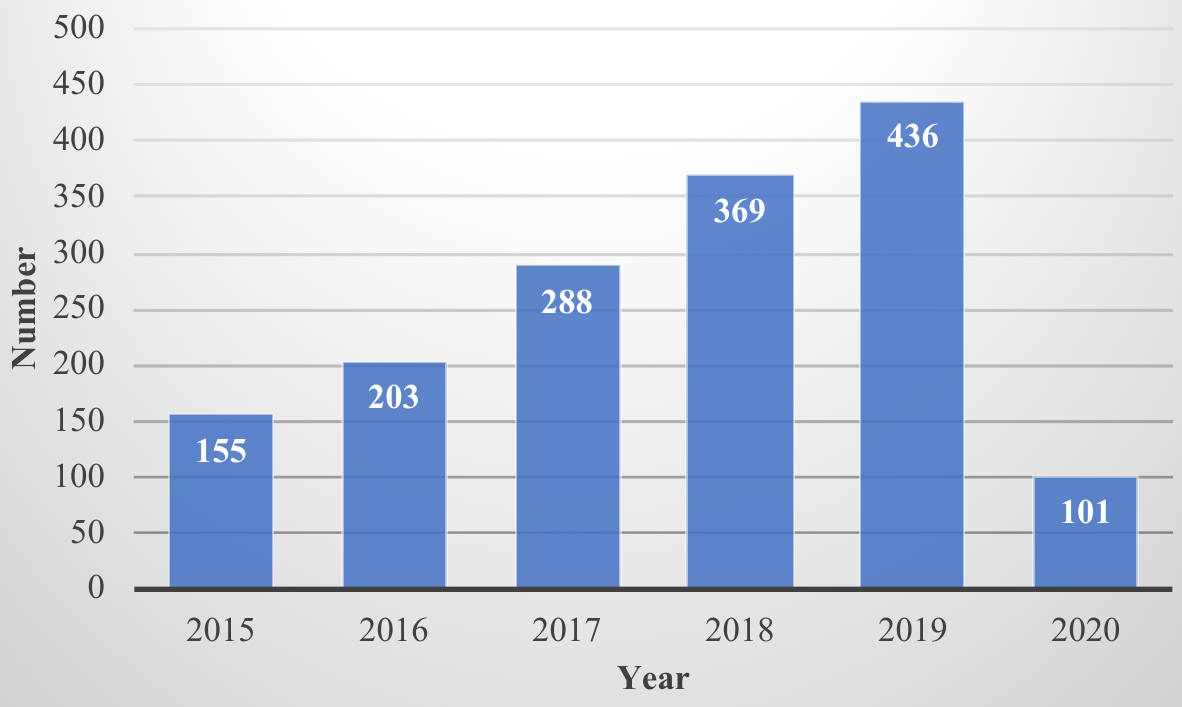}}
	\\ [-5pt]
	\caption{The relationship between the number of published papers and publication years covering the last six years (as of April 2020). The basic descriptor is “TI= ((deep OR autoencoder OR convolutional network* OR neural network*) AND (fault OR condition monitoring OR health management OR intelligent diagnosis))”.}
	\label{Fig1}
\end{figure}

Due to the efforts of many researchers, DL has become one of the most popular data-driven methods to perform intelligent diagnosis.
In general, DL-based methods can extract representative features adaptively without any manual intervention and can achieve a higher accuracy than traditional machine learning algorithms in most of the tasks when the dataset is large enough.
We conducted a literature search using Web of Science with a database called web of science core collection.
As shown in \figref{Fig1}, we can observe that the number of published papers related to DL-based intelligent algorithms increases year by year.

Another interesting observation is that many review papers on this topic have been published in the recent four years.
Therefore, in this paper, we first summarize the main contents of different review papers, allowing readers who just enter this field to find suitable review papers quickly.

In bearing fault diagnosis, Li et al. \cite{li2018systematic} provided a systematic review of fuzzy formalisms including combination with other machine learning algorithms.
Hoang et al. \cite{hoang2019survey} provided a comprehensive review of three popular DL algorithms (AE, DBN, and CNN) for bearing fault diagnosis.
Zhang et al. \cite{zhang2020deep} systematically reviewed the machine learning and DL-based algorithms for bearing fault diagnosis and also provided a comparison of the classification accuracy of CWRU with different DL-based methods.
Hamadache et al. \cite{hamadache2019comprehensive} reviewed different fault modes of rolling element bearings and described various health indexes for PHM.
Meanwhile, it also provided a survey of artificial intelligence (AI) methods for PHM including shallow learning and DL.

In rotating machinery intelligent diagnosis, Ali et al. \cite{ali2016acoustic} provided a review of AI-based methods using acoustic emission data for rotating machinery condition monitoring.
Liu et al. \cite{liu2018artificial} reviewed Al-based approaches including k-nearest neighbors (KNN), support vector machine (SVM), artificial neural networks (ANN), Naive Bayes, and DL for fault diagnosis of rotating machinery.
Wei et al. \cite{wei2019review} summarized early fault diagnosis of gears, bearings, and rotors through signal processing methods (adaptive decomposition methods, wavelet transform, and sparse decomposition) and AI-based methods (KNN, neural network, and SVM).

In machinery condition monitoring, Zhao et al. \cite{zhao2016research} and Duan et al. \cite{duan2018deep} reviewed diagnosis and prognosis of mechanical equipment based on DL algorithms such as DBN and CNN.
Zhang et al. \cite{zhang2017comprehensive} reviewed computational intelligent approaches including ANN, evolutionary algorithms, fuzzy logic, and SVM for machinery fault diagnosis.
Zhao et al. \cite{zhao2019deep} reviewed data-driven machine health monitoring through DL methods (AE, DBN, CNN, and RNN) and provided the data and codes (in Keras) about an experimental study.
Lei et al. \cite{lei2020applications} presented a systematical review to cover the development of intelligent diagnosis following the progress of machine learning and DL models and offer a future perspective called transfer learning theories.

In addition, Nasiri et al. \cite{nasiri2017fracture} surveyed the state-of-the-art AI-based approaches for fracture mechanics and provided the accuracy comparisons achieved by different machine learning algorithms for mechanical fault detection.
Tian et al. \cite{tian2018review} surveyed different modes of traction induction motor fault and their diagnosis algorithms including model-based methods and AI-based methods.
Khan et al. \cite{khan2018review} provided a comprehensive review of AI for system health management and emphasized the trend of DL-based methods with limitations and benefits.
Stetco et al. \cite{stetco2018machine} reviewed machine learning approaches applied to wind turbine condition monitoring and made a discussion of the possibility for future research.
Ellefsen et al. \cite{ellefsen2019comprehensive} reviewed four well-established DL algorithms including AE, CNN, DBN, and long short-term memory network (LSTM) for PHM applications and discussed challenges for the future studies, especially in the field of PHM in autonomous ships.
AI-based algorithms (traditional machine learning algorithms and DL-based approaches) and applications (smart sensors, intelligent manufacturing, PHM, and cyber-physical systems) were reviewed in \cite{ademujimi2017review, chang2018review, wang2018deep, sharp2018survey} for smart manufacturing and manufacturing diagnosis.

Due to the fact that there are already many review papers covering DL-based rotating machinery intelligent diagnosis published before 2020, we further review most of related papers published in 2020 and summarize their main contributions to fill the void.

In AE models, for model improvement, AE models were combined with some other data preprocessing methods, such as singular value decomposition \cite{ISI:000506605800013} and nonlinear frequency spectrum \cite{ISI:000500942200115}.
The ensemble learning strategy was also used to boost the performance of AE models in \cite{ISI:000500942200149, ISI:000500942200073}.
Meanwhile, the semi-supervised learning methods were also embedded into AE models by \cite{ISI:000513324100001, ISI:000518910900003}.
For imbalanced learning, generation adversarial network (GAN) was used to combine with AE models to generate new labeled samples in \cite{ISI:000519000400001, ISI:000501653900023, ISI:000510903200058}.
In \cite{ISI:000508908600056}, a model called deep Laplacian AE (DLapAE) was proposed by introducing the Laplacian regularization to improve the generalization performance.
For transfer learning, the pretrained and fine-tuned approach was applied to AE models to realize the knowledge transfer in \cite{ISI:000508908600107,ISI:000517663200035}.
Domain adaptation was also used to transfer the knowledge learned by AE models to the target domain in \cite{ISI:000513850100020, ISI:000518412200123}.

In CNN models, for model improvement, different input types, such as time-frequency images \cite{ISI:000521728800006}, vibration spectrum images \cite{ISI:000513183100012}, infrared thermal images \cite{ISI:000518910900004}, and two-dimensional images \cite{ISI:000518910900005}, were used as the inputs of CNN models.
Multiple wavelet regularizations \cite{ISI:000508908600067}, data augmentation methods \cite{ISI:000519899300012}, and information fusion technology \cite{ISI:000509205900001} were also applied to improve the performance of CNN models.
Hand-crafted features were combined with CNN features to boost the performance in \cite{ISI:000519983300041}.
For imbalanced learning, GAN was also used to combine with CNN models to generate new labeled samples in \cite{ISI:000510066300004, ISI:000508908600098}.
Focal loss, which can deal with severe imbalanced problems, was used by \cite{li2020adaptive} to allow CNN models to learn discriminative features.
For transfer learning, the pretrained and fine-tuned approach was used to leverage the prior knowledge from the source task in \cite{ISI:000506605800019, ISI:000508908600019,ISI:000508428900032}.
Domain adaptation methods were also applied to allow CNN models to learn transferable features in \cite{ISI:000519983300020, jiao2019unsupervised}.
In addition, layer-wise relevance propagation was also used to understand how CNN models learn to distinguish different patterns \cite{ISI:000520089000040}.

Beyond that, complex wavelet packet energy moment entropy \cite{ISI:000513295000018} and the grey wolf optimizer algorithm \cite{ISI:000503013100005} were combined with an enhanced deep gated recurrent unit to improve the security of rotating machinery.
Joint distribution adaptation was embedded into LSTM to realize learning transferable features in \cite{ISI:000500942200145}.
DBN models were also modified in \cite{ISI:000514158700006,ISI:000499374500001,ISI:000506068300015} to improve the diagnosis performance of rotating machinery.
Deep reinforcement learning was also used in intelligent fault diagnosis for rotating machinery in \cite{ding2019intelligent, dai2020fault}.
Meanwhile, a deep graph convolutional network (DGCN) was first applied to rolling bearing fault diagnosis based on acoustic signals \cite{ISI:000519983300033}.

Although a large body of DL-based methods and many related reviews have been published in the field of intelligent diagnosis, few studies thoroughly evaluate various DL-based intelligent diagnosis algorithms for most of the publicly available datasets, provide the benchmark accuracy, and release the code library for complete evaluation procedures.
For example, a simple code written in Keras was published in \cite{zhao2019deep}, which is not comprehensive enough for different datasets and models.
The accuracy comparisons were provided in \cite{zhang2020deep,nasiri2017fracture} according to existing papers, but they were not comprehensive enough due to different configurations.
Therefore, this paper is intended to make up for this gap and emphasize the importance of open source codes and the benchmark study.

\section{Evaluation Algorithms}
\label{S:3}
It is impossible to cover all the published models since there is currently no open source community in this field.
Therefore, we switch to test the performance of four models (including MLP, AE, CNN, and RNN) via embedding some advanced techniques.
It should be noted that DBN is also another commonly used DL model in intelligent diagnosis, but we do not add it to this code library due to that the fact the training way of DBN is much different from those four models.

\subsection{MLP}
\begin{figure}[!t]
	\centering
	\subfigure{\includegraphics[scale = 0.7]{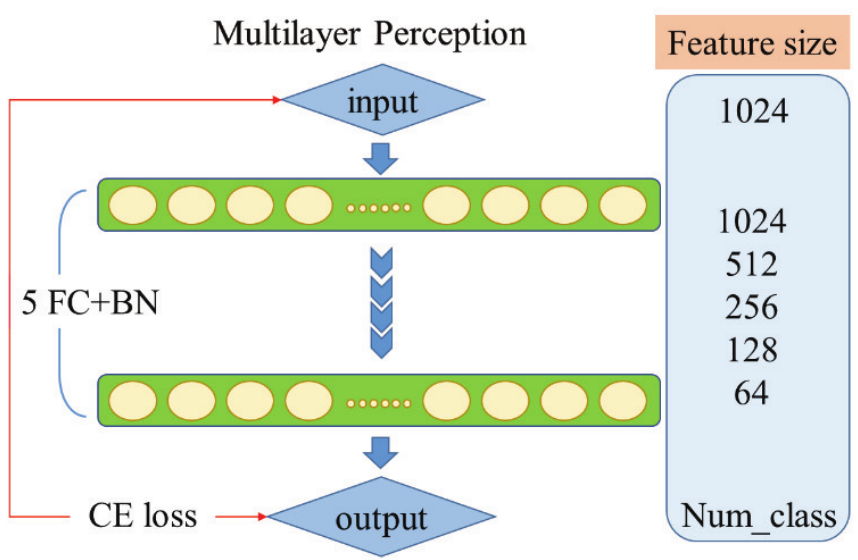}}
	\\ [-5pt]
	\caption{The structure of MLP.}
	\label{MLP}
\end{figure}

MLP \cite{rumelhart1985learning}, which was a fully connected network with multiple hidden layers, was proposed in 1987 as the prototype of ANN.
With such a simple structure, MLP can complete some easy classification tasks such as MNIST.
However, as the task becomes more complex, it would be hard to train MLP because of the huge amount of parameters.
MLP with five fully connected layers and five batch normalization layers is used in this paper for one dimension (1D) input data.
The structure and parameters of this model are shown in \figref{MLP}.
Besides, in \figref{MLP}, FC means the fully connected layer, BN means the Batch Normalization layer, and CE loss means the softmax cross-entropy loss.

\subsection{AE}
AE was first proposed in 2006, as a method for dimensionality reduction.
It can reduce the dimensionality of the input data while retaining most of the information.
AE consists of an encoder and a decoder, which tries to reconstruct the input from the output of an encoder, and the reconstruction error is used as a loss function. 
In detail, an encoder takes $ x $ as an input and transforms it into a hidden representaion $ h $ which can be formualted as:
\begin{align}
\label{eq:ae1}
h = \phi(W \cdot x+b )
\end{align}
$\phi(\cdot )$ denotes the nonlinear activation function (in this paper, we use ReLU), and $ W $ and $ b $ represent the weight and bias required to learn, respectively.
After that, a decoder generates the output $ x' $ from the hidden representaion $ h $, which can be formulated as:
\begin{align}
\label{eq:ae2}
x' = \phi(W' \cdot h+b')
\end{align}
where $ W' $ and $ b' $ represent the weight and bias required to learn, respectively.
For traditional AE, the mean squared error (MSE) loss is often used as a loss function, shown as follows:
\begin{align}
\label{eq:ae3}
L_{\text{MSE}} = 
\dfrac{1}{N} \sum\limits_{i = 1}^N || x_i - x'_i ||^2_2
\end{align}
where $ x_i $ and $ x'_i $ are the $ i $-th sample and its approximation, respectively, and $ N $ is the number of data samples.
In practice, we often stack multiple layers in both encoder and decoder to produce better learning results.

\begin{figure}[!t]
	\centering
	\subfigure{\includegraphics[scale = 1]{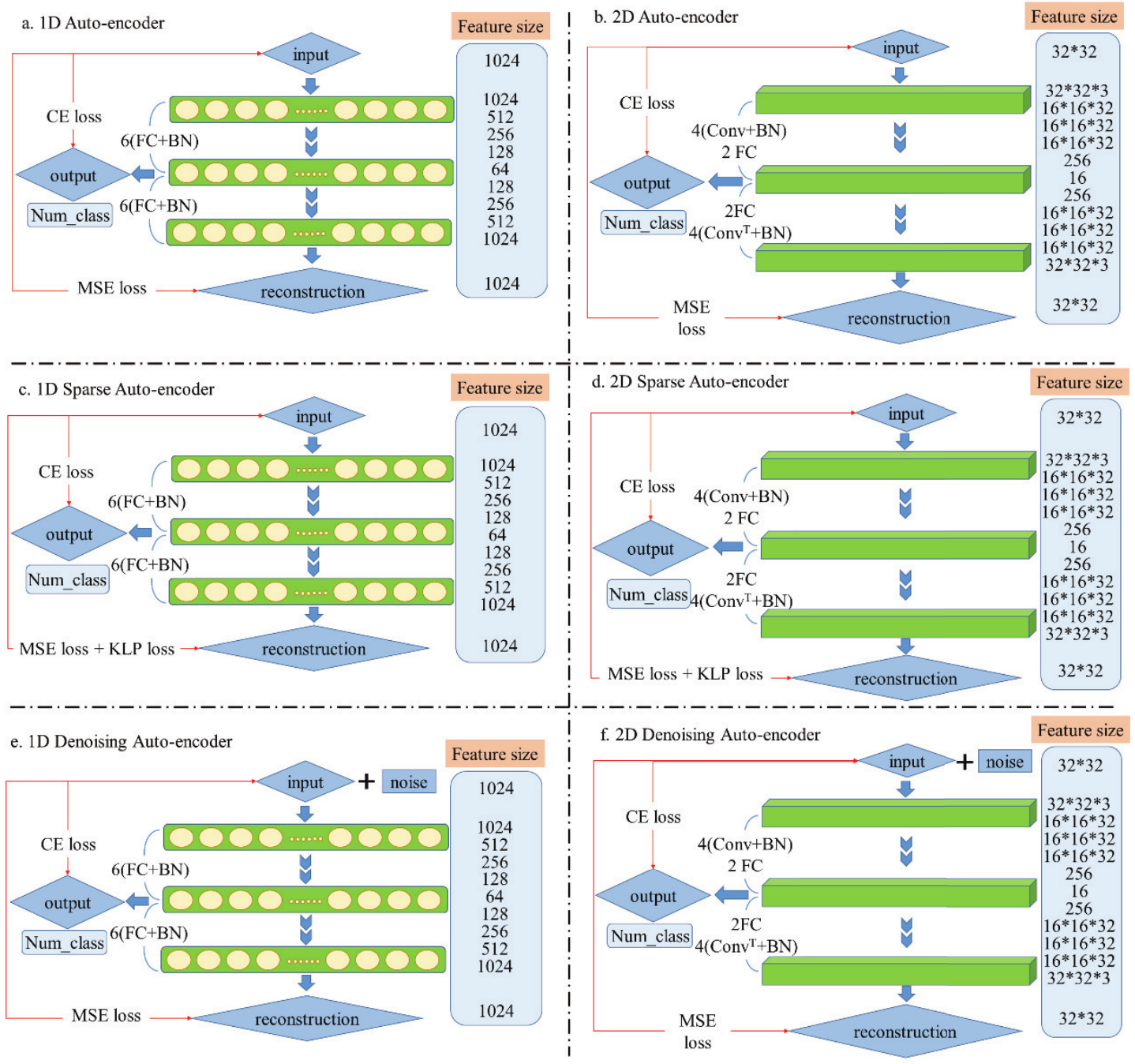}}
	\\ [-5pt]
	\caption{The structure of AE and its derivatives}
	\label{AE}
\end{figure}

Subsequently, various derivatives of AE were proposed by different researchers, such as denoising auto-encoder (DAE) \cite{vincent2008extracting} and sparse auto-encoder (SAE) \cite{ranzato2007efficient}.
In this paper, we design a deep AE and its derivatives for 1D input data (using the fully connected layer) and two dimension (2D) input data (using the convolutional layer), respectively.
Considering different features of neural networks, the structures and hyper-parameters of them shown in \figref{AE} change adaptively.
Specifically, the network structures of DAE and SAE are the same with AE, and their differences lie in the loss function and inputs.
DAE takes an input corrupted with noise and is trained to reconstruct the clean version of the input; SAE uses a MSE loss regularized with a sparsity constraint (the Kullback-Leibler divergence is often used) to train the AE model.
During the training of AE and its derivatives, the encoder and decoder are trained jointly to get the low-dimension features of data.
After that, the encoder and classifier are trained jointly using the softmax cross-entropy loss for the classification task.
The details of AE and its derivatives are shown in \figref{AE}.
In \figref{AE}, the MSE loss means the mean square error loss defined in \eqref{eq:ae3}, Conv means the convolutional layer, $ \text{Conv}^\text{T} $ means the transposed convolutional (e.g. inverse convolution) layer, and the KLP loss means the Kullback-Leibler divergence loss.

\subsection{CNN}
\begin{figure}[!t]
	\centering
	\subfigure{\includegraphics[scale = 1]{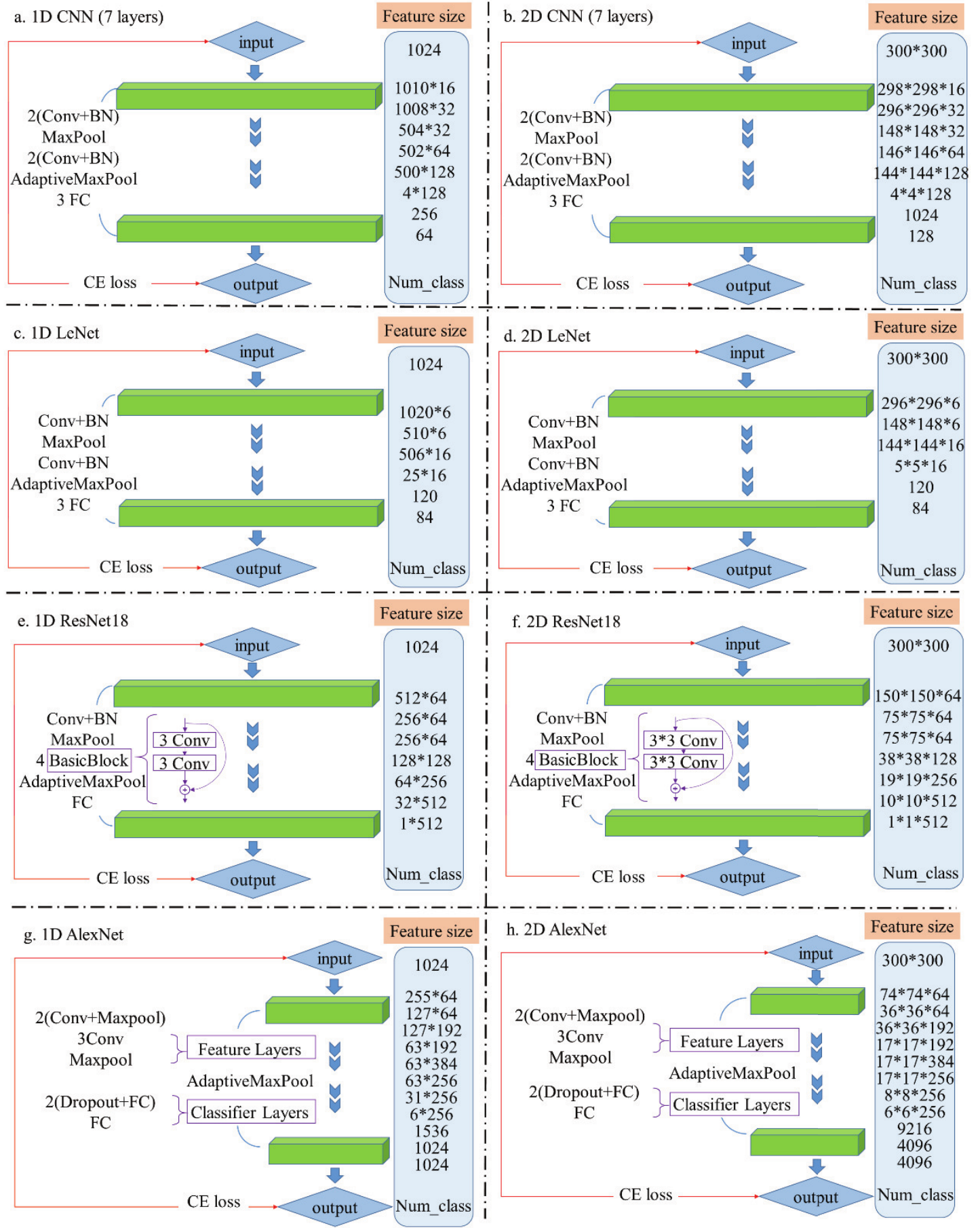}}
	\\ [-5pt]
	\caption{The structure of CNN and its derivatives}
	\label{CNN}
\end{figure}

CNN \cite{lecun1995convolutional} was first proposed in 1997 and the proposed network was also called LeNet.
CNN is a specialized kind of neural network for processing data that has a known grid-like topology.
Sparse interactions, parameter sharing, and equivalent representations are realized with convolution and pooling operations on CNN.
In 2012, AlexNet \cite{krizhevsky2012imagenet} won the title in the ImageNet competition by far surpassing the second place, and CNN has attracted wide attention.
Besides, in 2016, ResNet \cite{he2016deep} was proposed and its classification accuracy exceeded the human baseline.
In this paper, we design 5 layers 1D CNN and 2D CNN for 1D input data and 2D input data, respectively, and also adapt three well known CNN models (LeNet, ResNet18, and AlexNet) for two types of the input data.
The details of CNN and its derivatives are shown in \figref{CNN}.
In \figref{CNN}, MaxPool means the Max Pooling layer, AdaptiveMaxPool means the Adaptive Max Pooling layer, and Dropout means the Dropout layer.

As shown in \figref{CNN}, CNN mainly contains three kinds of layers, including the convolutional layer, the maxpooling layer, and the classifier layer.
Convolutional and maxpooling layers are utilized to perform feature learning, and the classifier layer classifies learned features into different classes.
For an input $ x $, the convolutional layer can be defined as a multiplication with a filter kernel $ w $, and the final feature map after the nonlinear activation can be formulated as:
\begin{align}
\label{eq:cnn1}
h_k^l = \phi(w_k^l*x+b_k^l )
\end{align}
where $ * $ denotes the convolution operator.
$ h_k^l$, $ w_k^l $, and $ b_k^l $ represent the obtained feature map, the weight and bias of $ k $-th convolutional kernel of $ l $-th layer, respectively. 

The maxpooling layer is set behind the convolution layer to extract the most significant local information in each feature map and to reduce the dimension of obtained features.
The maxpooling layer can be defined as:
\begin{align}
\label{eq:cnn2}
z_k^l = \text{down}(h_k^l;\;s )
\end{align}
where $\text{down}(\cdot )$ denotes the down-sampling function of the maxpooling layer, $ z_k^l $ is the output feature map of the maxpooling layer, and $ s $ is the pooling size.

After a number of stacked convolutional and pooling layers, the extracted high-level features of input data are input into the classifier layer.
In this paper, a fully connected layer is used to map features into different classes.

\subsection{RNN}
\begin{figure}[!t]
	\centering
	\subfigure{\includegraphics[scale = 1]{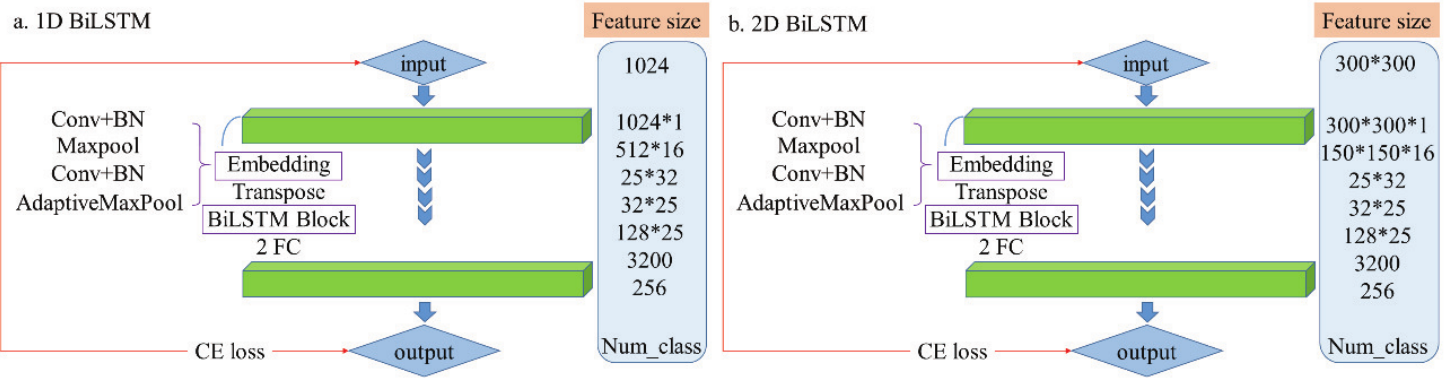}}
	\\ [-5pt]
	\caption{The structure of 1D BiLSTM and 2D BiLSTM}
	\label{LSTM}
\end{figure}

RNN can describe the temporal dynamic behavior and is very suitable for dealing with time series.
However, RNN often exists the gradient vanishing and exploding problems during the training procedure.
To overcome these problems, LSTM was proposed in 1997 \cite{hochreiter1997long} for processing continual input streams and has made great success in various fields.
Bi-directional LSTM (BiLSTM) can capture bidirectional dependencies over long distances and learn to remember and forget information selectively.
We utilize BiLSTM as the representation of RNN to deal with two types of input data (1D and 2D) for the classification task.
The details of 1D BiLSTM and 2D BiLSTM are shown in \figref{LSTM}.
Besides, in \figref{LSTM}, Transpose means transposing the channel and feature dimensions of the input data, and BiLSTM Block means the BiLSTM layer.

\begin{figure}[!t]
	\centering
	\subfigure{\includegraphics[scale = 0.6]{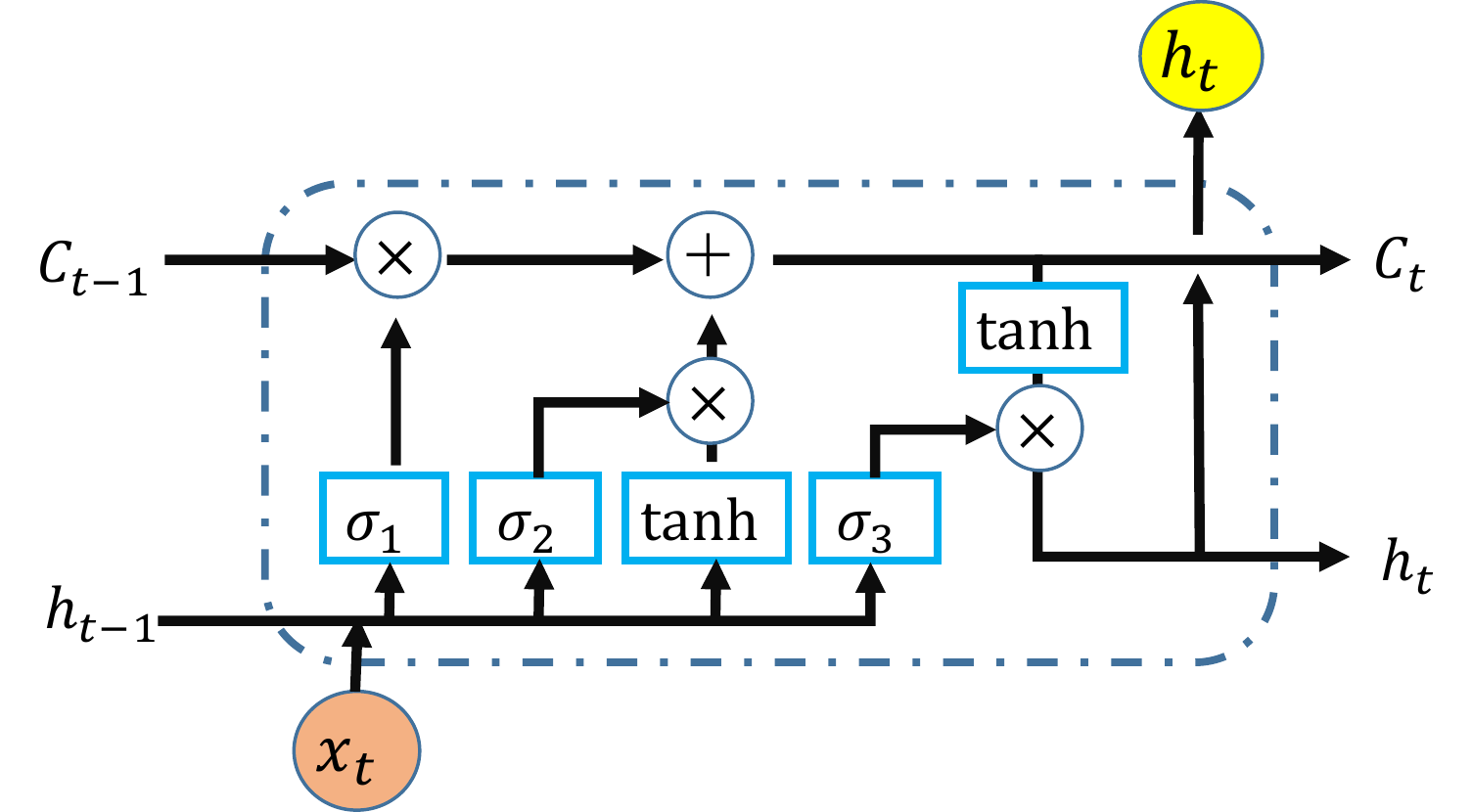}}
	\\ [-5pt]
	\caption{The structure of a LSTM cell}
	\label{LSTMcell}
\end{figure}

The structure of a LSTM cell is shown in \figref{LSTMcell}, including the forget gate layer $ \sigma_1 $, the input gate layer $ \sigma_2 $, the output gate layer $ \sigma_3 $, and the tanh layer. 
Firstly, the hidden state of a last cell $ h_{t-1} $ and the current input $ x_{t} $ are fed into the forget gate layer to decide whether we should forget the last cell state $ C_{t-1} $.
Secondly, $ h_{t-1} $ and $ x_{t} $ are fed into the input gate layer and the tanh layer to decide values we want to update.
Thirdly, $ h_{t-1} $ and $ x_{t} $ are fed into the output layer to decide what we should export for the last cell.
Based on the structure shown in \figref{LSTMcell}, the output cell state $ C_{t} $ can be calculated as follows:
\begin{align}
\label{eq:lstm2}
C_{t}  = \sigma_1(h_{t-1}, x_{t}) \otimes C_{t-1} \oplus \sigma_2(h_{t-1}, x_{t}) \otimes \text{tanh}(h_{t-1}, x_{t})
\end{align}
where $ \otimes $ and $ \oplus $ denote the element-wise multiplication and addition, respectively.
Meanwhile, $ \sigma_i(h_{t-1}, x_{t}) $ is defined as follows:
\begin{align}
\label{eq:lstm3}
\sigma_i(h_{t-1}, x_{t}) = \sigma(W_i\cdot[h_{t-1}, x_{t}] + b_i), \quad i=1,2,3
\end{align}
where $ W_i $ and $ b_i $ represent the weight and bias, respectively. $ \sigma(\cdot) $ denotes the sigmoid function.
Similarly, the tanh layer replaces the sigmoid function with the tanh function.
In addition, the output hidden state $ h_{t} $ can be calculate as follows:
\begin{align}
\label{eq:lstm5}
h_{t} = \sigma_3(h_{t-1}, x_{t})\otimes \text{tanh}(C_{t})
\end{align}

Many repeating cells are linked together to form a LSTM block, designed for capturing both long-term and short-term dependencies.
The BiLSTM layer is the combining of forward and backward LSTM blocks, in which information is bidirectionally transmitted.
For each input, information of the whole time series can be used simultaneously.

\section{Datasets}
\label{S:4}
In the field of intelligent diagnosis, publicly available datasets have not been investigated in depth.
Actually, for comprehensive performance comparisons, it is important to gather different kinds of representative datasets.
We collected nine commonly used datasets which all have specific labels and explanations in addition to the PHM 2012 bearing dataset and Intelligent Maintenance Systems (IMS) bearing dataset, so PHM 2012 and IMS are not suitable for fault classification that requires labels.
To sum up, this paper uses seven datasets to verify the performance of models introduced in Section \ref{S:3}.
The description of all these datasets is listed as follows.

\subsection{CWRU Bearing Dataset}
Case Western Reserve University (CWRU) datasets were provided by the Case Western Reserve University Bearing Data Center \cite{CWRU}.
Vibration signals were collected at 12 kHz or 48 kHz for normal bearings and damaged bearings with single-point defects under four different motor loads.
Within each working condition, single-point faults were introduced with fault diameters of 0.007, 0.014, and 0.021 inches on the rolling element, the inner ring, and the outer ring, respectively.
In this paper, we used the data collected from the drive end, and the sampling frequency was equivalent to 12 kHz.
In Table \ref{Tab_1}, one healthy bearing and three fault modes, including the inner ring fault, the rolling element fault, and the outer ring fault, were classified into ten categories (one health state and 9 fault states) according to different fault sizes.

\begin{table}[!t]
	\caption{Detailed description of CWRU datasets} 
	\centering
	\label{Tab_1}
	\begin{tabular}{| c | c |}
		\hline                 
		Fault Mode & Description\\
		\hline                
		Health State & the normal bearing at 1791 rpm and 0 HP\\
		\hline                
		Inner ring 1 & 0.007 inch inner ring fault at 1797 rpm and 0 HP\\
		\hline
		Inner ring 2 & 0.014 inch inner ring fault at 1797 rpm and 0 HP\\
		\hline                 
		Inner ring 3 & 0.021 inch inner ring fault at 1797 rpm and 0 HP\\
		\hline                
		Rolling Element 1 & 0.007 inch rolling element fault at 1797 rpm and 0 HP\\
		\hline                
		Rolling Element 2 & 0.014 inch rolling element fault at 1797 rpm and 0 HP\\
		\hline
		Rolling Element 3 & 0.021 inch rolling element fault at 1797 rpm and 0 HP\\
		\hline               
		Outer ring 1 & 0.007 inch outer ring fault at 1797rpm and 0 HP\\
		\hline                
		Outer ring 2 & 0.014 inch outer ring fault at 1797rpm and 0 HP\\
		\hline
		Outer ring 3 & 0.021 inch outer ring fault at 1797rpm and 0 HP\\
		\hline
	\end{tabular}
\end{table}

\subsection{MFPT Bearing Dataset}
Machinery Failure Prevention Technology (MFPT) datasets were provided by Society for Machinery Failure Prevention Technology \cite{MFPT}.
MFPT datasets consisted of three bearing datasets:
1) a baseline dataset sampled at 97656 Hz for six seconds in each file;
2) seven outer ring fault datasets sampled at 48828 Hz for three seconds in each file;
3) seven inner ring fault datasets sampled at 48828 Hz for three seconds in each file;
4) some other datasets which were not used in this paper (more detailed information can be referred to the website of MFPT datasets \cite{MFPT}).
In Table \ref{Tab_2}, one health state bearing and two fault bearings including the inner ring fault and the rolling element fault were classified into 15 categories (one health state and 14 fault states) according to different loads.

\begin{table}[!t]
	\caption{Detailed description of MFPT datasets} 
	\centering
	\label{Tab_2}  
	\begin{tabular}{| c | c |}
		\hline                 
		Fault Mode & Description\\
		\hline                
		Health State & Fault-free bearing working at 270 lbs\\
		\hline                
		Outer ring 1 & Outer ring fault bearing working at 25 lbs\\
		\hline
		Outer ring 2 & Outer ring fault bearing working at 50 lbs\\
		\hline                 
		Outer ring 3 & Outer ring fault bearing working at 100 lbs\\
		\hline                
		Outer ring 4 & Outer ring fault bearing working at 150 lbs\\
		\hline                
		Outer ring 5 & Outer ring fault bearing working at 200 lbs\\
		\hline
		Outer ring 6 & Outer ring fault bearing working at 250 lbs\\
		\hline               
		Outer ring 7 & Outer ring fault bearing working at 300 lbs\\
		\hline                
		Outer ring 1 & Inner ring fault bearing working at 0 lbs\\
		\hline
		Inner ring 2 & Inner ring fault bearing working at 50 lbs\\
		\hline                 
		Inner ring 3 & Inner ring fault bearing working at 100 lbs\\
		\hline                
		Inner ring 4 & Inner ring fault bearing working at 150 lbs\\
		\hline                
		Inner ring 5 & Inner ring fault bearing working at 200 lbs\\
		\hline
		Inner ring 6 &Inner ring fault bearing working at 250 lbs\\
		\hline               
		Inner ring 7 & Inner ring fault bearing working at 300 lbs\\
		\hline   
	\end{tabular}
\end{table}

\subsection{PU Bearing Dataset}
Paderborn University (PU) datasets were provided by the Paderborn University Bearing Data Center \cite{PU,lessmeier2016condition}, and PU datasets consisted of 32 sets of current signals and vibration signals.
As shown in Table \ref{Tab_3}, bearings were divided into:
1) six undamaged bearings;
2) twelve artificially damaged bearings;
3) fourteen bearings with real damages caused by accelerated lifetime tests.
Each dataset was collected under four working conditions as shown in Table \ref{Tab_PU}.

\begin{table}[!t]
	\caption{Detailed description of PU datasets (S: single damage; R: repetitive damage; M: multiple damage)} 
	\centering
	\label{Tab_3}  
	\begin{tabular}{|m{0.07\textwidth}<{\centering}|m{0.18\textwidth}<{\centering}|m{0.14\textwidth}<{\centering}|m{0.07\textwidth}<{\centering}|m{0.18\textwidth}<{\centering}|m{0.14\textwidth}<{\centering}|}
		\hline                 
		Bearing Code & Fault Mode & Description & Bearing Code & Fault Mode & Description\\
		\hline                
		K001 & Health state & Run-in 50 h before test &KI07 & Artificial inner ring fault (Level 2)&Made by electric engraver\\
		\hline                
		K002 & Health state &Run-in 19 h before test &KI08 & Artificial inner ring fault (Level 2) &Made by electric engraver\\
		\hline
		K003 & Health state & Run-in 1 h before test &KA04 & Outer ring damage (single point + S + Level 1) &Caused by fatigue and pitting\\
		\hline                 
		K004 & Health state & Run-in 5 h before test &KA15 & Outer ring damage (single point + S + Level 1) &Caused by plastic deform and  indentation\\
		\hline                
		K005 & Health state & Run-in 10 h before test &KA16 & Outer ring damage (single point + R + Level 2) &Caused by fatigue and pitting\\
		\hline                
		K006 & Health state & Run-in 16 h before test &KA22 & Outer ring damage (single point + S + Level 1)&Caused by fatigue and pitting\\
		\hline
		KA01 & Artificial outer ring fault (Level 1)  & Made by EDM &KA30 & Outer ring damage (distributed + R + Level 1) &Caused by plastic deform and indentation\\
		\hline               
		KA03 & Artificial outer ring fault (Level 2) & Made by electric engraver&KB23 & Outer ring and inner ring damage (single point + M + Level 2) &Caused by fatigue and pitting\\
		\hline                
		KA05 & Artificial outer ring fault (Level 1) & Made by electric engraver&KB24 & Outer ring and inner ring damage (distributed + M + Level 3) &Caused by fatigue and pitting\\
		\hline
		KA06 & Artificial outer ring fault (Level 2) &Made by electric engraver &KB27 & Outer ring and inner ring damage (distributed + M + Level 1) &Caused by plastic deform and  indentation\\
		\hline                 
		KA07 & Artificial outer ring fault (Level 1) & Made by drilling&KI04 & Inner ring damage (single point + M + Level 1) &Caused by fatigue and pitting\\
		\hline                
		KA08 & Artificial outer ring fault (Level 2) & Made by drilling&KI14 & Inner ring damage (single point + M + Level 1) &Caused by fatigue and pitting\\
		\hline                
		KA09 & Artificial outer ring fault (Level 2) & Made by drilling&KI16 & Inner ring damage (single point + S + Level 3) &Caused by fatigue and pitting\\
		\hline
		KI01 & Artificial inner ring fault (Level 1)  &Made by EDM &KI17 & Inner ring damage (single point + R + Level 1) &Caused by fatigue and pitting\\
		\hline               
		KI03 & Artificial inner ring fault (Level 1)  & Made by electric engraver &KI18 & Inner ring damage (single point + S + Level 2) &Caused by fatigue and pitting\\
		\hline   
		KI05 & Artificial inner ring fault (Level 1)  & Made by electric engraver &KI21 & Inner ring damage (single point + S + Level 1) &Caused by fatigue and pitting\\
		\hline   
	\end{tabular}
\end{table}

\begin{table}[!t]
	\caption{Four working conditions of PU datasets} 
	\centering
	\label{Tab_PU}  
	\begin{tabular}{| c | c | c | c | c |}
		\hline                 
		No. & Rotating speed (rpm) & Load torque (Nm) & Radial force (N) & Name of setting\\
		\hline                
		0 & 1500 & 0.7 & 1000 & N15\_M07\_F10\\
		\hline                
		1 & 900 & 0.7 & 1000 & N09\_M07\_F10\\
		\hline
		2 & 1500 & 0.1 & 1000 & N15\_M01\_F10\\
		\hline                 
		3 & 1500 & 0.7 & 400 & N15\_M07\_F04\\
		\hline                            
	\end{tabular}
\end{table}

In this paper, since using all the data would cause huge computational time, we only used the data collected from real damaged bearings (including KA04, KA15, KA16, KA22, KA30, KB23, KB24, KB27, KI14, KI16, KI17, KI18, and KI22) under the working condition N15\_M07\_F10 to carry out the performance verification.
Since KI04 was the same as KI14 completely shown in Table \ref{Tab_3}, we deleted KI04 and the total number of classes was thirteen.
Besides, only vibration signals were used for testing the models.

\subsection{UoC Gear Fault Dataset}
University of Connecticut (UoC) gear fault datasets were provided by the University of Connecticut \cite{UoC}, and UoC datasets were collected at 20 kHz.
In this dataset, nine different gear conditions were introduced to the pinions on the input shaft, including healthy condition, missing tooth, root crack, spalling, and chipping tip with 5 different levels of severity.
All the collected datasets were used and classified into nine categories (one health state and eight fault states) to test the performance.

\subsection{XJTU-SY Bearing Dataset}
XJTU-SY bearing datasets were provided by the Institute of Design Science and Basic Component at Xi’an Jiaotong University and the Changxing Sumyoung Technology Co. \cite{XJTU,wang2018hybrid}.
XJTU-SY datasets consisted of fifteen bearings run-to-failure data under three different working conditions.
The dataset was collected at 2.56 kHz.
A total of 32768 data points were recorded, and the sampling period was equal to one minute.
The details of bearing lifetime and fault elements were shown in Table \ref{Tab_4}.
In this paper, we used all the data described in Table \ref{Tab_5} and the total number of classes was fifteen.
It should be noticed that we used collected data at the end of run-to-failure experiments.

\begin{table}[!t]
	\caption{Detailed description of XJTU-SY datasets} 
	\centering
	\label{Tab_4}  
\begin{tabular}{|c|c|c|c|}
	\hline
	Condition                                                                                                 & File                              & Lifetime                       & Fault element                          \\ \hline
	\multirow{5}{*}{\begin{tabular}[c]{@{}c@{}}Speed: 35 Hz\\ Load: 12 kN\end{tabular}}                       & Bearing 1\_1                      & 2h 3min                        & Outer ring                             \\ \cline{2-4} 
	& Bearing 1\_2                      & 2h 41min                       & Outer ring                             \\ \cline{2-4} 
	& Bearing 1\_3                      & 2h 38min                       & Outer ring                             \\ \cline{2-4} 
	& Bearing 1\_4                      & 2h 2min                        & Cage                                   \\ \cline{2-4} 
	& Bearing 1\_5                      & 52 min                         & Inner ring and Outer ring              \\ \hline
	\multirow{5}{*}{\begin{tabular}[c]{@{}c@{}}Speed: 37.5 Hz\\ Load: 11 kN\end{tabular}}                     & Bearing 2\_1                      & 8h 11min                       & Inner ring                             \\ \cline{2-4} 
	& Bearing 2\_2                      & 2h 41min                       & Outer ring                             \\ \cline{2-4} 
	& Bearing 2\_3                      & 8h 53min                       & Cage                                   \\ \cline{2-4} 
	& Bearing 2\_4                      & 42min                          & Outer ring                            \\ \cline{2-4} 
	& Bearing 2\_5                      & 5h 39min                       & Outer ring                             \\ \hline
	\multicolumn{1}{|c|}{\multirow{5}{*}{\begin{tabular}[c]{@{}c@{}}Speed: 40 Hz\\ Load: 10 kN\end{tabular}}} & \multicolumn{1}{c|}{Bearing 3\_1} & \multicolumn{1}{c|}{42h 18min} & Outer ring                             \\ \cline{2-4} 
	\multicolumn{1}{|c|}{}                                                                                    & \multicolumn{1}{c|}{Bearing 3\_2} & \multicolumn{1}{c|}{41h 36min} & Inner ring, Rolling element, Cage, and Outer ring \\ \cline{2-4} 
	\multicolumn{1}{|c|}{}                                                                                    & \multicolumn{1}{c|}{Bearing 3\_3} & \multicolumn{1}{c|}{6h 11min}  & Inner ring                             \\ \cline{2-4} 
	\multicolumn{1}{|c|}{}                                                                                    & \multicolumn{1}{c|}{Bearing 3\_4} & \multicolumn{1}{c|}{25h 15min} & Inner ring                            \\ \cline{2-4} 
	\multicolumn{1}{|c|}{}                                                                                    & \multicolumn{1}{c|}{Bearing 3\_5} & \multicolumn{1}{c|}{1h 54min}  & Outer ring                             \\ \hline
\end{tabular}
\end{table}

\subsection{SEU Gearbox Dataset}
Southeast University (SEU) gearbox datasets were provided by Southeast University \cite{SEU,shao2018highly}.
SEU datasets contained two sub-datasets, including a bearing dataset and a gear dataset, which were both acquired on drivetrain dynamic simulator (DDS).
There were two kinds of working conditions with rotating speed - load configuration (RS-LC) set to be 20 Hz - 0 V and 30 HZ - 2 V shown in Table \ref{Tab_5}.
The total number of classes was equal to twenty according to Table \ref{Tab_5} under different working conditions.
Within each file, there were eight rows of vibration signals, and we used the second row of vibration signals.

\begin{table}[!t]
	\caption{Detailed description of SEU datasets} 
	\centering
	\label{Tab_5}  
	\begin{tabular}{| c | c | c | c |}
		\hline                 
		Fault Mode & RS-LC  & Fault Mode & RS-LC\\
		\hline                
		Health Gear & 20 Hz - 0 V & Health Bearing & 20 Hz - 0 V\\
		\hline                
		Health Gear & 30 Hz - 2 V & Health Bearing & 30 Hz - 2 V\\
		\hline
		Chipped Tooth & 20 Hz - 0 V & Inner ring & 20 Hz - 0 V\\
		\hline                 
		Chipped Tooth & 30 Hz - 2 V & Inner ring & 30 Hz - 2 V\\
		\hline                
		Missing Tooth & 20 Hz - 0 V & Outer ring & 20 Hz - 0 V\\
		\hline                
		Missing Tooth & 30 Hz - 2 V & Outer ring & 30 Hz - 2 V\\
		\hline
		Root Fault & 20 Hz - 0 V & Inner + Outer ring & 20 Hz - 0 V\\
		\hline                
		Root Fault & 30 Hz - 2 V & Inner + Outer ring & 30 Hz - 2 V\\
		\hline              
		Surface Fault & 20 Hz - 0 V & Rolling Element & 20 Hz - 0 V\\
		\hline                
		Surface Fault & 30 Hz - 2 V & Rolling Element & 30 Hz - 2 V\\
		\hline                    
	\end{tabular}
\end{table}

\subsection{JNU Bearing Dataset}
Jiangnan University (JNU) bearing datasets were provided by Jiangnan University \cite{JNU,li2013sequential}.
JNU datasets consisted of three bearing vibration datasets with different rotating speeds, and the data were collected at 50 kHz.
As shown in Table \ref{Tab_6}, JNU datasets contained one health state and three fault modes which include inner ring fault, outer ring fault, and rolling element fault.
Therefore, the total number of classes was equal to twelve according to different working conditions.

\begin{table}[!t]
	\caption{Detailed description of JNU datasets} 
	\centering
	\label{Tab_6}  
	\begin{tabular}{| c | c | c | c | c | c |}
		\hline                 
		Fault Mode & Rotating Speed & Fault Mode & Rotating Speed & Fault Mode & Rotating Speed\\
		\hline                
		Health State & 600 rpm & Health State & 800 rpm & Health State & 1000 rpm\\
		\hline                
		Inner ring & 600 rpm& Inner ring & 800 rpm & Inner ring & 1000 rpm\\
		\hline
		Outer ring & 600 rpm & Outer ring & 800 rpm &Outer ring & 1000 rpm\\
		\hline                 
		Rolling Element & 600 rpm & Rolling Element & 800 rpm & Rolling Element & 1000 rpm\\
		\hline                
	\end{tabular}
\end{table}

\subsection{PHM 2012 Bearing Dataset}
PHM 2012 bearing datasets were used for PHM IEEE 2012 Data Challenge \cite{PHM,nectoux2012pronostia}.
In PHM 2012 datasets, seventeen run-to-failure datasets were provided including six training sets and eleven testing sets.
Three different loads were considered.
Vibration and temperature signals were gathered during all those experiments.
Since no label on the types of failures was given, it was not used in this paper.

\subsection{IMS Bearing Dataset}
IMS bearing datasets were generated by the NSF I/UCR Center for Intelligent Maintenance Systems \cite{lee2007bearing}.
IMS datasets were made up of three bearing datasets, and each of them contained vibration signals of four bearings installed on the different locations.
At the end of the run-to-failure experiment, a defect occurred on one of the bearings.
The failure occurred in the different locations of bearings.
It is inappropriate to classify these failures simply using three classes, so IMS datasets were not evaluated in this paper.

\section{Data Prepreocessing}
\label{S:5}
The type of input data and the way of normalization have a great impact on the performance of DL models.
Types of input data determine the difficulty of feature extraction, and normalization methods determine the difficulty of calculation.
So, in this paper, the effects of five input types and three normalization methods on the performance of DL models are discussed.

\subsection{Input Types}
Many researchers use signal processing methods to map the time series to different domains to boost the performance.
However, which input type is more suitable for intelligent diagnosis is still an open question.
In this paper, the effects of different input types on model performance are discussed.

\subsubsection{Time Domain Input}
For the time domain input, vibration signals are directly used as the input without data preprocessing.
In this paper, the length of each sample is 1024 and the total number of samples can be obtained from \eqrefA{eq1}.
After generating samples, we take 80\% of total samples as the training set and 20\% of total samples as the testing set.
\begin{align}
\label{eq1}
N = \text{floor}(\frac{L}{1024})
\end{align}
where  $ L $ is the length of each signal, $ N $  is the total samples, and $ \text{floor} $ means rounding towards minus infinity.

\subsubsection{Frequency Domain Input}
For the frequency domain input, FFT is used to transform each sample $ x_i $ from the time domain into the frequency domain shown in \eqrefA{eq2}.
After this operation, the length of data will be halved and the new sample can be expressed as:
\begin{align}
\label{eq2}
x_i^{\text{FFT}} = \text{FFT}(x_i)
\end{align}
where the operator $ \text{FFT}(\cdot) $ represents transforming $ x_i $ into the frequency domain and taking the first half of the result.

\subsubsection{Time-Frequency Domain Input}
For the time-frequency domain input, short-time Fourier transform (STFT) is applied to each sample $ x_i $ to obtain the time-frequency representation shown in \eqrefA{eq3}.
The Hanning window is used and the window length is 64.
After this operation, the time-frequency representation (a 33x33 image) will be generated as:
\begin{align}
\label{eq3}
x_i^{\text{STFT}} = \text{STFT}(x_i), \quad i=1,2,...,N
\end{align}
where the operator $ \text{SFFT}(\cdot) $ represents transforming $ x_i $ into the time-frequency domain.

\subsubsection{Wavelet Domain Input}
For the wavelet domain input, continuous wavelet transform (CWT) is applied to each sample $ x_i $ to obtain the wavelet domain representation shown in \eqrefA{eq4}.
Because CWT is time-consuming, the length of each sample $ x_i $ is set to 100.
After this operation, the wavelet coefficients (an 100x100 image) will be obtained as:
\begin{align}
\label{eq4}
x_i ^{\text{CWT}} = \text{CWT}(x_i), \quad i=1,2,...,N
\end{align}
where the operator $ \text{CWT}(\cdot) $ represents transforming $ x_i $ into the wavelet domain.

\subsubsection{Slicing Image Input}
For slicing image input, each sample $ x_i $ is reshaped into a 32x32 image.
After this operation, the new sample can be denoted as:
\begin{align}
\label{eq5}
x_i ^{\text{Reshape}}= \text{Reshape}(x_i), \quad i=1,2,...,N
\end{align}
where the operator $ \text{Reshape}(\cdot) $ represents reshaping $ x_i $ into a 32x32 image.

However, above data preprocessing methods have some problems for training AE models and CNN models in the following two aspects:
1) if AE models input a large 2D signal, it will lead the decoder to have difficulty in the reconstruction procedure and the reconstruction error is very large;
2) if CNN models input a small 2D signal, it will make CNN unable to extract appropriate features. 

Therefore, we have made a compromise on the data size obtained by the above data preprocessing methods.
The sizes of the time domain and the frequency domain input are unchanged as shown in \eqrefA{eq1} and \eqrefA{eq2}.
For the AE class, sizes of all 2D inputs are adjusted to 32x32, while for CNN models, sizes of signals after CWT, STFT, and slicing image are adjusted to 300x300, 330x330, and 320x320, respectively.
It should be noted that input sizes of CNN models can be different since we use the AdaptiveMaxPooling layer to adapt different input sizes.

\subsection{Normalization}
Input normalization is the basic step in data preparation, which can facilitate the subsequent data processing and accelerate the convergence of DL models.
Therefore, we discuss the effects of three normalization methods on the performance of DL models.

\textbf{Maximum-Minimum Normalization}: This normalization method can be implemented by
\begin{align}
\label{eq6}
x^{normalize-1}_i = \frac{x_i - x_i^{min}}{x_i^{max} - x_i^{min}}, \quad i=1,2,...,N
\end{align}
where $ x_i $ is the input sample, $ x_i^{min} $ is the minimum value in $ x_i $, and $ x_i^{max} $ is the maximum value in $ x_i $.

\textbf{[-1-1] Normalization}: This normalization method can be implemented by
\begin{align}
\label{eq7}
x^{normalize-2}_i = -1 + 2 * \frac{x_i - x_i^{min}}{x_i^{max} - x_i^{min}}, \quad i=1,2,...,N
\end{align}

\textbf{Z-score Normalization}: This normalization method can be implemented by
\begin{align}
\label{eq8}
x^{normalize-3}_i = \frac{x_i - x_i^{mean}}{x_i^{std}}, \quad i=1,2,...,N
\end{align}
where $ x_i^{mean} $ is the mean value of $ x_i $, and $ x_i^{std} $ is the standard deviation of $ x_i $.

\section{Data Augmentation}
\label{S:6}
Data augmentation is important to make the training datasets more diverse and to alleviate the learning difficulties caused by small sample problems.
However, data augmentation for intelligent diagnosis has not been investigated in depth.
The key challenge for data augmentation is to create the label-corrected samples from existing samples, and this procedure mainly depends on the domain knowledge.
However, it is difficult to determine whether the generated samples are label-corrected.
So, this paper provides some data augmentation techniques to increase the concerns of other scholars.
In addition, these data augmentation strategies are only a simple test and their applications still need to be studied in depth.
\subsection{One Dimension Input Augmentation}
\textbf{RandomAddGaussian}: this strategy randomly adds Gaussian noise into the input signal formulated as follows:
\begin{align}
\label{eq9}
x := x + n
\end{align}
where $ x $ is the 1D input signal, and $ n $ is generated by Gaussian distribution $\mathcal N (0, 0.01) $.

\textbf{RandomScale}: this strategy randomly multiplies the input signal with a random factor which is formulated as follows:
\begin{align}
\label{eq10}
x := \beta * x
\end{align}
where $ x $ is the 1D input signal,  and $ \beta $ is a scaler following the distribution $ \mathcal N (1, 0.01) $.

\textbf{RandomStretch}: this strategy resamples the signal into a random proportion and ensures the equal length by nulling and truncating.

\textbf{RandomCrop}: this strategy randomly covers partial signals which is formulated as follows:
\begin{align}
\label{eq11}
x := mask  * x
\end{align}
where $ x $ is the 1D input signal, and $ mask $ is the binary sequence whose subsequence of random position is zero.
In this paper, the length of the subsequence is equal to 10.

\subsection{Two Dimension Input Augmentation}
\textbf{RandomScale}: this strategy randomly multiplies the input signal with a random factor which is formulated as follows:
\begin{align}
\label{eq12}
x := \beta * x
\end{align}
where $ x $ is the 2D input signal,  and $ \beta $ is a scaler following the distribution $ \mathcal N (1, 0.01) $.

\textbf{RandomCrop}: this strategy randomly covers partial signals, which is formulated as follows:
\begin{align}
\label{eq13}
x := mask  * x
\end{align}
where $ x $ is the 2D input signal, and $ mask $ is the binary sequence whose subsequence of random position is zero.
In this paper, the length of the subsequence is equal to 20.

\section{Data Split}
\label{S:7}
\begin{figure}[!t]
	\centering
	\subfigure{\includegraphics[scale = 0.3]{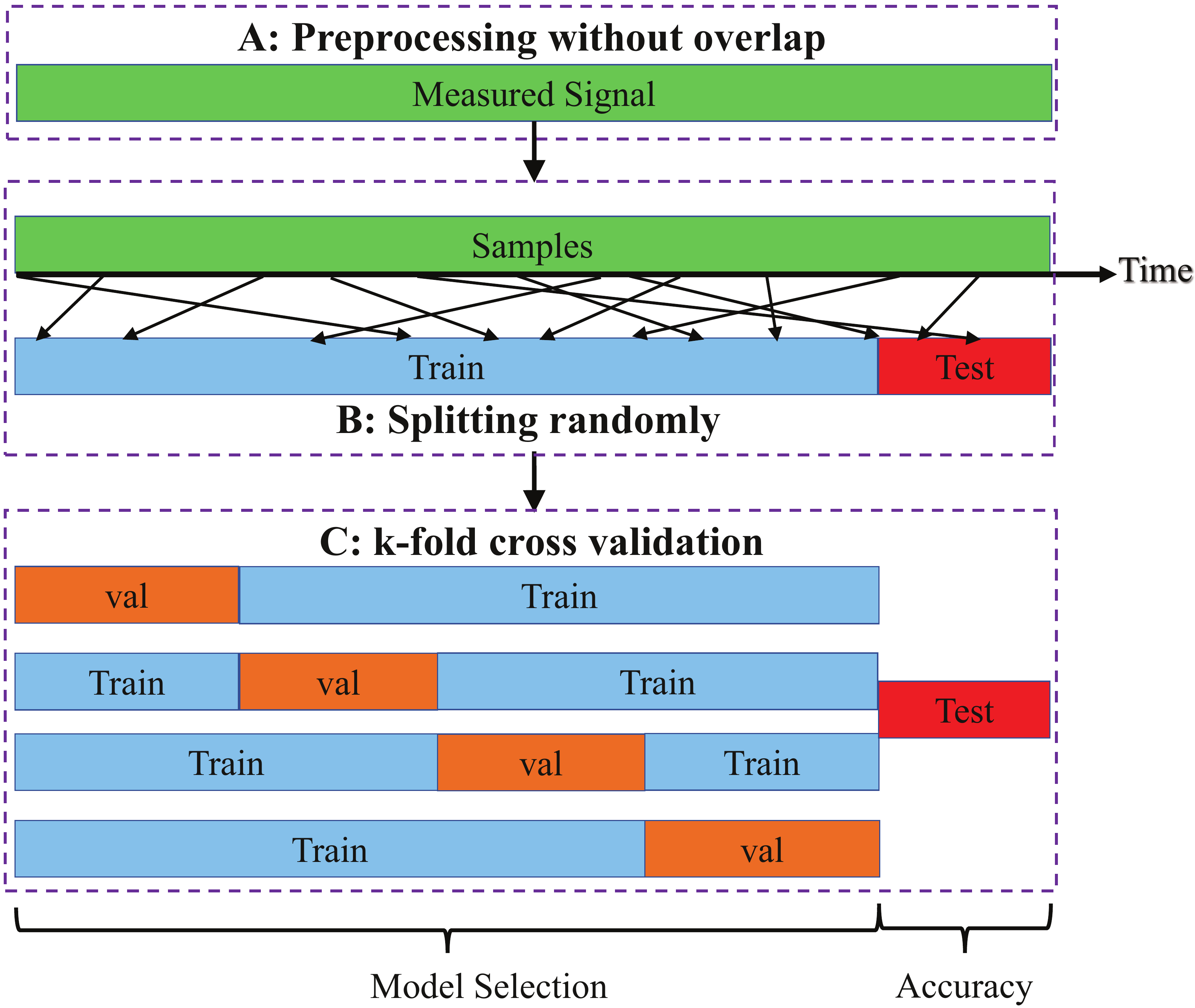}}
	\\ [-5pt]
	\caption{Random data splitting strategy with preprocessing without overlap.}
	\label{Fig-data-split0}
\end{figure}

\begin{figure}[!t]
	\centering
	\subfigure{\includegraphics[scale = 0.3]{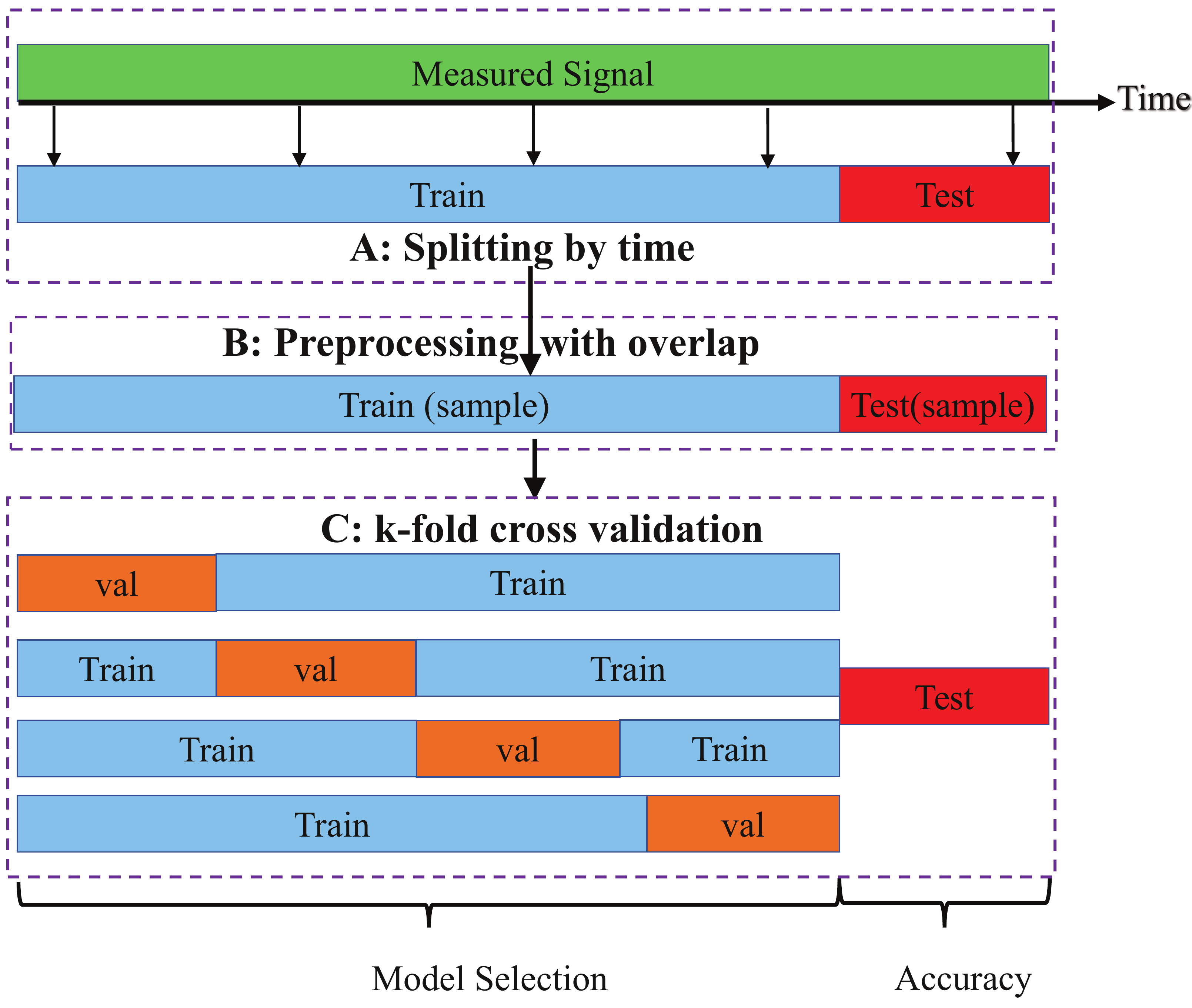}}
	\\ [-5pt]
	\caption{Another condition with the training and testing sets split as the first step.}
	\label{Fig-data-split1}
\end{figure}

One common practice of data split in intelligent diagnosis is the random split strategy, and the diagram of this strategy is shown in \figref{Fig-data-split0}. 
From this diagram, we stress the preprocessing step without overlap due to the fact that if the sample preparation process exists any overlap, the evaluation of classification algorithms may have test leakage (if users split the training set and the testing set from the beginning of the preprocessing step, then they can use any processing to simultaneously deal with the training and testing sets, as shown in \figref{Fig-data-split1}).

The formal way is that the training set is further split into the training set and the validation set for the model selection.
\figref{Fig-data-split0} shows the condition of 4-fold cross-validation, and we often use the average accuracy of 4-fold cross-validation to represent the generalization accuracy, if there is no testing set.
In this paper, for testing convenience and time saving, we only use 1-fold validation and use the last epoch accuracy to represent the testing accuracy (we also list the maximum accuracy in the whole epochs for comparisons).
It is worth noting that some papers use the maximum accuracy of the validation set, and this strategy is also dangerous because the validation set is used to select the parameters accidentally.

\begin{figure}[!t]
	\centering
	\subfigure{\includegraphics[scale = 0.3]{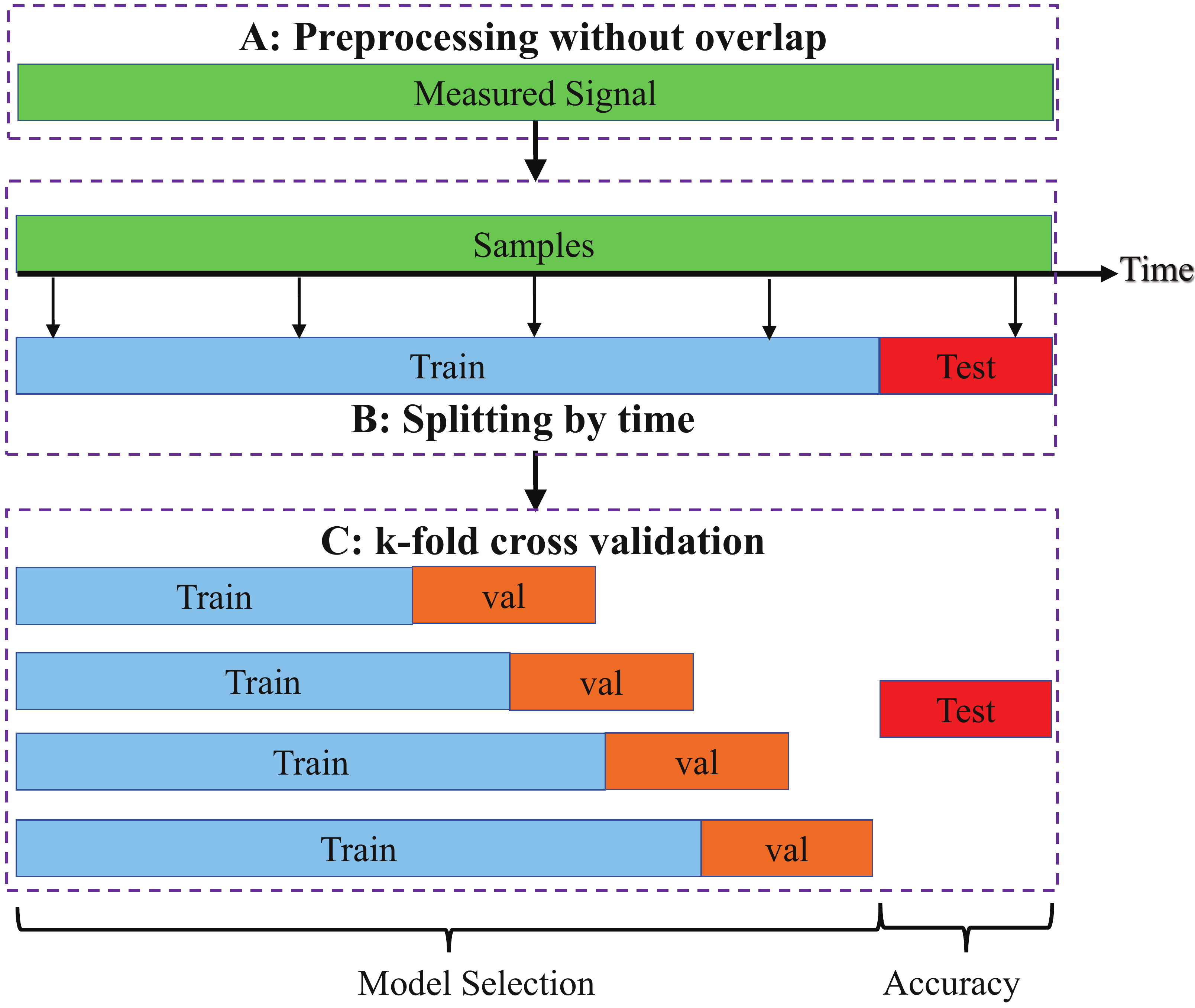}}
	\\ [-5pt]
	\caption{Data split according to time sequences.}
	\label{Fig-data-split2}
\end{figure}

For industrial data, they are rarely random and are always sequential (they might contain trends or other temporal correlation).
Therefore, it is more appropriate to split data according to time sequences (order split).
The diagram of data split strategy according to time sequences is shown in \figref{Fig-data-split2}.
From this diagram, it can be observed that we split the training and testing sets with the time phase instead of splitting the data randomly.
In addition, \figref{Fig-data-split2} also shows the condition of 4-fold cross-validation with time.
In the following study, we will compare the results of this strategy with the random split strategy.

\section{Evaluation Methodology}
\label{S:8}

\subsection{Evaluation Metrics}
It is a rather challenging task to evaluate the performance of intelligent diagnosis algorithms with suitable evaluation metrics.
It has three standard evaluation metrics, which have been widely used, including the overall accuracy, the average accuracy, and the confusion matrix.
In this paper, we only use the overall accuracy to evaluate the performance of algorithms.
The overall accuracy is defined as the number of correctly classified samples divided by the total number of samples. 
The average accuracy is defined as the average classification accuracy of each category.

Since the performance of DL-based intelligent diagnosis algorithms fluctuates during the training process, to obtain reliable results and show the best overall accuracy that the model can achieve, we repeat each experiment five times.
Four indicators are used to assess the performance of models, including the mean and maximum values of the overall accuracy obtained by the last epoch (the accuracy in the last epoch can represent the real accuracy without any test leakage), and the mean and maximum values of the maximal overall accuracy.
For simplicity, they can be denoted as Last-Mean, Last-Max, Best-Mean, and Best-Max.

\subsection{Experimental Setting}
In the preparation stage, we use two strategies, including random split and order split to divide the dataset into training and testing sets.
For random split, a sliding window is used to truncate the vibration signal without any overlap and each data sample contains 1024 points.
After the preparation, we randomly take 80\% of samples as the training set and 20\% of samples as the testing set.
For order split, we take the former 80\% of time series as the training set and the last 20\% as the testing set.
Then, in two time series, a sliding window is used to truncate the vibration signal without any overlap, and each sample contains 1024 points. 

In order to verify how input types, data normalization methods, and data split methods affect the performance of models, we set up three configurations of experiments (shown in Table \ref{Tab_E1}, Table \ref{Tab_E2}, and Table \ref{Tab_E3}.).
During model training, we use Adam as the optimizer. The learning rate and the batch size of each experiment are set to 0.001 and 64, respectively. 
Each model is trained for 100 epochs, and during the training procedure, model training and model testing are alternated.
In addition, all the experiments are executed under Windows 10 and Pytorch 1.1 through running on a computer with an Intel Core i7-9700K, GeForce RTX 2080Ti, and 16G RAM.

\begin{table}[!t]
	\caption{Experiment setup 1} 
	\centering
	\label{Tab_E1}  
	\begin{tabular}{c}      
		\includegraphics[scale = 0.9]{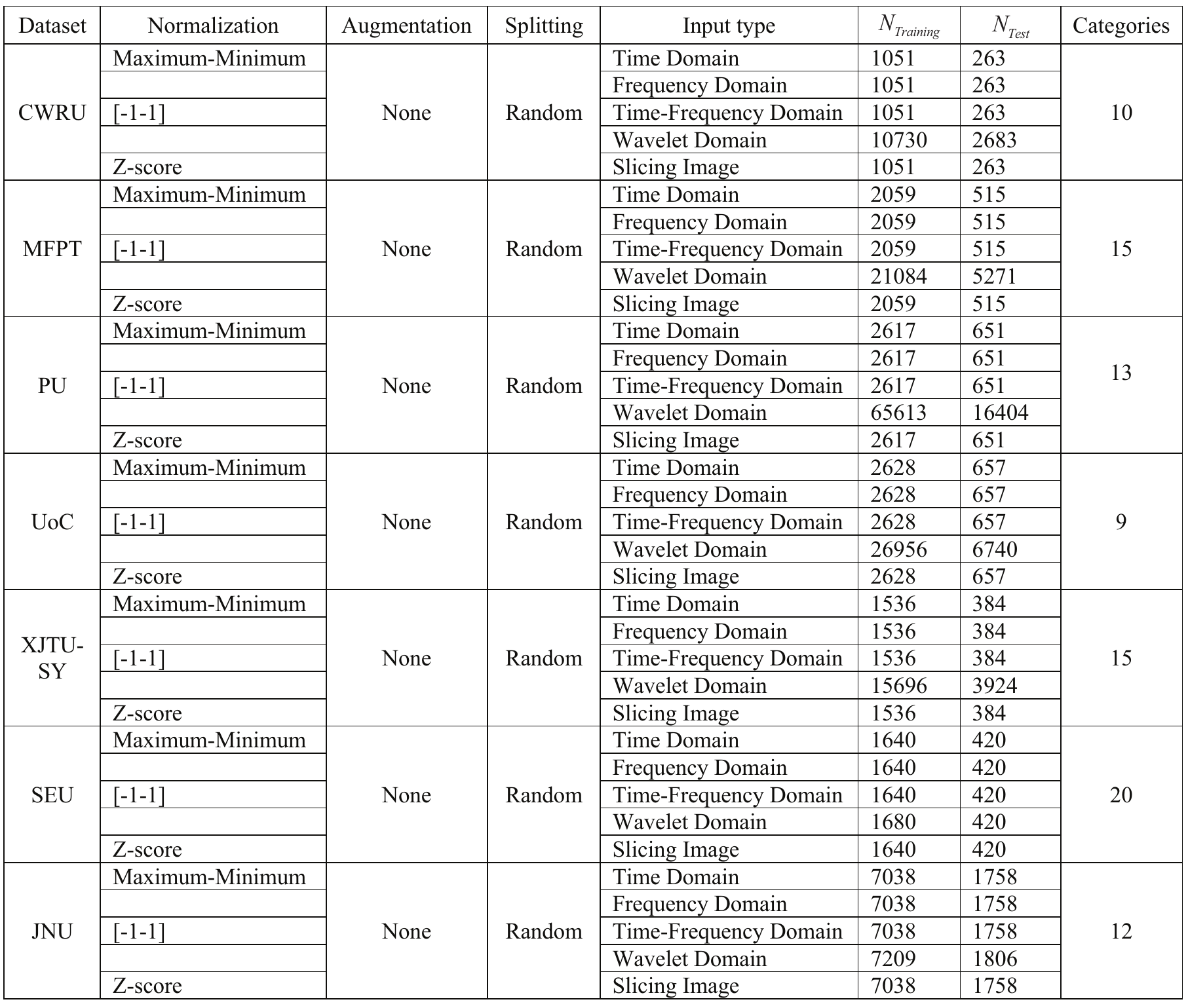}
	\end{tabular}
\end{table}

\begin{table}[!t]
	\caption{Experiment setup 2} 
	\centering
	\label{Tab_E2}  
	\begin{tabular}{c}      
		\includegraphics[scale = 0.9]{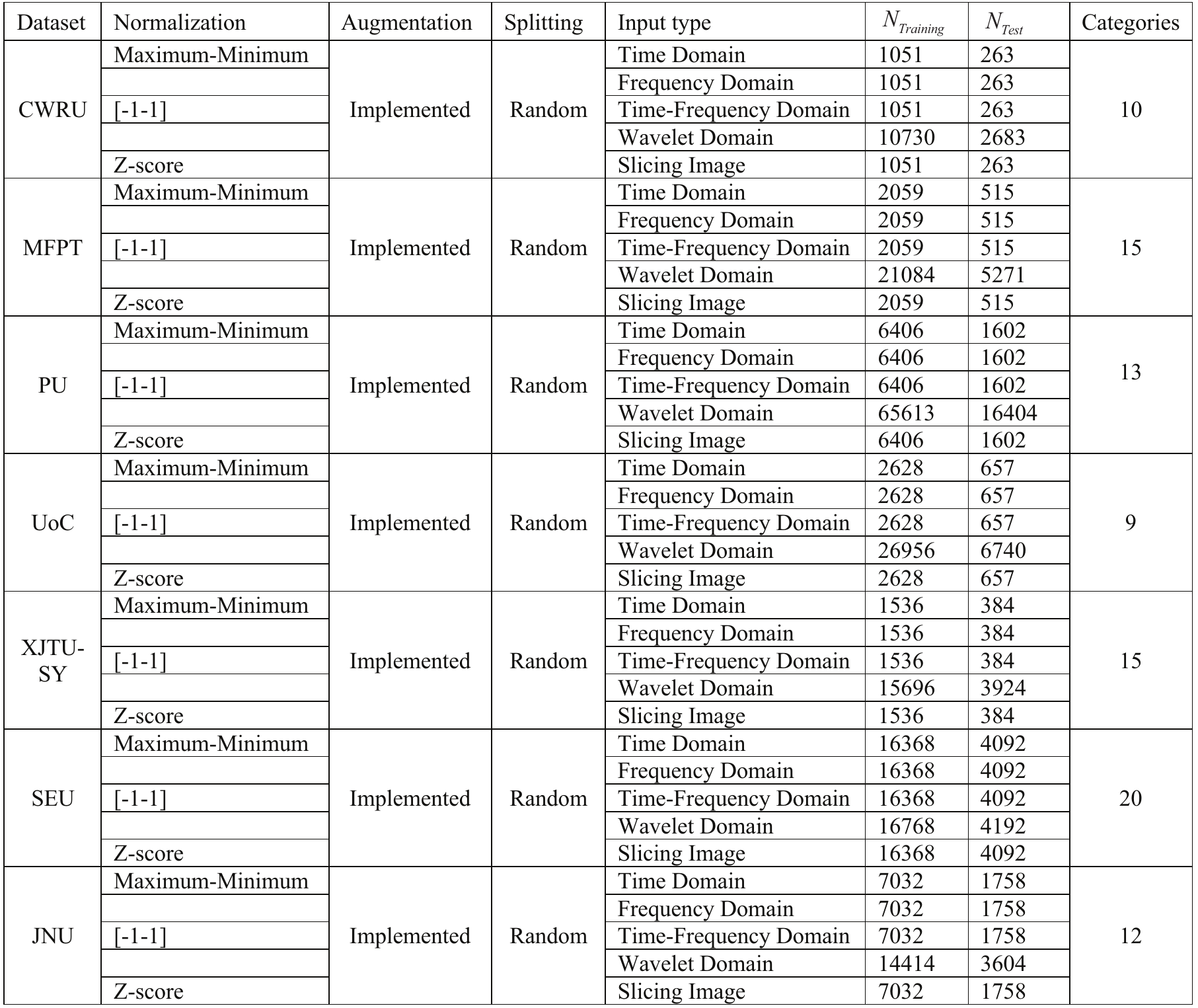}
	\end{tabular}
\end{table}

\begin{table}[!t]
	\caption{Experiment setup 3} 
	\centering
	\label{Tab_E3}  
	\begin{tabular}{c}      
		\includegraphics[scale = 0.9]{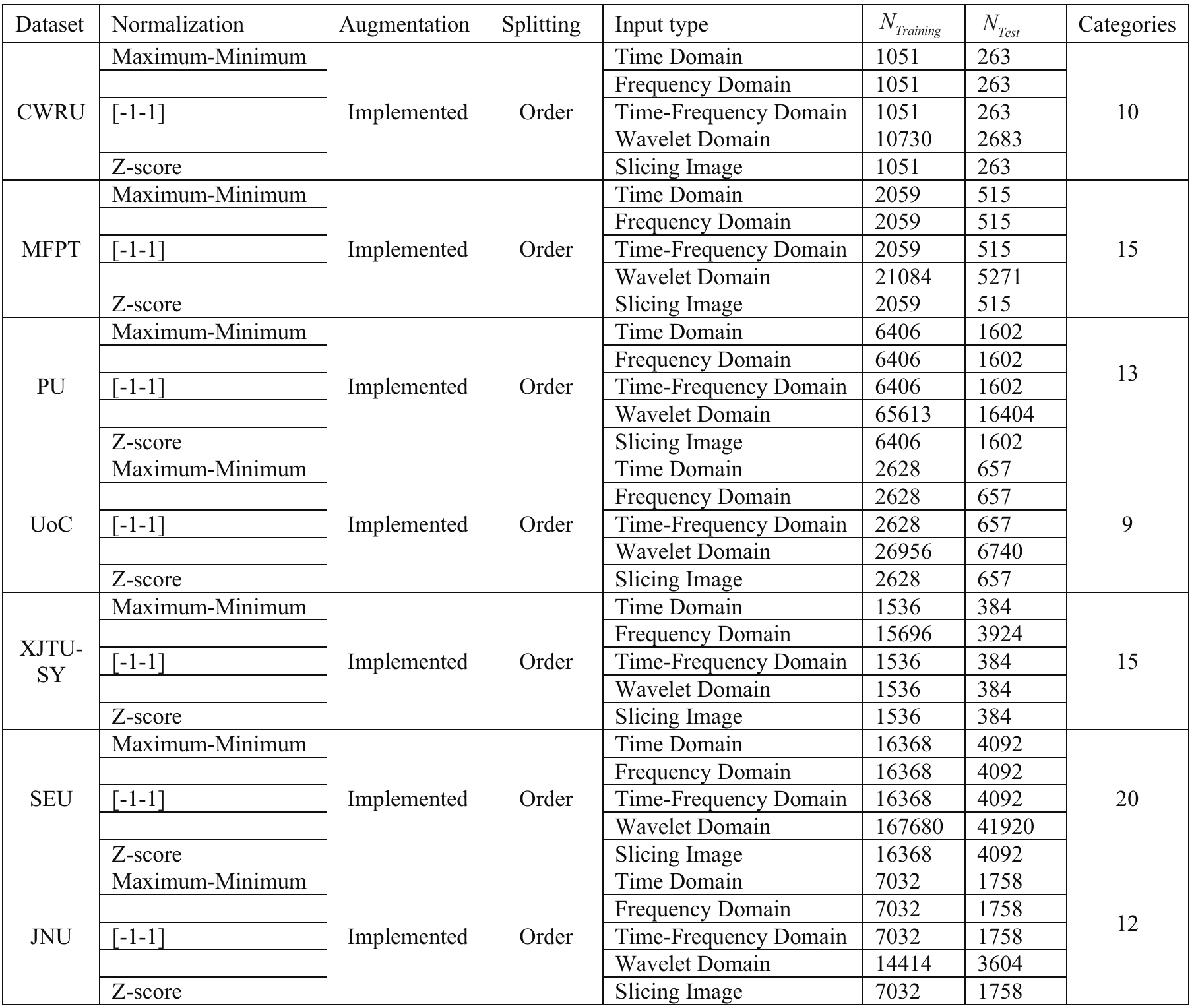}
	\end{tabular}
\end{table}

\section{Evaluation Results}
\label{S:9}
In this section, we will first discuss the experimental results of different datasets in depth.
After that, the results of datasets, input types, models, input normalization, data augmentation, and data splitting will be summarized separately.
Complete results are shown in \textbf{Appendix A.} and each accuracy which is larger than 95\% are bold.

\subsection{Detailed Analysis of Different Datasets}
\subsubsection*{A. CWRU Dataset}
The results of CWRU dataset are shown in \textbf{Appendix A} from Table 1 to Table 3.
From those results, we can observe that the accuracy of CNN models is generally higher than that of AE models.
In addition, using FFT and STFT to process the signal allows models to achieve better accuracy among five kinds of input.
CNN models with Z-score normalization can get better accuracy while using -1-1 normalization allows AE models to achieve higher accuracy.
Using data augmentation does not improve the accuracy of AE models, but it can improve the accuracy of CNN models.
The order split would slightly reduce the accuracy.

\subsubsection*{B. JNU Dataset}
The results of JNU dataset are shown in \textbf{Appendix A} from Table 4 to Table 6.
As can be seen from those tables, using FFT to process the raw signal allows models to achieve better accuracy among five types of input.
CNN models with Z-score normalization can get better accuracy while using -1-1 normalization enables AE models to achieve higher accuracy.
Using data augmentation can improve the accuracy of CNN models and AE models.
The order split would highly reduce the accuracy.

\subsubsection*{C. MFPT Dataset}
The results of MFPT dataset are shown in \textbf{Appendix A} from Table 7 to Table 9.
We can observe that models with time or wavelet domain as the input would have the worse performance.
However, using FFT to process the signal allows models to achieve better accuracy, and the accuracy of AE models is even higher than CNN models in this dataset.
CNN models with Z-score normalization can get better accuracy while using -1-1 normalization enables the AE models to achieve higher accuracy.
Using data augmentation can improve the accuracy of CNN models and AE models.
The order split would heavily reduce the accuracy.

\subsubsection*{D. PU Dataset}
The results of PU dataset are shown in \textbf{Appendix A} from Table 10 to Table 12.
It is shown that the accuracy of CNN models is generally higher than that of AE models.
Besides, the accuracy is worse when using the wavelet domain as the input, while using FFT and STFT to process the signal allows models to achieve better accuracy.
Using Z-score normalization enables AE models and CNN models to achieve higher accuracy.
Data augmentation does not help AE models improve the accuracy, while it can increase the accuracy of CNN models.
Similarly, the order split would heavily reduce the accuracy.

\subsubsection*{E. SEU Dataset}
The results of SEU dataset are shown in \textbf{Appendix A} from Table 13 to Table 15.
We can observe that when using the time domain or wavelet domain as the input, models would achieve worse accuracy.
However, using FFT to process the signal allows models to achieve better accuracy and the accuracy of AE models is even higher than that of CNN models.
Using Z-score normalization allows AE models and CNN models to achieve higher accuracy. 
Data augmentation can improve the accuracy of both CNN and AE models. 
In this case, the order split would slightly reduce the accuracy.

\subsubsection*{F. UoC Dataset}
The results of UoC dataset are shown in \textbf{Appendix A} from Table 16 to Table 18.
We can observe that most models do not perform well in this case, and among them, the performance of AlexNet is relatively worse.
Besides, using FFT to process the signal allows models to achieve bette accuracy, and the accuracy of AE models is higher than that of CNN models.
AE models and CNN models with Z-score normalization would achieve higher accuracy. 
Data augmentation can help different models improve the final accuracy. 
The order split would heavily reduce the accuracy.

\subsubsection*{G. XJTU-SY Dataset}
The results of XJTU-SY dataset are shown in \textbf{Appendix A} from Table 19 to Table 21.
We can observe that most models perform well in this dataset.
Besides, we can find that using FFT and STFT to process the signal allows models to achieve the better accuracy, and the accuracy of CNN models is higher than that of AE models, generally.
AE models and CNN models with Z-score normalization would achieve higher accuracy.
Data augmentation can help different models improve the final accuracy.
The order split would quietly reduce the accuracy.

\subsection{Results of Datasets}
It can be seen from the results that with the exception of the UoC dataset, the accuracy of both AE and CNN models on those datasets exceeds 95\%.
In addition, the accuracy of CWRU, SEU and XJTU-SY datasets can reach to 100\%.
The accuracy of UoC is much lower than others in all conditions.
Besides, the diagnostic difficulty of seven datasets can be ranked according to the number of  diagnostic accuracy exceeding 95\% in each dataset.
As shown in Fig. 9, we can split the datasets into four levels of difficulty.

\begin{figure}[!t]
	\centering
	\subfigure{\includegraphics[scale = 0.3]{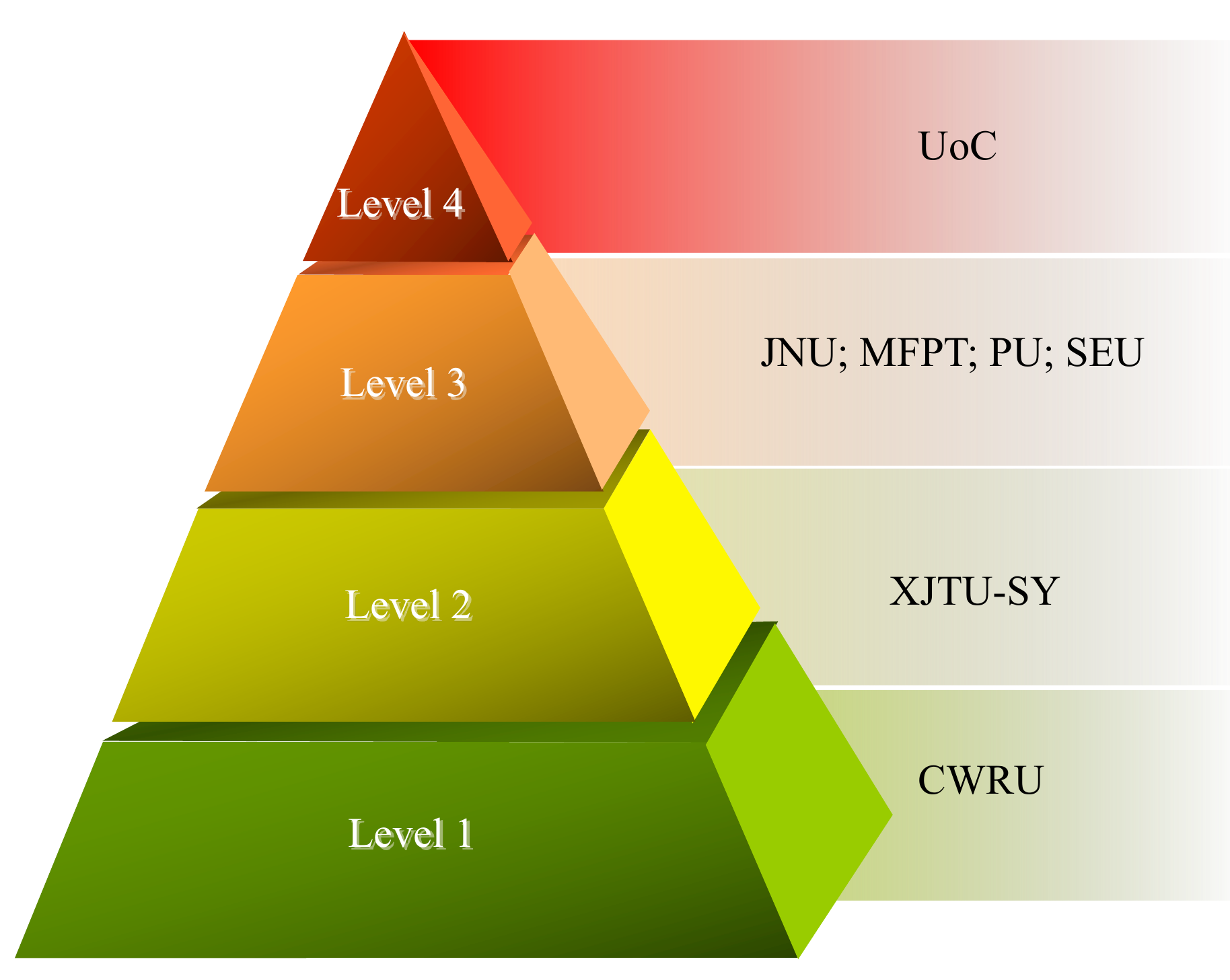}}
	\\ [-5pt]
	\caption{The level of dataset difficulty.}
	\label{Datasets}
\end{figure}

\subsection{Results of Input Types}
In all datasets, the frequency domain input can always achieve the highest accuracy followed by the time-frequency domain input since in the frequency domain, the noise is spread over the full frequency band and the fault information is much easier to be distinguished than that in the time domain.
According to the computational load of CWT, we use the short length of samples to perform CWT and then upsample the wavelet coefficients.
These steps may degrade the classification accuracy of CWT.

\subsection{Results of Models}
From the results, we can observe that models, especially ResNet18, can achieve the best accuracy
in most of datasets including CWRU, JNU, PU, SEU, and XJTU-SY.
However, for MFPT and UoC, models belonging to AE can perform better than other models.
This phenomenon may be caused by the size of datasets and the overfitting problem.
Therefore, not every dataset can get better results using a more complex model.

\subsection{Results of Data Normalization}
It is hard to conclude which data normalization method is the best one, and from the results, we can observe that the accuracy of different data normalization methods also depends on the used models and datasets.
In general, Z-score normalization can make models achieve better accuracy.

\subsection{Results of Data Augmentation}
We can conclude that when the accuracy of datasets is already high enough, data augmentation methods may slightly degrade the performance because models have already fitted original datasets well.
More augmentation methods may change the distribution of the original data and make the learning process harder.
However, when the accuracy of datasets is not very high, data augmentation methods improve the performance of models, especially for the time domain input.
Therefore, researchers can design other various data augmentation methods for their specific inputs.

\subsection{Results of Data Splitting}
When the datasets are easy to handle (CWRU and XJTU-SY), the results between random split and order split are quite similar.
However, the accuracy of some datasets (PU and UoC) decreases sharply when using order split.
What we should pay more attention to is that whether randomly splitting these datasets has the risk of test leakage.
It may be more suitable for splitting the datasets according to time sequences to verify the performance.

According to the above discussion, we summarize the following conclusion coming from the evaluation results.
First, not all datasets are suitable for comparing the classification effectiveness of the proposed methods since basic models can achieve very high accuracy on these datasets, like CWRU and XJTU-SY.
Second, the frequency domain input can achieve the highest accuracy in all datasets, so researchers should first try to use the frequency domain as the input.
Third, it is not necessary for CNN models to get the best results in all cases, and we also should consider the overfitting problem.
Fourth, when the accuracy of datasets is not very high, data augmentation methods improve the performance of models, especially for the time domain input.
Thus, more effective data augmentation methods need to be investigated.
Finally, in some cases, it may be more suitable for splitting the datasets according to time sequences (order split) since random split may provide virtually high accuracy.
We also release a code library to evaluate DL-based intelligent diagnosis algorithms and provide the benchmark accuracy (a lower bound) to avoid useless improvement.
Meanwhile, we use specific-designed cases to discuss existing issues, including class imbalance, generalization ability, interpretability, few-shot learning, and model selection.
Through these works, we aim to allow comparisons fairer and quicker, emphasize the importance of open source codes, and provide deep discussions of existing issues.
To the best of our knowledge, this is the first work to comprehensively perform the benchmark study and release the code library to the public.

\section{Discussion}
\label{S:10}
Although intelligent diagnosis algorithms can achieve high classification accuracy in many datasets, there are still many issues that need to be discussed.
In this paper, we further discuss the following five issues including class imbalance, generalization ability, interpretability, few-shot learning, and model selection using experimental cases.

\subsection{Class Imbalance}
Most of measured signals are in the normal state, and only a few of them are in the fault state, which means that fault modes often have different probabilities of happening.
Therefore, the class imbalance issue will occur when using intelligent algorithms in real applications.
Recently, although some researchers have published some related papers using traditional imbalanced learning methods \cite{zhang2018imbalanced} or GAN \cite{mao2019imbalanced} to solve this problem, these studies are far from enough.
In this paper, PU Bearing Datasets are used to simulate the class imbalance issue.
In this experiment, we adopt ResNet18 as the experimental model and only use two kinds of input types (the time domain input and the frequency domain input).
Besides, data augmentation methods are used and the normalization method is the Z-score normalization, while the dataset is randomly split.
Three groups of datasets with different imbalance ratios are constructed, which are shown in Table \ref{Tab_CI}.
\begin{table}[!t]
	\caption{Number of samples in three groups of imbalanced datasets} 
	\centering
	\label{Tab_CI}  
	\begin{tabular}{|c|c|c|c|c|}
		\hline
		\multirow{2}{*}{Fault mode} & \multicolumn{3}{c|}{Training samples} & Testing samples \\ \cline{2-5} 
		& Group1      & Group2     & Group3     & Group1/2/3      \\ \hline
		KA04                        & 125         & 125        & 125        & 125             \\ \hline
		KA15                        & 125         & 75         & 50         & 125             \\ \hline
		KA16                        & 125         & 75         & 50         & 125             \\ \hline
		KA22                        & 125         & 75         & 50         & 125             \\ \hline
		KA30                        & 125         & 37         & 25         & 125             \\ \hline
		KB23                        & 125         & 37         & 25         & 125             \\ \hline
		KB24                        & 125         & 37         & 25         & 125             \\ \hline
		KB27                        & 125         & 25         & 6          & 125             \\ \hline
		KI14                        & 125         & 25         & 6          & 125             \\ \hline
		KI16                        & 125         & 25         & 6          & 125             \\ \hline
		KI17                        & 125         & 12         & 2          & 125             \\ \hline
		KI18                        & 125         & 12         & 2          & 125             \\ \hline
		KI21                        & 125         & 12         & 2          & 125             \\ \hline
	\end{tabular}
\end{table}

As shown in Table \ref{Tab_CI}, three datasets (Group1, Group2, and Group3) are constituted with different imbalanced ratios.
Group1 is a balanced dataset, and there is no imbalance for each state.
In real applications, it is almost impossible to let the number of data samples be the same.
We reduce the training samples of some fault modes in Group1 to construct Group2, and then the imbalanced classification is simulated.
In Group3, the imbalanced ratio between fault modes increases further.
Group2 can be considered as a moderately imbalanced dataset, while Group3 can be considered as a highly imbalanced dataset.

\begin{figure}[!t]\centering
	\centering
	\subfigure{\includegraphics[scale = 0.7]{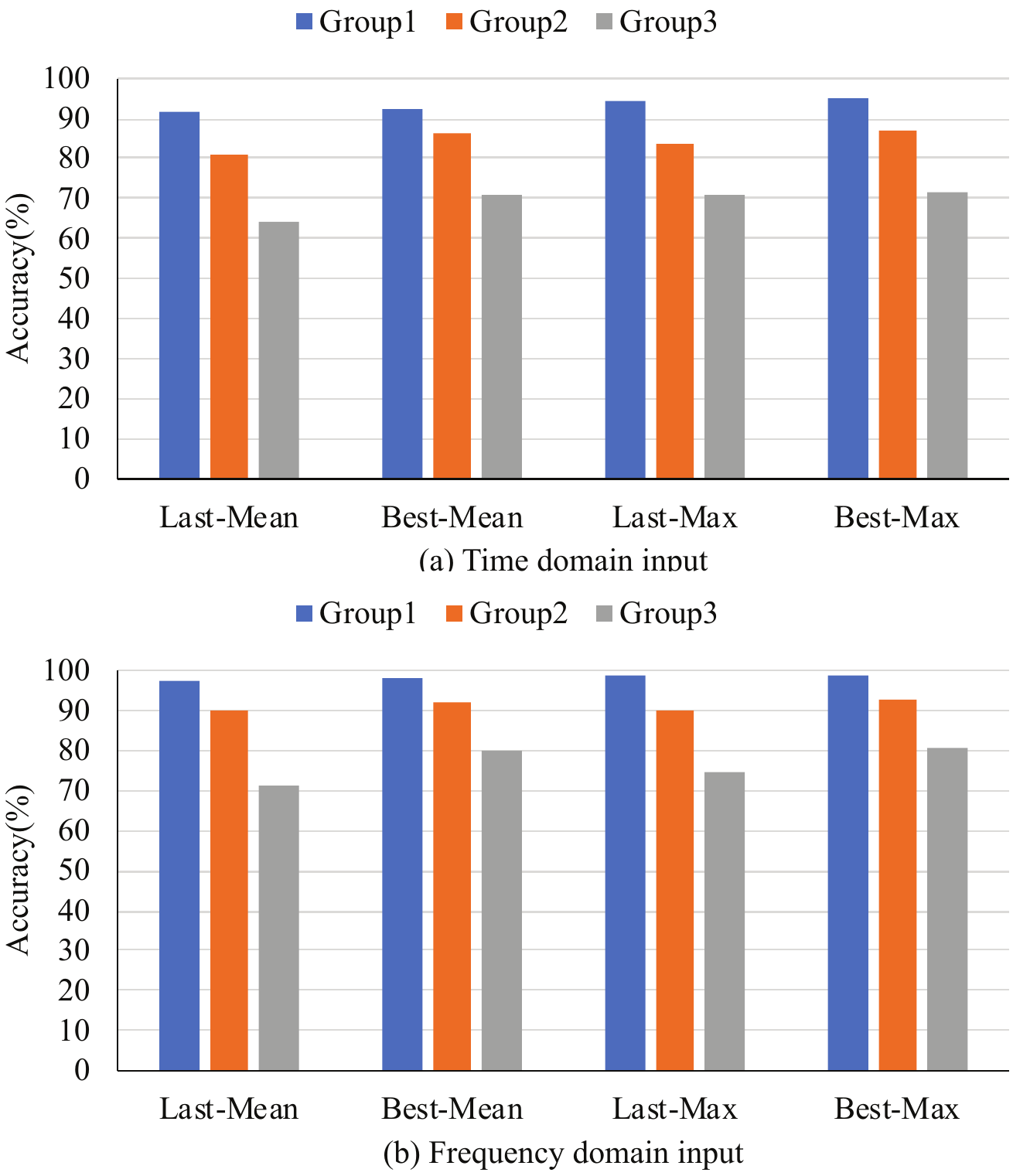}}
	\\ [-10pt]
	\caption{Experimental results of three groups of datasets. (a) time domain input, and (b) frequency domain input.}
	\label{FIG1}
\end{figure}

Experimental results are shown in \figref{FIG1}, and it can be observed that the overall accuracy in Group3 is much lower than that of Group1, which indicates that the class imbalance will greatly degrade the performance of models.
To address the problem of class imbalance, data-level methods and classifier-level methods can be used \cite{buda2018systematic}.
Oversampling and undersampling methods are the most commonly used data-level methods, and some methods for generating samples based on GAN have also been studied recently.
For the classifier-level methods, thresholding-based methods are applied in the test phase to adjust the decision threshold of the classifier.
Besides, cost-sensitive learning methods assign different weights to different classes to avoid the suppression of categories with a small number of samples.
In the field of intelligent diagnosis, other methods based on physical meanings and fault attention need to be explored.

\subsection{Generalization ability}
Many existing intelligent algorithms perform very well on one working condition, but the diagnostic performance tends to drop significantly on another working condition, and here, we call it the generalization problem.
Recently, many researchers have used algorithms based on transfer learning strategies to solve this problem, and a comparative study with open source codes was performed in \cite{zhao2019unsupervised}.
To illustrate the weak generalization ability of the intelligent diagnosis algorithms, experiments are also carried out on the PU bearing dataset.
Experiments use the data under three working conditions (N15\_M07\_F10, N09\_M07\_F10, N15\_M01\_F10).
In these experiments, data under one working condition is used to train models, and data under another working condition is used to test the performance.
A total of six groups are performed, and the detailed information is shown in Table \ref{Tab_GP}.

\begin{table}[!t]
	\caption{Training data and testing data for each experiment} 
	\centering
	\label{Tab_GP}  
	\begin{tabular}{| c | c | c |}
		\hline                 
		Group & Data for training & Data for testing\\
		\hline                
		Group1 & N15\_M07\_F10 & N09\_M07\_F10 \\
		\hline                
		Group2 &  N15\_M07\_F10 & N15\_M01\_F10\\
		\hline
		Group3 & N09\_M07\_F10  & N15\_M07\_F10\\
		\hline                 
		Group4 &  N09\_M07\_F10 & N15\_M01\_F10\\
		\hline
		Group5 &  N15\_M01\_F10 & N15\_M07\_F10\\
		\hline 
		Group6 &  N15\_M01\_F10 & N09\_M07\_F10 \\
		\hline                                   
	\end{tabular}
\end{table}

\begin{figure}[!t]\centering
	\centering
	\subfigure{\includegraphics[scale = 0.7]{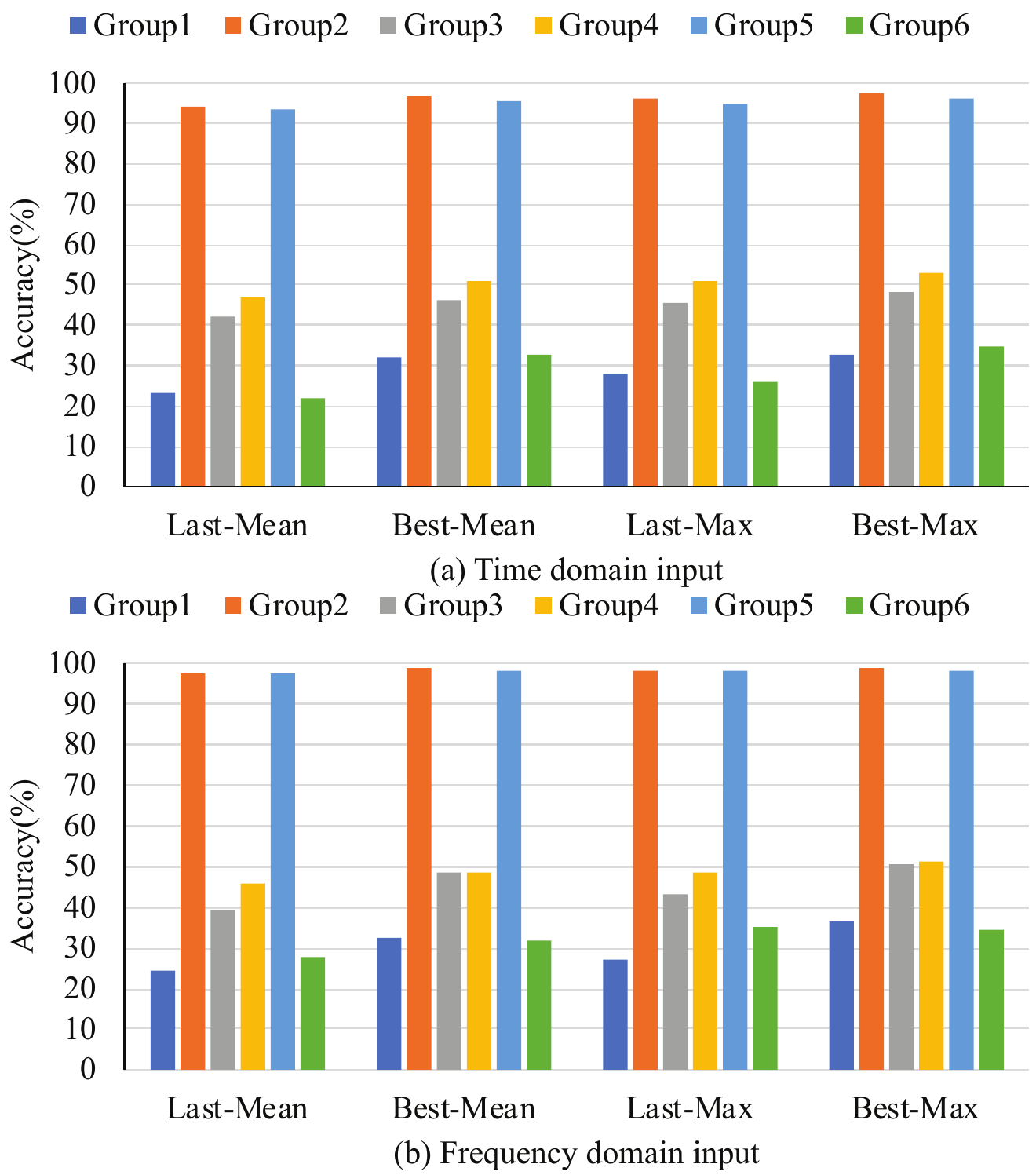}}
	\\ [-10pt]
	\caption{Experimental results of working conditions transfer. (a) time domain input, and (b) frequency domain input.}
	\label{FIG2}
\end{figure}

The experimental results are shown in \figref{FIG2}.
It can be concluded that in most cases, intelligent diagnosis algorithms trained on one working condition cannot perform well on another working condition, which means the generalization ability of algorithms is insufficient. 
In general, we expect that our algorithms could adapt to the changes in working conditions or measurement situations since these changes occur frequently in real applications.
Therefore, studies are still required on how to transfer the trained algorithms to different working conditions effectively.

Two excellent review papers \cite{zheng2019cross,yan2019knowledge} and other applications \cite{han2019deep,han2019learning} published recently pointed out several potential research directions which could be considered and studied further to improve the generalization ability.

\subsection{Interpretability}
Although intelligent diagnosis algorithms can achieve high diagnostic accuracy in their tasks, the interpretability of these models is often insufficient and these black-box models would generate high-risk results \cite{rudin2019stop}, which greatly limits their applications. 
Actually, some papers in intelligent diagnosis have noted this problem and attempted to propose some interpretable models \cite{li2019understanding,li2019waveletkernelnet}.

To point out that intelligent diagnostic algorithms lack interpretability, we perform three sets of experiments on the PU bearing dataset, and the datasets are shown in Table \ref{Tab_IP}.
In each set of experiments, we use two different data, which have the same fault pattern and are acquired under the same condition.

\begin{table}[!t]
	\caption{The bearing code and the number of samples used in each experiment} 
	\centering
	\label{Tab_IP}  
	\begin{tabular}{|c|c|c|c|}
		\hline
		Group                   & Bearing code & Training samples & Testing samples \\ \hline
		\multirow{2}{*}{Group1} & KA03         & 200              & 50              \\ \cline{2-4} 
		& KA06         & 200              & 50              \\ \hline
		\multirow{2}{*}{Group2} & KA08         & 200              & 50              \\ \cline{2-4} 
		& KA09         & 200              & 50              \\ \hline
		\multirow{2}{*}{Group3} & KI07         & 200              & 50              \\ \cline{2-4} 
		& KI08         & 200              & 50              \\ \hline
	\end{tabular}
\end{table}

\begin{figure}[!t]\centering
	\centering
	\subfigure{\includegraphics[scale = 0.7]{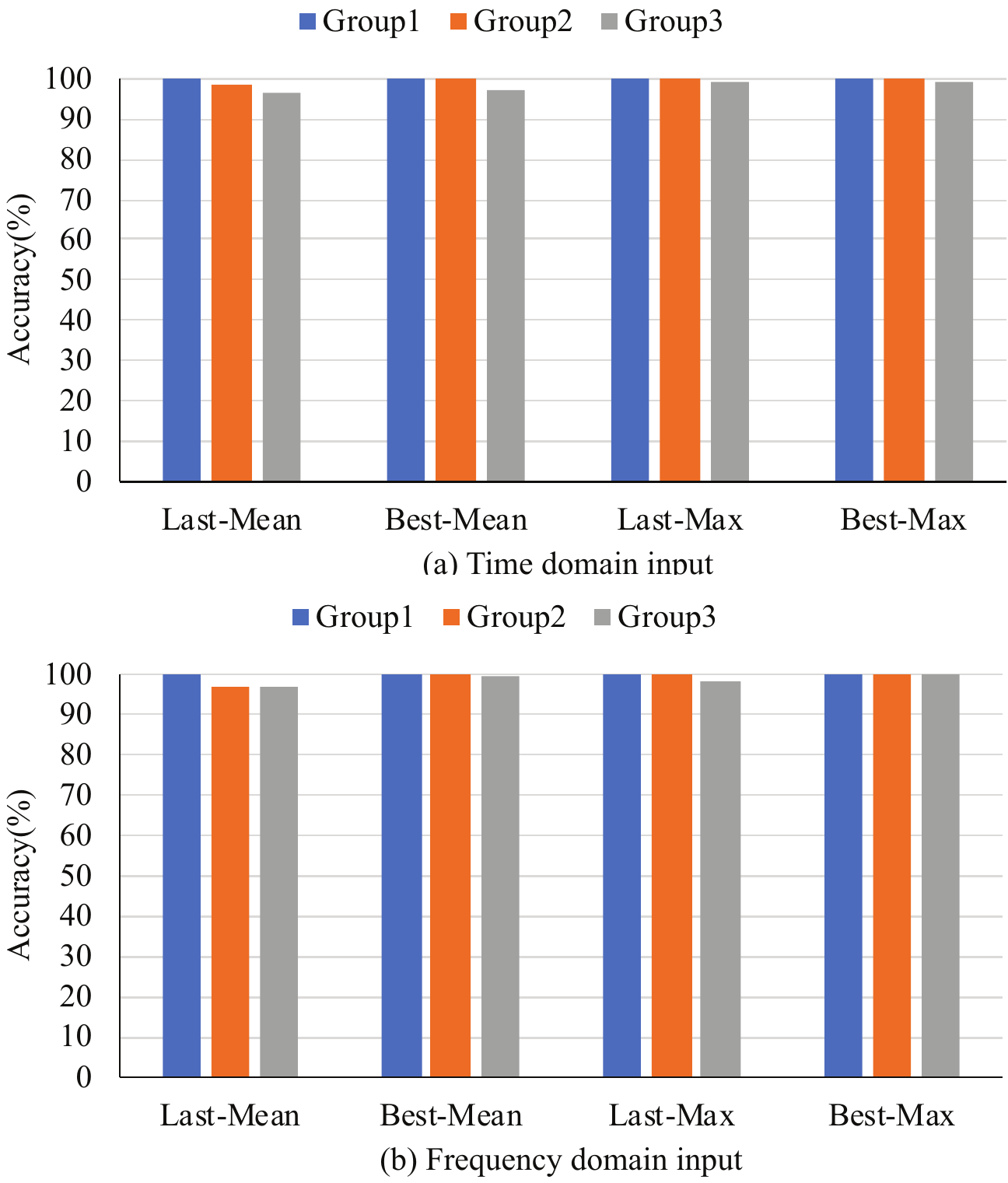}}
	\\ [-10pt]
	\caption{Experimental results of three groups. (a) time domain input, and (b) frequency domain input.}
	\label{FIG3}
\end{figure}

The results, in which intelligent algorithms can get high accuracy in each set of experiments, are shown in \figref{FIG3}.
Nevertheless, for each binary classification task, since the fault mode and the working condition at the time of acquisition are the same between two classes, theoretically, methods should not be able to achieve such high accuracy.
These expected results are exactly contrary to those of the experiment, which shows that models only learn the discrimination of different collection points and do not learn how to extract the essential characteristics of fault signals.
Therefore, it is very important to figure out whether models can learn essential fault characteristics or just classify the different conditions of collected signals.

According to the development of interpretability, we might be able to study the interpretability of DL-based models from the following aspects: (1) visualize the results of neurons to analyze the attention points of models \cite{zeiler2014visualizing};  (2) add physical constraints to the loss function \cite{tang2018regularized} to meet specific needs of fault feature extraction; (3) add prior knowledge to network structures and convolutions \cite{ravanelli2018interpretable} or unroll the existing optimization algorithms \cite{gregor2010learning} to extract corresponding fault features.

\subsection{Few-Shot Learning}
In intelligent diagnosis, the amount of data is far from big data because of the preciousness of fault data and the high cost of fault simulation experiments, especially for the key components.
To manifest the influence of the sample number on the classification accuracy, we use the PU bearing dataset to design a few-shot training pattern with six groups of different sample numbers in each class.

\begin{figure}[!t]\centering
	\centering
	\subfigure{\includegraphics[scale = 0.7]{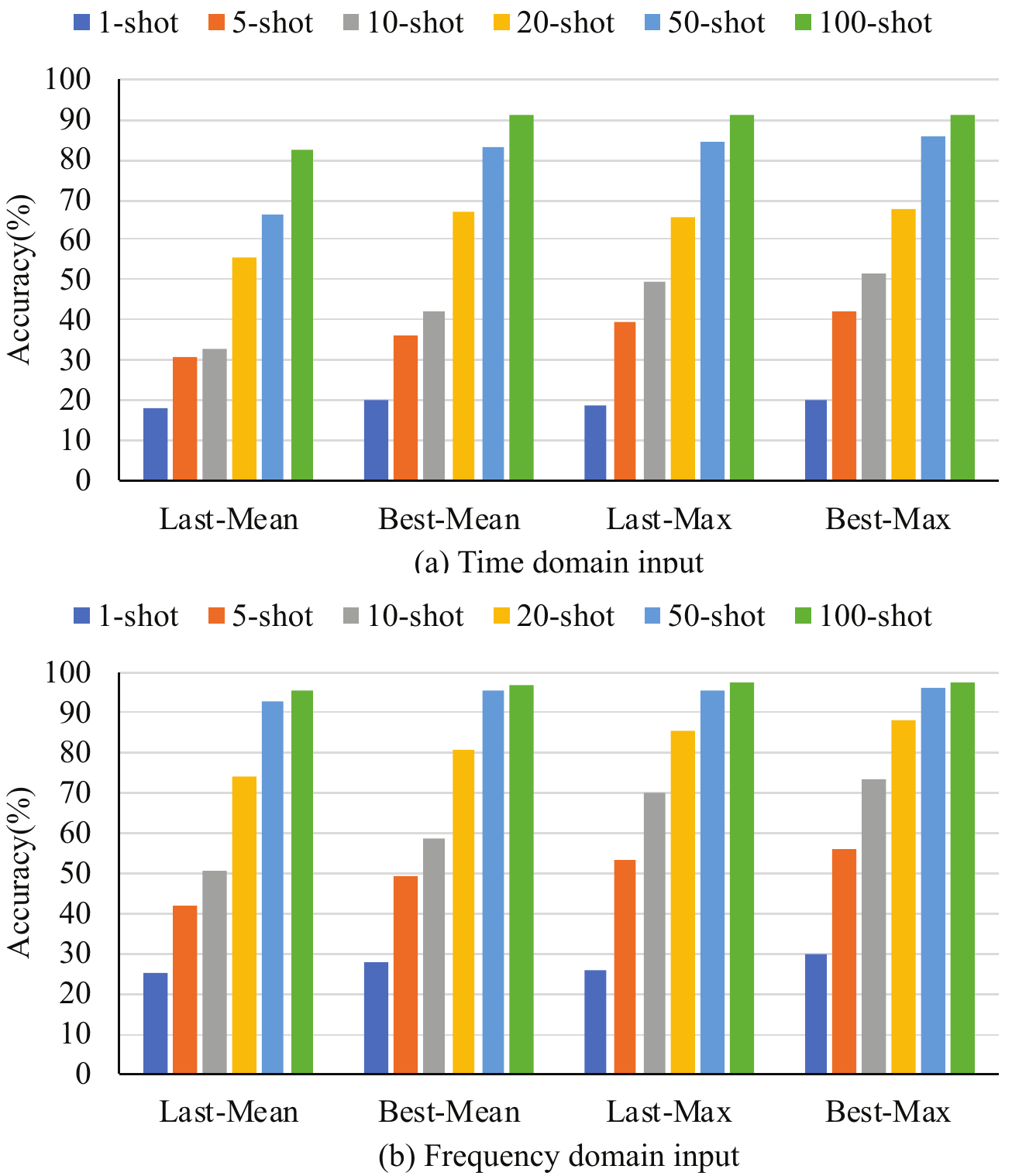}}
	\\ [-10pt]
	\caption{Experimental results of different few-shot training patterns. (a) time domain input, and (b) frequency domain input.}
	\label{FIG4}
\end{figure}

Results of the time domain input and the frequency domain input are shown in \figref{FIG4}.
It is shown that with the decrease of the sample number, the accuracy decreases sharply.
As shown in \figref{FIG4}, for the time domain input, the Best-Max accuracy decreases from 91.46\% to 20.39\% as the sample number decreases from 100 to 1.
Meanwhile, the Best-Max accuracy decreases from 97.73\% to 29.67\% as the sample number decreases from 100 to 1 with the frequency domain input.

Although the accuracy can be increased after using FFT, it is still too low to be accepted when the number of samples is extremely small.
It is necessary to develop methods based on few-shot learning to cope with application scenarios with limited samples.

Many DL-based few-shot learning models have been proposed in recent years \cite{wang2019generalizing}, most of these methods adopt a meta-learning paradigm by training networks with a large number of tasks, which means that the big data in other related fields is necessary for these methods.
In the field of fault diagnosis, there is no relevant data with such a big size available, so methods embedding with physical mechanisms are required to address this problem effectively.

\subsection{Model selection}
For intelligent diagnosis, designing a neural network is not the final goal, and our task is to apply the model to real industrial applications while designing a neural network is only a small part of our task.
However, to achieve a good effect, we have to spend considerable time and energy on designing the corresponding networks. 
Because building a neural network is an iterative process consisting of repeated trial and error, and the performance of models should be fed back to us to adjust models.
The single trial and error cost multiplied by the number of trial and error can easily reach a huge cost.
Besides, reducing this cost is also the partial purpose of this benchmark study which provides some guidelines to choose a baseline model.

Actually, there is another way called neural architecture search (NAS) \cite{elsken2019neural} to avoid the huge cost of trial and error.
NAS can allow designing a neural network automatically through searching for a specific network based on a specific dataset.
Limited search space of the network is first constructed according to the physical prior.
After that, a neural network matching a specific dataset is sampled from the search space through reinforcement learning, the evolutionary algorithm or the gradient strategy.
Besides, the whole construction process does not require manual participation, which greatly reduces the cost of building a neural network and allows us to focus on specific engineering applications.

\section{Conclusion}
\label{S:11}
In this paper, we collect nine publicly available datasets to evaluate the performance of MLP, AE, CNN, and RNN models.
This work mainly focuses on evaluating DL-based intelligent diagnosis algorithms from different perspectives and providing the benchmark accuracy (a lower bound) to avoid useless improvement.
In addition, we release a code library for other researchers to test the performance of their own DL-based intelligent diagnosis models of these datasets. 
We hope that the evaluation results and the code library could promote a better understanding of DL-based models and provide a unified framework for generating more effective models.
For further studies, we will focus on five listed issues (class imbalance, generalization ability, interpretability, few-shot learning, and model selection) to propose more customized works.

\section*{Acknowledgment}
This work was supported by Natural Science Foundation of China (No. 51835009, No. 51705398).

\bibliographystyle{References/model1-num-names}
\bibliography{References/Reference}

\begin{landscape}
	

\section*{Supplementary material}

# Appendix A. Evaluation Results

CWRU
Results with random split and data augmentation

| Nor | Input | Loc | AE Mean | AE Max | DAE Mean | DAE Max | SAE Mean | SAE Max | MLP Mean | MLP Max | CNN Mean | CNN Max | LeNet Mean | LeNet Max | AlexNet Mean | AlexNet Max | ResNet18 Mean | ResNet18 Max | LSTM Mean | LSTM Max |
|---|---|---|---|---|---|---|---|---|---|---|---|---|---|---|---|---|---|---|---|---|
| A | 1 | Last | 54.71 | 66.67 | 54.33 | 64.37 | 67.13 | 69.73 | 52.95 | 67.43 | **98.47** | **99.62** | 96.4 | **99.23** | 91.11 | **96.17** | 88.28 | **98.47** | **98.47** | **98.85** |
|   |   | Best | 71.42 | 73.95 | 78.24 | 80.84 | 72.95 | 76.25 | 73.49 | 75.48 | **98.93** | **100** | **99.08** | **99.62** | 91.11 | **96.17** | **100** | **100** | **98.47** | **98.85** |
|   | 2 | Last | **100** | **100** | **100** | **100** | 99.92 | **100** | **100** | **100** | 99.89 | **100** | **100** | **100** | **100** | **100** | 99.08 | **100** | **100** | **100** |
|   |   | Best | **100** | **100** | **100** | **100** | **100** | **100** | **100** | **100** | 99.92 | **100** | **100** | **100** | **100** | **100** | **100** | **100** | **100** | **100** |
|   | 3 | Last | 86.95 | 89.89 | 82.33 | 84.27 | 86.02 | 91.39 |   |   | 95.28 | 98.88 | 98.59 | 99.33 | 97.89 | 98.73 | 98.08 | 99.52 | **98.11** | **98.43** |
|   |   | Best | 90.95 | 92.51 | 86.14 | 88.39 | 90.01 | 91.76 |   |   | 97.03 | 99.18 | 99.17 | 99.33 | 97.89 | 98.73 | **99.66** | **99.85** | 98.11 | 98.43 |
|   | 4 | Last | **100** | **100** | **100** | **100** | **100** | **100** |   |   | 98.7 | **99.23** | 98.7 | **100** | 99.69 | **100** | **100** | **100** | 99.85 | **100** |
|   |   | Best | **100** | **100** | **100** | **100** | **100** | **100** |   |   | 99.46 | 99.62 | **100** | **100** | 99.69 | **100** | **100** | **100** | 99.85 | **100** |
|   | 5 | Last | 94.71 | **97.32** | 92.19 | 94.64 | 94.10 | **97.32** |   |   | 86.74 | 95.02 | 74.94 | 89.27 | 46.36 | 94.25 | 90.19 | **100** | 84.29 | 89.27 |
|   |   | Best | **96.86** | **98.47** | 95.94 | **97.32** | 96.09 | **98.85** |   |   | 94.86 | 96.93 | 89.81 | 90.8 | 46.36 | 94.25 | **100** | **100** | 84.29 | 89.27 |
| B | 1 | Last | 74.79 | 77.01 | 80.31 | 83.14 | 76.09 | 77.78 | 76.02 | 78.93 | **99.45** | **100** | 98.08 | 99.23 | 98.31 | 99.23 | **99.46** | **100** | 98.85 | **100** |
|   |   | Best | 78.85 | 80.46 | 82.68 | 83.52 | 77.78 | 78.54 | 76.71 | 79.69 | **99.67** | **100** | 98.83 | 99.88 | 98.54 | 99.62 | **99.46** | **100** | **99.92** | **100** |
|   | 2 | Last | 99.92 | **100** | **100** | **100** | 99.54 | **100** | 99.92 | **100** | 99.39 | **100** | **100** | **100** | 99.92 | **100** | 99.92 | **100** | 99.85 | **100** |
|   |   | Best | **100** | **100** | **100** | **100** | **100** | **100** | 99.92 | **100** | 99.39 | **100** | **100** | **100** | 99.92 | **100** | 99.92 | **100** | **100** | **100** |
|   | 3 | Last | 85.89 | 90.64 | 85.27 | 88.01 | 87.58 | 89.14 |   |   | 98.94 | 99.22 | 98.84 | 99.07 | 98.6 | 98.84 | **99.17** | **99.78** | 93.71 | 94.76 |
|   |   | Best | 88.83 | 92.13 | 89.01 | 92.13 | 90.7 | 92.13 |   |   | 99.12 | 99.29 | 99.15 | 99.36 | 98.71 | 99.03 | **99.6** | **99.95** | 95.36 | 95.88 |
|   | 4 | Last | **100** | **100** | **100** | **100** | **100** | **100** |   |   | 97.62 | 99.62 | 99 | 99.62 | 97.85 | **100** | 99.92 | **100** | 99.77 | **100** |
|   |   | Best | **100** | **100** | **100** | **100** | **100** | **100** |   |   | 98.93 | **100** | 99.23 | **100** | 98.31 | **100** | 99.92 | **100** | **100** | **100** |
|   | 5 | Last | 93.1 | 95.79 | 94.56 | 96.55 | 89.65 | 95.02 |   |   | 86.9 | 93.87 | 86.05 | 88.51 | 91.03 | **97.32** | **97.09** | **99.23** | 91.65 | 93.87 |
|   |   | Best | 95.33 | 96.17 | **97.01** | **98.85** | 93.95 | 97.32 |   |   | 89.46 | 95.64 | 86.67 | 90.04 | 92.27 | 97.32 | **97.62** | **100** | 93.33 | 94.64 |
| C | 1 | Last | 66.13 | 72.03 | 74.87 | 77.78 | 68.81 | 74.71 | 69.88 | 74.71 | **99.16** | **100** | 99.15 | 99.23 | **99.46** | **100** | 99.16 | **100** | 99.16 | 99.62 |
|   |   | Best | 70.8 | 72.8 | 77.32 | 79.69 | 71.8 | 74.71 | 71.95 | 74.71 | **100** | **100** | 99.69 | **100** | 99.92 | **100** | **100** | **100** | **100** | **100** |
|   | 2 | Last | **100** | **100** | **100** | **100** | **100** | **100** | **100** | **100** | 99.85 | **100** | **100** | **100** | **100** | **100** | **100** | **100** | **100** | **100** |
|   |   | Best | **100** | **100** | **100** | **100** | **100** | **100** | **100** | **100** | **100** | **100** | **100** | **100** | **100** | **100** | **100** | **100** | **100** | **100** |
|   | 3 | Last | 88.49 | 91.39 | 88.97 | 90.64 | 88.39 | 92.13 |   |   | 99.17 | 99.37 | 99.07 | 99.25 | 98.69 | 98.92 | **99.59** | **99.81** | 98.3 | 98.58 |
|   |   | Best | 91.35 | 93.63 | 91.25 | 92.51 | 90.94 | 93.63 |   |   | 99.43 | 99.55 | 99.23 | 99.37 | 99.13 | 99.25 | **99.86** | **99.89** | 98.86 | 99.07 |
|   | 4 | Last | **100** | **100** | **100** | **100** | **100** | **100** |   |   | 99.85 | **100** | 97.85 | **100** | **100** | **100** | **100** | **100** | 99.85 | **100** |
|   |   | Best | **100** | **100** | **100** | **100** | **100** | **100** |   |   | **100** | **100** | 99.92 | **100** | **100** | **100** | **100** | **100** | **100** | **100** |
|   | 5 | Last | 95.25 | **98.85** | 94.86 | 97.7 | 97.09 | 98.47 |   |   | 92.03 | 93.87 | 87.2 | 91.57 | 87.2 | 98.85 | 91.19 | **100** | 92.26 | 96.17 |
|   |   | Best | 97.93 | 99.62 | 96.47 | 98.47 | 98.62 | **100** |   |   | 94.41 | 95.02 | 92.11 | 93.1 | 98.08 | 99.23 | **100** | **100** | 96.01 | 97.7 |

*A is the max-min normalization; B is the -1-1 normalization; C is the Z-score normalization
1 is the time domain input; 2 is the frequency domain input; 3 is the wavelet domain input; 4 is the time domain sample after STFT; 5 is the time domain sample reshape to a 2D matrix
In the same input, the first line are the results of last epoch, and the second line are is the results of the best epoch.

CWRU
Results with random split and without data augmentation

| | | | AE | | SAE | | DAE | | MLP | | CNN | | LeNet | | AlexNet | | ResNet18 | | LSTM | |
|---|---|---|---|---|---|---|---|---|---|---|---|---|---|---|---|---|---|---|---|---|
| Nor | Input | Loc | Mean | Max | Mean | Max | Mean | Max | Mean | Max | Mean | Max | Mean | Max | Mean | Max | Mean | Max | Mean | Max |
| A | 1 | Last | 60.08 | 65.9 | 54.48 | 68.97 | 69.58 | 71.26 | 64.29 | 70.11 | **99.54** | **99.62** | **99.46** | **100** | 95.79 | 98.08 | **99.77** | **100** | **99.92** | **100** |
| | | Best | 70.96 | 73.56 | 71.34 | 76.63 | 71.57 | 73.18 | 74.41 | 76.63 | **99.85** | **100** | **99.77** | **100** | 97.4 | 98.47 | **100** | **100** | **100** | **100** |
| | 2 | Last | **100** | **100** | **100** | **100** | **100** | **100** | **100** | **100** | 99.85 | 100 | **100** | **100** | **100** | **100** | 99.77 | 100 | 99.92 | 100 |
| | | Best | **100** | **100** | **100** | **100** | **100** | **100** | **100** | **100** | 100 | 100 | 100 | 100 | 100 | 100 | 100 | 100 | 100 | 100 |
| | 3 | Last | 85.14 | 91.01 | 85.33 | 89.89 | 89.45 | 91.76 | | | 98.85 | 99.07 | 98.61 | 99.18 | 82.17 | 98.77 | 96.6 | 98.66 | 97.71 | 98.66 |
| | | Best | 88.08 | 91.39 | 87.08 | 91.39 | 91.01 | 93.63 | | | 99.35 | 99.4 | 99.16 | 99.18 | 82.59 | 98.84 | 99.55 | 99.63 | 98.67 | 98.84 |
| | 4 | Last | **100** | **100** | **100** | **100** | **100** | **100** | | | 99.31 | 100 | 99.69 | 100 | **100** | **100** | **100** | **100** | **100** | **100** |
| | | Best | **100** | **100** | **100** | **100** | **100** | **100** | | | 99.46 | 100 | 99.92 | 100 | 100 | 100 | 100 | 100 | 100 | 100 |
| | 5 | Last | 92.95 | **95.79** | 90.11 | **96.55** | 85.29 | **96.17** | | | 85.52 | 93.87 | 79.62 | 85.44 | 63.68 | 95.79 | 96.55 | 100 | 86.82 | 91.19 |
| | | Best | 93.87 | **96.93** | 93.41 | **98.47** | 94.94 | **97.7** | | | 94.64 | 96.55 | 88.81 | 90.04 | 67.2 | 97.32 | 100 | 100 | 90.96 | 93.1 |
| B | 1 | Last | 69.96 | 71.65 | 67.28 | 73.18 | 70.8 | 73.56 | 71.1 | 73.18 | **99.4** | **99.62** | **99.07** | **100** | 97.7 | 98.85 | **99.51** | **100** | **99.78** | **100** |
| | | Best | 74.48 | 76.63 | 71.11 | 73.95 | 73.18 | 76.25 | 75.26 | 78.54 | **99.62** | **100** | **99.56** | **100** | 98.58 | 99.62 | **100** | **100** | **100** | **100** |
| | 2 | Last | 99.62 | 100 | 99.92 | 100 | **100** | **100** | **100** | **100** | 99.95 | 100 | **100** | **100** | 99.89 | 100 | 97.37 | 100 | **100** | **100** |
| | | Best | **100** | **100** | **100** | **100** | **100** | **100** | **100** | **100** | 100 | 100 | 100 | 100 | 100 | 100 | 100 | 100 | 100 | 100 |
| | 3 | Last | 89.45 | 91.39 | 85.39 | 89.51 | 90.38 | 93.63 | | | 98.87 | 99.29 | 97.79 | 98.99 | 98.58 | 99.14 | 99.28 | 99.74 | 98.27 | 98.81 |
| | | Best | 90.32 | 92.13 | 88.14 | 90.26 | 91.2 | 94.01 | | | 99.21 | 99.4 | 99.15 | 99.33 | 99.04 | 99.33 | 99.58 | 99.74 | 98.82 | 98.96 |
| | 4 | Last | 90.65 | **100** | **100** | **100** | **100** | **100** | | | 99.4 | 100 | 99.45 | 100 | 99.78 | 100 | **100** | **100** | 99.78 | 100 |
| | | Best | **100** | **100** | **100** | **100** | **100** | **100** | | | 99.73 | 100 | 99.78 | 100 | 100 | 100 | 100 | 100 | 100 | 100 |
| | 5 | Last | 93.03 | **97.7** | 94.33 | **96.17** | 95.25 | **96.55** | | | 91.57 | 96.93 | 85.77 | 91.95 | 83.2 | 96.93 | 95.57 | 99.62 | 85.11 | 92.34 |
| | | Best | **96.09** | **97.7** | **96.63** | **97.32** | **96.32** | **97.7** | | | 95.46 | 96.93 | 89.82 | 94.25 | 86.75 | 97.32 | 100 | 100 | 94.25 | **95.4** |
| C | 1 | Last | 64.75 | 67.43 | 66.9 | 69.73 | 62.68 | 69.35 | 62.53 | 63.98 | **99.62** | **100** | **98.7** | **99.62** | 99 | 100 | **99.46** | **100** | **99.77** | **100** |
| | | Best | 68.28 | 70.5 | 69.81 | 71.26 | 67.36 | 70.88 | 68.43 | 71.26 | **99.7** | **100** | **99.85** | **100** | 99.69 | 100 | **100** | **100** | **100** | **100** |
| | 2 | Last | **100** | **100** | **100** | **100** | **100** | **100** | **100** | **100** | 100 | 100 | 100 | 100 | 100 | 100 | 100 | 100 | 100 | 100 |
| | | Best | **100** | **100** | **100** | **100** | **100** | **100** | **100** | **100** | 100 | 100 | 100 | 100 | 100 | 100 | 100 | 100 | 100 | 100 |
| | 3 | Last | 88.91 | 91.76 | 88.46 | 89.51 | 90.56 | 92.88 | | | 99.17 | 99.52 | 98.57 | 99.22 | 98.6 | 98.88 | 99.36 | 99.78 | 98.56 | 98.7 |
| | | Best | 89.44 | 92.88 | 88.99 | 90.64 | 92.06 | 94.38 | | | 99.46 | 99.55 | 99.35 | 99.52 | 99.14 | 99.4 | 99.87 | 99.89 | 99.05 | 99.14 |
| | 4 | Last | **100** | **100** | **100** | **100** | **100** | **100** | | | 99.69 | 100 | 99.77 | 100 | **100** | **100** | **100** | **100** | **100** | **100** |
| | | Best | **100** | **100** | **100** | **100** | **100** | **100** | | | 99.85 | 100 | 99.77 | 100 | 100 | 100 | 100 | 100 | 100 | 100 |
| | 5 | Last | **95.4** | **97.32** | 90.12 | **99.62** | 93.79 | **97.32** | | | 93.79 | 94.64 | 91.95 | 93.49 | 90.8 | 95.4 | 95.17 | 100 | 91.26 | 96.17 |
| | | Best | **96.47** | **98.47** | 93.94 | **100** | **97.09** | **98.08** | | | 94.94 | 95.4 | 93.03 | 94.25 | 97.62 | 98.47 | 100 | 100 | 95.79 | 97.32 |

CWRU
Results with order split and data augmentation

| Nor | Input | Loc | AE Mean | AE Max | SAE Mean | SAE Max | DAE Mean | DAE Max | MLP Mean | MLP Max | CNN Mean | CNN Max | LeNet Mean | LeNet Max | AlexNet Mean | AlexNet Max | ResNet18 Mean | ResNet18 Max | LSTM Mean | LSTM Max |
|---|---|---|---|---|---|---|---|---|---|---|---|---|---|---|---|---|---|---|---|---|
| A | 1 | Last | 58.71 | 70.08 | 55.84 | 69.7 | 64.55 | 68.18 | 65.23 | 73.11 | **95.99** | **98.11** | **97.04** | **98.86** | 91.21 | **95.83** | 82.65 | **99.62** | **97.27** | **98.48** |
|   |   | Best | 71.67 | 72.73 | 74.77 | 76.89 | 71.29 | 73.48 | 69.39 | 78.41 | **98.18** | **100** | **97.57** | **99.24** | 92.5 | **96.97** | **98.41** | **99.62** | **98.1** | **98.86** |
|   | 2 | Last | **100** | **100** | **100** | **100** | **100** | **100** | **100** | **100** | 99.54 | **100** | 99.85 | **100** | 99.92 | **100** | 98.79 | **100** | 99.85 | **100** |
|   |   | Best | **100** | **100** | **100** | **100** | **100** | **100** | **100** | **100** | 99.7 | **100** | **100** | **100** | **100** | **100** | 98.79 | **100** | 99.85 | **100** |
|   | 3 | Last | 72.69 | 78.1 | 73.9 | 76.64 | 76.64 | 79.2 |  |  | 87.15 | 88.32 | 96.67 | 97.54 | 95.12 | 96.76 | 93.16 | 99.29 | 95.17 | 96.76 |
|   |   | Best | 76.82 | 80.29 | 77.07 | 80.66 | 79.5 | 80.66 |  |  | 89.2 | 90.51 | 96.86 | 97.85 | 96.34 | 96.8 | 98.66 | 99.29 | 95.81 | 97.54 |
|   | 4 | Last | **99.92** | **100** | 99.77 | **100** | **100** | **100** |  |  | 97.73 | 98.11 | 99.47 | **100** | 97.27 | **100** | **100** | **100** | 99.01 | **100** |
|   |   | Best | **100** | **100** | **100** | **100** | **100** | **100** |  |  | 98.79 | 99.24 | 99.62 | **100** | 97.35 | **100** | **100** | **100** | 99.85 | **100** |
|   | 5 | Last | 92.35 | **97.73** | 92.35 | **98.11** | 92.2 | 96.21 |  |  | 78.56 | 95.45 | 83.18 | 90.53 | 93.47 | 96.59 | **100** | **100** | 87.35 | 90.15 |
|   |   | Best | **96.74** | **98.11** | **96.06** | **99.62** | **95.07** | **99.62** |  |  | 94.85 | 96.59 | 85.08 | 92.8 | 94.98 | 96.59 | **100** | **100** | 88.26 | 92.42 |
| B | 1 | Last | 70.46 | 72.73 | 76.36 | 78.79 | 70.91 | 74.24 | 72.24 | 73.48 | **98.7** | **99.24** | 96.81 | **98.86** | **97.08** | **97.73** | **99.24** | **100** | **98.41** | **99.24** |
|   |   | Best | 74.54 | 76.89 | 78.86 | 79.92 | 74.85 | 76.52 | 72.84 | 73.48 | **98.97** | **99.26** | **97.24** | **98.86** | **97.62** | **98.11** | **99.57** | **100** | **99.09** | **99.24** |
|   | 2 | Last | **100** | **100** | 99.85 | **100** | **100** | **100** | **100** | **100** | 98.81 | 99.62 | **100** | **100** | 99.56 | **100** | 97.73 | **100** | **100** | **100** |
|   |   | Best | **100** | **100** | **100** | **100** | **100** | **100** | **100** | **100** | 99.13 | **100** | **100** | **100** | 99.56 | **100** | 99.95 | **100** | **100** | **100** |
|   | 3 | Last | 76.95 | 79.93 | 74.21 | 80.29 | 76.83 | 78.83 |  |  | 97.04 | 97.92 | 96.68 | 97.65 | 96.88 | 97.65 | 98.19 | 98.92 | 86.72 | 88.32 |
|   |   | Best | 78.83 | 80.29 | 77.31 | 80.29 | 79.5 | 80.66 |  |  | 97.23 | 98.1 | 96.8 | 97.8 | 97.29 | 98.29 | 98.53 | 98.92 | 88.69 | 89.78 |
|   | 4 | Last | **99.92** | **100** | 99.92 | **100** | 98.94 | **100** |  |  | 98.05 | 98.48 | 99.24 | 99.62 | 97.24 | **100** | 99.57 | **100** | 98.94 | 99.62 |
|   |   | Best | **100** | **100** | **100** | **100** | **100** | **100** |  |  | 98.37 | 98.48 | 99.57 | **100** | 99.3 | **100** | 99.57 | **100** | 99.92 | **100** |
|   | 5 | Last | 94.32 | **97.73** | 92.35 | **98.11** | 91.21 | 95.08 |  |  | 88.53 | 96.59 | 92.64 | 94.32 | 93.02 | **97.73** | 96.91 | **100** | 89.32 | 94.32 |
|   |   | Best | **97.12** | **98.86** | **95.53** | **98.86** | **95.83** | **96.97** |  |  | 88.8 | 96.59 | 93.51 | 95.83 | 93.35 | **97.73** | 96.91 | **100** | 92.88 | 94.32 |
| C | 1 | Last | 64.24 | 68.94 | 73.79 | 77.27 | 65.46 | 67.05 | 66.16 | 68.18 | **98.7** | **99.62** | 97.98 | **98.86** | **98.17** | **99.24** | 97.98 | **100** | **98.56** | **98.86** |
|   |   | Best | 67.73 | 70.83 | 75.15 | 77.27 | 67.12 | 68.94 | 67.17 | 72.35 | **99.02** | **99.62** | **98.23** | **99.24** | **98.29** | **99.24** | **100** | **100** | **98.79** | **99.24** |
|   | 2 | Last | **100** | **100** | **100** | **100** | **100** | **100** | **100** | **100** | 99.3 | **100** | 99.94 | **100** | 99.94 | **100** | 99.68 | **100** | **100** | **100** |
|   |   | Best | **100** | **100** | **100** | **100** | **100** | **100** | **100** | **100** | 99.78 | **100** | 99.94 | **100** | **100** | **100** | **100** | **100** | **100** | **100** |
|   | 3 | Last | 78.29 | 79.93 | 75.79 | 78.47 | 77.07 | 81.75 |  |  | 96.54 | 97.99 | 97.05 | 97.77 | 95.45 | 98.21 | 98.93 | 99.55 | 96.44 | 97.02 |
|   |   | Best | 81.63 | 84.31 | 80.17 | 83.21 | 80.84 | 84.67 |  |  | 97.61 | 98.36 | 97.29 | 98.18 | 96.32 | 98.21 | 99.28 | 99.55 | 96.69 | 97.32 |
|   | 4 | Last | **99.92** | **100** | **100** | **100** | **100** | **100** |  |  | 98.97 | 99.62 | 99.56 | **100** | 99.77 | **100** | **100** | **100** | 99.77 | **100** |
|   |   | Best | **100** | **100** | **100** | **100** | **100** | **100** |  |  | 99.08 | 99.62 | 99.68 | **100** | 99.77 | **100** | **100** | **100** | 99.92 | **100** |
|   | 5 | Last | 93.71 | **97.35** | 87.5 | 93.18 | 94.09 | **98.11** |  |  | 93.34 | 97.35 | 93.56 | 95.08 | 89.01 | 93.94 | 98.18 | **100** | 93.64 | 95.83 |
|   |   | Best | **96.21** | **97.73** | **92.42** | **97.35** | **96.74** | **98.11** |  |  | 93.56 | 97.35 | 94.07 | 95.83 | 89.92 | 95.45 | 98.41 | **100** | 94.52 | 95.83 |

JNU
Results with random split and data augmentation

| Nor | Input | Loc | AE Mean | AE Max | DAE Mean | DAE Max | SAE Mean | SAE Max | MLP Mean | MLP Max | CNN Mean | CNN Max | LeNet Mean | LeNet Max | AlexNet Mean | AlexNet Max | ResNet18 Mean | ResNet18 Max | LSTM Mean | LSTM Max |
|---|---|---|---|---|---|---|---|---|---|---|---|---|---|---|---|---|---|---|---|---|
| A | 1 | Last | 46.74 | 49.03 | 46.01 | 50.51 | 43.8 | 45.9 | 46.97 | 49.32 | 80.57 | 83.28 | 73.23 | 77.25 | 82.97 | 83.9 | 77.14 | 91.47 | 82.14 | 83.62 |
|   |   | Best | 52.6 | 54.72 | 51.19 | 53.58 | 53.41 | 54.66 | 52.48 | 53.81 | 82.28 | 85.32 | 77.23 | 79.18 | 83.86 | 85.32 | 93.07 | 93.69 | 83.2 | 84.7 |
|   | 2 | Last | **95.93** | **96.3** | **95.2** | **95.73** | 63.47 | **95.51** | **95.89** | **96.76** | 93.83 | 94.48 | 94.59 | **95.45** | 95.82 | **96.19** | 95.61 | 96.19 | 95.02 | 95.56 |
|   |   | Best | **96.62** | **96.93** | **95.58** | **96.13** | 64.52 | **96.64** | **97.21** | **97.38** | 94.36 | 94.99 | **95.86** | **96.02** | 96.23 | 96.7 | 96.76 | 96.99 | 95.36 | 95.56 |
|   | 3 | Last | 33.11 | 36.2 | 30.47 | 31.76 | 33.99 | 37.92 |  |  | 43.5 | 45.81 | 44.93 | 46.56 | 37.04 | 44.87 | 52.28 | 55.19 | 42.93 | 44.17 |
|   |   | Best | 37.42 | 41.42 | 36.48 | 36.92 | 38.98 | 40.92 |  |  | 46.66 | 48.25 | 48.66 | 49.28 | 37.59 | 44.87 | 55.51 | 57.16 | 43.55 | 45.27 |
|   | 4 | Last | 78.63 | 80.34 | 78.92 | 80.63 | 77.55 | 82.91 |  |  | 69.69 | 73.22 | 66.84 | 69.52 | 79.77 | 81.77 | 87.12 | 90.88 | 71.11 | 73.22 |
|   |   | Best | 82.91 | 83.76 | 82.68 | 84.05 | 81.99 | 86.32 |  |  | 73.16 | 75.21 | 70.66 | 72.36 | 80.46 | 84.05 | 91.17 | 92.02 | 72.71 | 74.07 |
|   | 5 | Last | 54.81 | 57.83 | 50.26 | 54.7 | 45.24 | 51.57 |  |  | 39.94 | 43.3 | 43.65 | 50.71 | 50.71 | 55.27 | 75.44 | 80.91 | 54.3 | 56.7 |
|   |   | Best | 57.44 | 60.11 | 55.96 | 57.26 | 47.98 | 54.99 |  |  | 48.15 | 50.43 | 49.12 | 52.14 | 52.31 | 57.83 | 81.94 | 83.76 | 56.53 | 60.4 |
| B | 1 | Last | 60.36 | 61.6 | 16.67 | 16.67 | 61.14 | 62 | 61.97 | 63.54 | 84.96 | 86.41 | 78.2 | 80.49 | 87.2 | 88.4 | 92.43 | 93.63 | 83.24 | 84.36 |
|   |   | Best | 61.93 | 63.14 | 16.67 | 16.67 | 62.26 | 62.8 | 62.88 | 64.11 | 86.2 | 86.86 | 78.55 | 80.89 | 87.99 | 88.51 | 92.83 | 93.63 | 84.46 | 84.93 |
|   | 2 | Last | 94.97 | **96.59** | 94.63 | **96.08** | 95.69 | **96.76** | 92.73 | **96.53** | 93.94 | 94.6 | 94.11 | **95.22** | 95.56 | 95.79 | 95.73 | 95.96 | 94.73 | 95.34 |
|   |   | Best | **96.7** | **96.93** | **96.59** | **97.1** | **96.66** | **96.76** | 94.31 | **97.04** | 94.38 | 94.88 | 94.91 | **95.73** | 95.71 | 96.25 | 95.92 | 96.13 | 95.64 | 95.9 |
|   | 3 | Last | 33.33 | 34.65 | 33 | 34.87 | 33.96 | 37.09 |  |  | 46.69 | 47.67 | 46.82 | 48.03 | 48.63 | 52.83 | 51.37 | 52.44 | 37.79 | 38.37 |
|   |   | Best | 39 | 39.81 | 38.42 | 40.2 | 39.09 | 40.64 |  |  | 47.5 | 50.53 | 47.42 | 48.64 | 48.78 | 53.3 | 52.58 | 53.39 | 43.07 | 43.75 |
|   | 4 | Last | 76.69 | 77.78 | 80.46 | 81.48 | 77.49 | 79.77 |  |  | 65.93 | 72.08 | 65.81 | 71.51 | 76.47 | 78.06 | 85.58 | 90.03 | 68.03 | 73.22 |
|   |   | Best | 81.03 | 82.62 | 84.44 | 86.89 | 82.56 | 84.9 |  |  | 67.41 | 76.64 | 68.49 | 71.51 | 77.89 | 83.48 | 87.29 | 92.31 | 73.67 | 75.21 |
|   | 5 | Last | 48.43 | 56.98 | 48.03 | 52.14 | 39.03 | 44.73 |  |  | 43.87 | 47.86 | 44.21 | 46.15 | 56.52 | 58.69 | 77.72 | 79.49 | 55.67 | 57.26 |
|   |   | Best | 55.16 | 62.11 | 52.71 | 56.7 | 44.33 | 51.28 |  |  | 44.96 | 48.15 | 45.58 | 47.01 | 57.78 | 61.25 | 78.83 | 80.63 | 60.28 | 60.68 |
| C | 1 | Last | 65.75 | 66.78 | 52.23 | 62.17 | 67.14 | 68.15 | 67.91 | 69 | 89.78 | 90.39 | 85.71 | 86.29 | 90.89 | 91.7 | 94.15 | 94.88 | 87.65 | 88.57 |
|   |   | Best | 66.66 | 67.35 | 52.96 | 62.74 | 67.71 | 68.71 | 68.98 | 69.28 | 91.24 | 91.47 | 86.73 | 87.88 | 91.99 | 92.72 | **95.68** | **95.85** | 89.2 | 89.65 |
|   | 2 | Last | **96.35** | **96.7** | **96.34** | **96.7** | **96.22** | **96.36** | **96.58** | **97.16** | 93.94 | 94.65 | **95.03** | **95.73** | **96.02** | **96.59** | 95.64 | 96.53 | 94.9 | **95.45** |
|   |   | Best | **97.01** | **97.21** | **97.19** | **97.38** | **96.92** | **97.1** | **97.14** | **97.27** | 95.05 | 95.62 | **95.94** | **96.25** | **96.64** | **96.99** | 96.59 | 96.87 | 95.62 | 95.85 |
|   | 3 | Last | 34.14 | 37.2 | 34.14 | 36.81 | 32.89 | 35.37 |  |  | 48.69 | 50.25 | 48.19 | 52.64 | 50.41 | 52.14 | 52.34 | 53.77 | 42.24 | 43.62 |
|   |   | Best | 38.92 | 41.03 | 37.88 | 39.98 | 38.99 | 40.09 |  |  | 52.21 | 53.22 | 51.16 | 52.64 | 53.13 | 54.08 | 55.29 | 55.6 | 46.91 | 48.34 |
|   | 4 | Last | 77.09 | 78.06 | 80.91 | 84.33 | 77.32 | 80.91 |  |  | 69.69 | 73.5 | 69.86 | 71.23 | 81.37 | 82.91 | 91.28 | 93.45 | 73.22 | 75.5 |
|   |   | Best | 82.74 | 83.76 | 84.16 | 86.04 | 83.65 | 84.62 |  |  | 74.76 | 76.64 | 72.42 | 74.07 | 86.44 | 87.46 | 93.11 | 93.73 | 76.75 | 77.49 |
|   | 5 | Last | 63.13 | 66.1 | 61.2 | 64.1 | 62.45 | 66.67 |  |  | 54.3 | 58.4 | 50.14 | 51.57 | 70.09 | 73.5 | 84.73 | 87.18 | 61.71 | 64.67 |
|   |   | Best | 65.98 | 68.38 | 65.81 | 68.66 | 67.12 | 69.8 |  |  | 58.4 | 60.68 | 55.67 | 58.4 | 75.21 | 76.35 | 86.38 | 87.46 | 65.93 | 67.52 |

*A is the max-min normalization; B is the -1-1 normalization; C is the Z-score normalization

1 is the time domain input, 2 is the frequency domain input; 3 is the wavelet domain input; 4 is the time domain sample after STFT; 5 is the time domain sample reshape to a 2D matrix

JNU
Results with random split and without data augmentation

| Nor | Input | Loc | AE Mean | AE Max | SAE Mean | SAE Max | DAE Mean | DAE Max | MLP Mean | MLP Max | CNN Mean | CNN Max | LeNet Mean | LeNet Max | AlexNet Mean | AlexNet Max | ResNet18 Mean | ResNet18 Max | LSTM Mean | LSTM Max |
|---|---|---|---|---|---|---|---|---|---|---|---|---|---|---|---|---|---|---|---|---|
| A | 1 | Last | 41.64 | 45.11 | 39.06 | 42.78 | 38.39 | 41.92 | 40.6 | 44.94 | 80.79 | 82.31 | 63.06 | 69.4 | 77.57 | 79.12 | 87.51 | 90.96 | 76.08 | 77.42 |
| | | Best | 44.68 | 45.96 | 43.12 | 44.2 | 45.83 | 48.75 | 45.17 | 46.25 | 83.56 | 84.81 | 71.96 | 72.81 | 80.19 | 81.51 | 90.26 | 91.24 | 78.88 | 81.8 |
| | 2 | Last | **95.77** | **96.47** | 95.51 | 97.04 | 80.21 | **96.3** | 96.52 | 97.21 | 92.88 | 94.25 | **95.05** | **95.62** | 95.37 | 95.9 | 96.49 | 97.38 | 95.23 | 95.79 |
| | | Best | **96.78** | **96.99** | 96.84 | 97.1 | 80.79 | 96.99 | 97.44 | 97.78 | 94.01 | 94.6 | **95.49** | **95.96** | 96.55 | 96.76 | 97.4 | 97.72 | 95.93 | 96.53 |
| | 3 | Last | 32.96 | 34.81 | 29.17 | 29.93 | 35.01 | 37.26 | | | 43.7 | 48.11 | 40.85 | 47.23 | 44.6 | 47.25 | 45.43 | 51.69 | 40.74 | 44.34 |
| | | Best | 37.06 | 38.59 | 36.8 | 38.76 | 38.51 | 40.53 | | | 47.16 | 51.66 | 47.36 | 49.67 | 46.64 | 49.42 | 52.81 | 56.22 | 45.7 | 47.5 |
| | 4 | Last | 81.31 | 86.04 | 81.14 | 87.18 | 82.45 | 84.62 | | | 69.74 | 72.36 | 67.24 | 68.95 | 78.58 | 81.2 | 90.19 | 91.74 | 71.32 | 76.07 |
| | | Best | 83.76 | 87.18 | 85.76 | 88.32 | 83.88 | 85.47 | | | 72.31 | 75.5 | 69.83 | 72.08 | 83.25 | 84.9 | 91.77 | 92.59 | 74.74 | 76.64 |
| | 5 | Last | 47.52 | 56.41 | 50.14 | 56.13 | 45.01 | 50.14 | | | 42.85 | 46.44 | 42.48 | 45.87 | 52.25 | 54.42 | 72.62 | 79.49 | 53.72 | 56.41 |
| | | Best | 52.48 | 58.69 | 54.59 | 58.4 | 48.2 | 53.56 | | | 48.66 | 49.57 | 48.35 | 52.14 | 56.52 | 58.69 | 80.69 | 82.91 | 59.42 | 60.68 |
| B | 1 | Last | 50.4 | 52.1 | 16.67 | 16.67 | 51.09 | 52.84 | 51.87 | 53.53 | 81.02 | 83.67 | 70.67 | 72.24 | 81.64 | 83.45 | 86.97 | 88.74 | 77.93 | 78.67 |
| | | Best | 53.31 | 54.1 | 16.67 | 16.67 | 53.29 | 54.21 | 55.49 | 55.75 | 82.31 | 83.67 | 72.41 | 73.15 | 82.94 | 84.41 | 89.94 | 90.39 | 78.43 | 80.03 |
| | 2 | Last | **95.58** | **96.87** | 94.29 | 96.42 | 95.65 | 96.19 | 88.35 | 94.77 | 92.08 | 94.25 | **95.38** | **96.42** | 95.31 | 96.08 | 96.81 | 97.38 | 94.73 | 95.73 |
| | | Best | **96.96** | **97.44** | 97.11 | 97.27 | 96.91 | 97.38 | 97.18 | 97.61 | 93.13 | 94.54 | **95.86** | **96.42** | 95.43 | 96.08 | 97.51 | 97.67 | 94.98 | 95.73 |
| | 3 | Last | 33.73 | 36.76 | 35.28 | 36.92 | 35.46 | 37.98 | | | 46.72 | 48.31 | 47.58 | 48.5 | 50.17 | 51.22 | 51.6 | 52.66 | 41.3 | 42.34 |
| | | Best | 38.16 | 39.76 | 39.05 | 39.98 | 39.29 | 40.48 | | | 47.7 | 49.67 | 50.51 | 51.55 | 50.82 | 51.44 | 56.01 | 56.38 | 41.75 | 43.23 |
| | 4 | Last | 84.1 | 90.03 | 81.82 | 86.32 | 79.03 | 81.2 | | | 72.08 | 72.65 | 68.26 | 70.94 | 76.18 | 78.06 | 85.13 | 90.31 | 69.29 | 70.66 |
| | | Best | 85.47 | 90.6 | 83.76 | 87.75 | 80.23 | 82.34 | | | 72.93 | 73.79 | 70.09 | 72.36 | 77.55 | 78.92 | 91.51 | 92.02 | 69.92 | 72.08 |
| | 5 | Last | 54.64 | 60.11 | 52.76 | 59.83 | 43.82 | 45.58 | | | 44.3 | 44.44 | 46.38 | 48.43 | 54.87 | 61.25 | 78.52 | 80.34 | 54.13 | 56.98 |
| | | Best | 56.58 | 61.82 | 54.7 | 61.54 | 45.24 | 49 | | | 50.14 | 52.71 | 51.28 | 52.99 | 56.47 | 61.25 | 79.54 | 81.48 | 55.84 | 57.55 |
| C | 1 | Last | 55.22 | 56.43 | 45.39 | 47.72 | 55.09 | 55.86 | 53.66 | 54.72 | 87.29 | 88 | 77.53 | 78.84 | 86.65 | 88.79 | 92.67 | 93.57 | 87.55 | 87.88 |
| | | Best | 57.18 | 57.74 | 47.48 | 48.41 | 56.88 | 57.74 | 56.22 | 56.71 | 88.33 | 89.19 | 79.92 | 82.82 | 88.77 | 89.19 | 94.43 | 94.88 | 88.71 | 89.25 |
| | 2 | Last | **96.19** | **96.53** | 96.24 | 96.81 | 95.79 | 96.25 | 96.03 | 96.47 | 93.72 | 94.71 | **95.14** | **95.51** | 95.45 | 96.08 | 96.1 | 97.16 | 95.26 | 95.79 |
| | | Best | **96.8** | **96.99** | 97.34 | 97.38 | 96.9 | 97.04 | 97.19 | 97.33 | 94.34 | 95.05 | **95.53** | **95.96** | 96.1 | 96.53 | 97.53 | 97.9 | 95.92 | 96.13 |
| | 3 | Last | 35.82 | 37.03 | 34.59 | 36.92 | 33.34 | 38.03 | | | 47.91 | 49.42 | 46.34 | 47.84 | 42.51 | 50.08 | 49.11 | 52.36 | 42.67 | 42.98 |
| | | Best | 38.95 | 40.26 | 37.91 | 39.64 | 38.52 | 41.31 | | | 52.77 | 53.63 | 49.21 | 49.97 | 44.43 | 53.33 | 54.55 | 55.22 | 46.91 | 47.14 |
| | 4 | Last | 79.48 | 85.75 | 81.58 | 82.34 | 84.14 | 84.9 | | | 71.62 | 73.5 | 69.57 | 72.36 | 84.05 | 86.04 | 91.4 | 93.73 | 73.96 | 74.93 |
| | | Best | 83.26 | 86.89 | 83.19 | 83.48 | 85.57 | 86.61 | | | 73.16 | 74.93 | 71.62 | 74.07 | 86.27 | 88.32 | 93.62 | 94.87 | 77.09 | 78.92 |
| | 5 | Last | 62.89 | 65.24 | 60.87 | 65.24 | 64.96 | 66.95 | | | 54.87 | 58.97 | 50.88 | 54.7 | 69.86 | 72.36 | 85.93 | 88.03 | 62.16 | 65.81 |
| | | Best | 64.6 | 67.81 | 63.44 | 66.67 | 66.86 | 69.23 | | | 57.95 | 59.54 | 55.67 | 58.4 | 73.05 | 74.36 | 86.72 | 88.32 | 65.53 | 66.95 |

JNU
Results with order split and data augmentation

| | | | AE | | SAE | | DAE | | MLP | | CNN | | LeNet | | AlexNet | | ResNet18 | | LSTM | |
|---|---|---|---|---|---|---|---|---|---|---|---|---|---|---|---|---|---|---|---|---|
| Nor | Input | Loc | Mean | Max | Mean | Max | Mean | Max | Mean | Max | Mean | Max | Mean | Max | Mean | Max | Mean | Max | Mean | Max |
| A | 1 | Last | 42.16 | 43.88 | 42.67 | 45.18 | 44.72 | 48.64 | 46.76 | 49.77 | 79.91 | 83.62 | 72.99 | 77.55 | 81.54 | 82.43 | 90.84 | 93.59 | 82.22 | 85.09 |
| | | Best | 52.07 | 52.27 | 49.67 | 51.3 | 51.63 | 54.25 | 51.68 | 54.25 | 85.33 | 86.05 | 77.61 | 78.46 | 81.54 | 82.43 | 93.11 | 93.59 | 84.2 | 85.71 |
| | 2 | Last | 94.92 | **95.46** | 94.66 | **95.92** | 48.11 | **95.29** | **95.95** | **96.2** | 93.84 | 94.56 | 94.47 | 94.73 | **96.1** | **96.54** | 95.83 | **96.2** | **95.28** | **95.63** |
| | | Best | **96.03** | **96.09** | **95.16** | **96.26** | 48.37 | **95.98** | **96.8** | **97.05** | 94.81 | **95.07** | **95.42** | **95.58** | **96.1** | **96.54** | **97.1** | **97.34** | **95.99** | **96.2** |
| | 3 | Last | 32.66 | 35.66 | 29.53 | 30.84 | 32.6 | 34.39 | | | 42.73 | 47.81 | 46.23 | 48.56 | 36.18 | 43.47 | 52.2 | 53.65 | 42.51 | 44.96 |
| | | Best | 36.93 | 38.82 | 36.05 | 37.94 | 37.83 | 38.44 | | | 48.73 | 49.42 | 49.84 | 50.75 | 37.87 | 43.47 | 54.94 | 55.68 | 47.75 | 49.03 |
| | 4 | Last | 75.28 | 79.55 | 76.19 | 77.31 | 75.41 | 76.75 | | | 68.57 | 72.55 | 69.25 | 70.31 | 79.41 | 82.35 | 81.57 | 89.08 | 74.37 | 75.35 |
| | | Best | 80.11 | 84.31 | 80.56 | 82.35 | 79.77 | 82.07 | | | 71.82 | 75.07 | 71.54 | 72.27 | 82.84 | 84.59 | 90.31 | 90.76 | 76.89 | 77.87 |
| | 5 | Last | 53.36 | 55.18 | 49.36 | 57.98 | 42.52 | 50.98 | | | 38.38 | 45.38 | 45.72 | 49.3 | 54.23 | 56.02 | 63.7 | 80.11 | 50.59 | 56.02 |
| | | Best | 57.42 | 61.62 | 55.4 | 60.5 | 48.96 | 52.38 | | | 49.41 | 51.26 | 50.7 | 51.54 | 56.75 | 58.82 | 80.78 | 84.31 | 57.42 | 58.26 |
| B | 1 | Last | 60.38 | 61.62 | 37.86 | 54.37 | 59.24 | 60.94 | 62.01 | 63.15 | 83.89 | 86.34 | 77.22 | 79.59 | 86.24 | 88.1 | 92.8 | 93.65 | 75.64 | 83.73 |
| | | Best | 61.32 | 61.85 | 39.07 | 55.27 | 60.48 | 61.39 | 62.01 | 63.15 | 85.63 | 86.85 | 77.22 | 79.59 | 86.95 | 88.76 | 93.11 | 93.65 | 76.66 | 84.86 |
| | 2 | Last | **95.5** | **96.09** | **95.12** | **95.86** | **95.09** | **95.92** | 94.43 | **96.15** | 93.32 | 93.82 | 93.59 | 94.27 | **95.77** | **96.43** | **95.63** | **96.71** | 94.55 | 94.9 |
| | | Best | **96.75** | **97.05** | **96.37** | **96.6** | **96.45** | **96.54** | 94.8 | **96.15** | 93.58 | 94.33 | 93.83 | **95.12** | **95.99** | **96.77** | **96.18** | **96.71** | **95.62** | **95.75** |
| | 3 | Last | 32.75 | 35.16 | 32.6 | 34.39 | 32.78 | 34.33 | | | 47.91 | 48.64 | 47.3 | 49 | 51.27 | 53.88 | 52.05 | 53.38 | 40.85 | 41.82 |
| | | Best | 38.26 | 40.27 | 38.25 | 39.66 | 38.47 | 38.99 | | | 48.29 | 49.22 | 48.04 | 49 | 51.42 | 53.88 | 52.98 | 53.49 | 45.28 | 46.15 |
| | 4 | Last | 72.1 | 81.51 | 75.24 | 77.87 | 73.78 | 77.59 | | | 68.13 | 69.75 | 67.45 | 71.15 | 74.23 | 75.91 | 85.49 | 86.27 | 69.75 | 71.71 |
| | | Best | 74.96 | 85.99 | 77.93 | 81.23 | 77.81 | 82.07 | | | 68.6 | 70.03 | 67.45 | 71.15 | 74.96 | 78.15 | 87.06 | 89.64 | 73.45 | 76.47 |
| | 5 | Last | 50.76 | 52.94 | 51.99 | 54.62 | 43.64 | 50.42 | | | 45.27 | 47.06 | 47.23 | 50.14 | 55.24 | 56.86 | 76.41 | 81.23 | 54.34 | 56.86 |
| | | Best | 57.48 | 58.82 | 56.52 | 57.98 | 47 | 54.06 | | | 45.6 | 47.62 | 47.96 | 50.98 | 56.59 | 60.22 | 78.99 | 81.23 | 58.88 | 59.94 |
| C | 1 | Last | 65.76 | 67.35 | 60.07 | 62.81 | 65.53 | 67.74 | 66.79 | 68.54 | 87.94 | 90.99 | 85.48 | 86.28 | 90.7 | 91.33 | 94.06 | **95.18** | 87.52 | 88.21 |
| | | Best | 66.39 | 67.35 | 61.04 | 63.15 | 66.59 | 67.74 | 67.35 | 68.54 | 88.96 | 90.99 | 85.61 | 86.28 | 91.32 | 92.12 | 94.27 | **95.18** | 87.78 | 88.55 |
| | 2 | Last | **95.74** | **96.6** | **96.01** | **96.77** | **96.27** | **96.54** | **96.12** | **96.54** | 93.59 | 94.22 | 94.72 | **95.58** | **95.83** | **97.05** | **95.34** | **96.43** | 94.62 | **95.35** |
| | | Best | **96.71** | **96.88** | **96.87** | **97.22** | **96.69** | **96.88** | **96.34** | **96.54** | 93.81 | 95.01 | 94.72 | **95.58** | **96.08** | **97.05** | **95.52** | **96.43** | 94.84 | **95.35** |
| | 3 | Last | 33.84 | 36 | 32.95 | 34 | 33.25 | 35.39 | | | 48.94 | 50.28 | 49.1 | 50.55 | 44.06 | 51.88 | 51.17 | 55.84 | 43.03 | 44.63 |
| | | Best | 38.43 | 39.93 | 37.11 | 38.44 | 37.42 | 39.93 | | | 49.22 | 50.28 | 49.29 | 50.55 | 44.38 | 52.77 | 52.65 | 55.84 | 43.42 | 44.63 |
| | 4 | Last | 78.43 | 82.07 | 75.35 | 77.59 | 79.66 | 81.23 | | | 70.12 | 71.71 | 69.28 | 72.55 | 80.67 | 82.07 | 87 | 90.76 | 72.66 | 75.35 |
| | | Best | 81.96 | 85.43 | 77.65 | 80.39 | 83.92 | 84.31 | | | 70.73 | 71.71 | 69.94 | 72.55 | 81.4 | 84.03 | 90.26 | 90.76 | 73.73 | 78.43 |
| | 5 | Last | 64.76 | 69.19 | 60.05 | 63.87 | 62.13 | 64.15 | | | 55.97 | 58.54 | 53.08 | 55.18 | 67.85 | 70.03 | 66.38 | 87.39 | 61.4 | 63.59 |
| | | Best | 69.02 | 72.83 | 67.34 | 72.83 | 65.72 | 71.43 | | | 56.49 | 58.54 | 53.38 | 55.18 | 68.01 | 70.03 | 66.72 | 87.39 | 62.19 | 63.59 |

MFPT
Results with random split and data augmentation

| Nor | Input | Loc | AE Mean | AE Max | DAE Mean | DAE Max | SAE Mean | SAE Max | MLP Mean | MLP Max | CNN Mean | CNN Max | LeNet Mean | LeNet Max | AlexNet Mean | AlexNet Max | ResNet18 Mean | ResNet18 Max | LSTM Mean | LSTM Max |
|---|---|---|---|---|---|---|---|---|---|---|---|---|---|---|---|---|---|---|---|---|
| A | 1 | Last | 35.84 | 39.22 | 35.07 | 40.97 | 35.92 | 38.45 | 30.63 | 36.12 | 86.72 | 87.57 | 65.9 | 67.77 | 77.16 | 79.61 | 93.86 | **96.5** | 76.82 | 78.64 |
|   |   | Best | 41.83 | 45.83 | 41.09 | 42.72 | 41.59 | 42.91 | 40.29 | 41.36 | 89.36 | 91.07 | 69.63 | 71.46 | 78.6 | 80.97 | **95.92** | **96.89** | 78.52 | 79.61 |
|   | 2 | Last | 94.21 | 94.76 | 94.76 | **95.92** | 94.6 | **95.92** | 94.84 | **95.15** | 85.63 | 88.54 | 91.81 | 93.4 | 92.04 | 92.43 | 91.8 | 93.4 | 91.88 | 92.82 |
|   |   | Best | **95.53** | **95.92** | **95.84** | **96.12** | **95.61** | **95.92** | **96.23** | **96.5** | 88.39 | 89.51 | 92.97 | 93.4 | 92.97 | 93.59 | 94.6 | **95.34** | 93.09 | 93.59 |
|   | 3 | Last | 23.8 | 26.62 | 23.91 | 27.76 | 26.62 | 29.66 |  |  | 37.98 | 40.3 | 38.63 | 41.83 | 41.25 | 43.92 | 46.96 | 47.91 | 40 | 41.44 |
|   |   | Best | 28.03 | 28.9 | 27.68 | 29.85 | 30.34 | 33.84 |  |  | 42.36 | 43.35 | 42.32 | 43.73 | 43.58 | 45.25 | 48.86 | 49.81 | 43.39 | 43.92 |
|   | 4 | Last | 88.47 | 90.68 | 87.23 | 88.74 | 88.35 | 90.68 |  |  | 70.21 | 75.15 | 77.09 | 78.45 | 88.23 | 90.29 | 91.26 | **95.15** | 80.97 | 83.3 |
|   |   | Best | 90.52 | 92.04 | 90.56 | 92.62 | 90.45 | 92.04 |  |  | 75.03 | 76.12 | 79.3 | 80.78 | 89.55 | 90.87 | **95.34** | **95.92** | 84.54 | 84.85 |
|   | 5 | Last | 46.83 | 52.23 | 48.86 | 51.26 | 46.25 | 49.71 |  |  | 50.22 | 52.43 | 50.29 | 53.2 | 55.54 | 63.69 | 85.9 | 91.26 | 54.18 | 58.06 |
|   |   | Best | 51.18 | 52.23 | 53.75 | 55.73 | 50.02 | 54.76 |  |  | 54.33 | 56.12 | 53.51 | 55.34 | 62.06 | 66.21 | 90.25 | 91.26 | 57.9 | 61.36 |
| B | 1 | Last | 47.03 | 49.71 | 46.56 | 48.93 | 47.53 | 48.54 | 49.79 | 51.26 | 87.26 | 89.13 | 67.3 | 68.54 | 84.97 | 87.57 | 92.97 | **95.73** | 77.09 | 79.03 |
|   |   | Best | 50.21 | 52.04 | 49.09 | 50.1 | 50.41 | 51.46 | 50.18 | 51.46 | 87.62 | 90.12 | 68.54 | 69.9 | 85.67 | 87.96 | 93.94 | **96.7** | 78.87 | 79.42 |
|   | 2 | Last | 93.59 | **96.31** | 93.44 | **95.92** | 92.15 | **95.15** | 94.14 | 94.95 | 85.59 | 86.8 | 91.19 | 92.62 | 91.49 | 92.23 | 92.08 | 93.4 | 92.47 | 93.2 |
|   |   | Best | **95.73** | **96.31** | **96.19** | **96.89** | **95.84** | **96.12** | 94.52 | 94.95 | 86.02 | 87.38 | 91.58 | 92.63 | 92.08 | 92.82 | 92.47 | 93.98 | 93.79 | 94.37 |
|   | 3 | Last | 22.7 | 25.67 | 23.8 | 27.19 | 26.96 | 29.47 |  |  | 39.39 | 41.06 | 38.59 | 41.63 | 42.09 | 44.11 | 46.92 | 48.86 | 39.58 | 40.87 |
|   |   | Best | 27.23 | 28.52 | 27.87 | 29.85 | 30.38 | 31.56 |  |  | 39.54 | 41.63 | 39.65 | 41.63 | 42.17 | 44.11 | 47.87 | 48.86 | 43.12 | 43.92 |
|   | 4 | Last | 87.42 | 88.93 | 86.06 | 88.93 | 87.96 | 90.29 |  |  | 72.04 | 73.59 | 72.04 | 76.7 | 86.79 | 87.96 | 91.57 | **95.53** | 79.46 | 80.97 |
|   |   | Best | 89.17 | 90.49 | 88.31 | 90.1 | 89.83 | 91.84 |  |  | 73.01 | 74.56 | 73.94 | 77.86 | 87.1 | 87.96 | **95.07** | **95.53** | 82.21 | 83.11 |
|   | 5 | Last | 48.97 | 53.79 | 50.91 | 55.73 | 42.83 | 47.18 |  |  | 50.8 | 53.59 | 51.41 | 54.17 | 62.18 | 65.05 | 89.01 | 90.68 | 57.17 | 58.64 |
|   |   | Best | 53.75 | 56.89 | 52.97 | 58.06 | 47.22 | 53.4 |  |  | 51.26 | 54.56 | 52.07 | 54.17 | 62.56 | 65.44 | 89.32 | 90.68 | 60.08 | 61.17 |
| C | 1 | Last | 49.71 | 51.07 | 49.36 | 52.62 | 48.47 | 51.84 | 47.69 | 51.07 | 88.85 | 90.29 | 69.98 | 72.43 | 91.38 | 92.43 | 93.86 | **97.67** | 82.21 | 83.5 |
|   |   | Best | 50.72 | 52.62 | 50.52 | 53.2 | 49.9 | 52.82 | 49.47 | 51.07 | 90.76 | 91.84 | 70.14 | 73.2 | 92.54 | 93.4 | **96.04** | **97.67** | 84.55 | 87.38 |
|   | 2 | Last | 94.56 | 95.53 | 95.03 | 95.53 | 95.49 | **96.5** | 95.15 | **96.12** | 86.41 | 88.35 | 91.03 | 92.43 | 91.34 | 92.23 | 90.87 | 91.65 | 91.61 | 93.01 |
|   |   | Best | **96.16** | **96.5** | **96.58** | **96.89** | **96.62** | **97.09** | **95.88** | **97.28** | 87.85 | 88.93 | 91.77 | 92.62 | 92.54 | 93.4 | 92.2 | 93.01 | 92.54 | 93.4 |
|   | 3 | Last | 26.01 | 30.42 | 24.4 | 27.95 | 28.03 | 30.23 |  |  | 40 | 41.44 | 39.62 | 41.44 | 40.41 | 41.25 | 45.02 | 48.67 | 40.19 | 42.4 |
|   |   | Best | 28.82 | 32.51 | 27.95 | 29.09 | 30.27 | 31.75 |  |  | 41.52 | 43.16 | 41.18 | 42.78 | 42.47 | 44.11 | 48.29 | 49.43 | 41.79 | 42.4 |
|   | 4 | Last | 88.23 | 89.71 | 87.5 | 88.74 | 88.58 | 90.68 |  |  | 75.07 | 78.64 | 77.94 | 79.61 | 88.04 | 90.29 | 94.83 | **96.12** | 79.81 | 82.91 |
|   |   | Best | 90.21 | 91.46 | 89.83 | 91.07 | 90.91 | 92.23 |  |  | 76.04 | 78.83 | 78.87 | 80 | 89.98 | 91.46 | **95.38** | **96.5** | 83.57 | 91.84 |
|   | 5 | Last | 56.04 | 61.36 | 51.54 | 57.09 | 50.33 | 57.48 |  |  | 54.6 | 60 | 54.33 | 55.92 | 60.51 | 71.07 | 91.53 | 93.4 | 61.71 | 64.08 |
|   |   | Best | 60.47 | 66.41 | 55.11 | 64.47 | 54.45 | 61.36 |  |  | 57.01 | 60 | 55.69 | 57.48 | 63.42 | 71.07 | 93.2 | 93.98 | 63.26 | 64.08 |

\*A is the max-min normalization; B is the -1-1 normalization; C is the Z-score normalization

1 is the time domain input, 2 is the frequency domain input; 3 is the wavelet domain input; 4 is the time domain sample after STFT; 5 is the time domain sample reshape to a 2D matrix

MFPT
Results with random split and without data augmentation

| | | | AE | | SAE | | DAE | | MLP | | CNN | | LeNet | | AlexNet | | ResNet18 | | LSTM | |
|---|---|---|---|---|---|---|---|---|---|---|---|---|---|---|---|---|---|---|---|---|
| Nor | Input | Loc | Mean | Max | Mean | Max | Mean | Max | Mean | Max | Mean | Max | Mean | Max | Mean | Max | Mean | Max | Mean | Max |
| A | 1 | Last | 20.93 | 27.38 | 17.59 | 22.33 | 22.91 | 26.02 | 22.49 | 25.05 | 83.22 | 86.8 | 62.25 | 62.91 | 76.08 | 78.64 | 78.18 | 85.44 | 69.48 | 71.07 |
| | | Best | 28.54 | 32.04 | 26.52 | 28.35 | 27.81 | 30.29 | 27.77 | 28.74 | 87.07 | 87.96 | 65.24 | 67.18 | 78.37 | 82.33 | 86.88 | 87.96 | 72.58 | 75.15 |
| | 2 | Last | 94.95 | **95.92** | 94.14 | **95.15** | 94.76 | **95.34** | 94.33 | **95.53** | 79.96 | 84.27 | 93.75 | 94.76 | 92.04 | 93.4 | 92.27 | **95.92** | 93.09 | 93.79 |
| | | Best | **96.23** | 96.5 | 96 | 96.7 | **96.27** | 96.89 | 95.96 | 96.5 | 82.68 | 84.66 | 93.86 | 94.95 | 93.36 | 94.17 | **95.85** | 96.31 | 94.06 | 94.76 |
| | 3 | Last | 26.12 | 29.09 | 23.08 | 24.52 | 25.59 | 28.14 | | | 38.1 | 40.11 | 40.76 | 41.44 | 29.46 | 41.06 | 44.22 | 48.67 | 40.49 | 43.16 |
| | | Best | 28.14 | 29.66 | 27.49 | 28.71 | 29.17 | 30.42 | | | 42.02 | 43.73 | 43.16 | 44.11 | 30.53 | 43.35 | 49.05 | 50.38 | 43.35 | 46.01 |
| | 4 | Last | 89.01 | 90.49 | 89.79 | 91.46 | 88.74 | 90.1 | | | 73.12 | 76.31 | 78.1 | 79.03 | 88.39 | 90.1 | 94.87 | **95.34** | 80.43 | 83.69 |
| | | Best | 89.98 | 91.26 | 91.26 | 92.23 | 89.9 | 91.26 | | | 74.99 | 76.89 | 79.34 | 80.97 | 90.41 | 91.65 | **95.42** | 95.73 | 83.5 | 86.02 |
| | 5 | Last | 47.49 | 51.07 | 48.23 | 53.01 | 46.6 | 51.26 | | | 50.84 | 53.98 | 50.87 | 54.76 | 52.2 | 65.44 | 88.31 | 89.9 | 53.98 | 58.25 |
| | | Best | 49.98 | 54.76 | 51.42 | 57.09 | 48.82 | 55.53 | | | 53.67 | 57.09 | 54.68 | 55.92 | 53.32 | 65.44 | 90.37 | 91.07 | 59.11 | 61.36 |
| B | 1 | Last | 25.32 | 27.96 | 22.33 | 23.3 | 25.52 | 28.16 | 25.4 | 27.18 | 85.2 | 86.99 | 63.57 | 66.21 | 80.62 | 82.91 | 86.99 | 87.96 | 65.94 | 72.04 |
| | | Best | 28.58 | 30.1 | 26.17 | 27.18 | 28.89 | 30.87 | 25.4 | 27.18 | 85.2 | 86.99 | 64.04 | 67.57 | 80.93 | 82.91 | 87.11 | 87.96 | 66.18 | 72.04 |
| | 2 | Last | 92.93 | **95.34** | 92.19 | 94.95 | 93.67 | **95.92** | 91.96 | 94.76 | 80.04 | 80.97 | 93.09 | 93.4 | 90.99 | 92.23 | 94.68 | **95.92** | 93.2 | 94.56 |
| | | Best | **95.61** | 95.73 | 95.42 | 95.53 | **96.12** | 96.7 | 95.54 | 96.12 | 81.13 | 82.91 | 94.1 | 94.37 | 93.09 | 93.98 | **96.23** | 96.7 | 94.06 | 94.56 |
| | 3 | Last | 26.47 | 29.85 | 25.59 | 26.43 | 27.07 | 30.8 | | | 39.24 | 40.87 | 38.86 | 40.11 | 39.92 | 42.59 | 44.22 | 46.58 | 39.54 | 40.11 |
| | | Best | 28.94 | 30.42 | 29.13 | 30.04 | 30 | 34.22 | | | 39.92 | 42.59 | 39.58 | 41.44 | 41.1 | 42.59 | 44.83 | 47.53 | 40.44 | 41.63 |
| | 4 | Last | 87.88 | 89.51 | 86.83 | 89.9 | 88.39 | 90.29 | | | 71.84 | 76.5 | 77.59 | 79.22 | 86.56 | 87.77 | 94.02 | **95.34** | 80.58 | 82.52 |
| | | Best | 89.01 | 90.49 | 89.63 | 92.04 | 89.79 | 90.87 | | | 75.15 | 77.09 | 78.95 | 79.81 | 88.85 | 89.71 | **95.69** | 96.31 | 82.41 | 83.11 |
| | 5 | Last | 50.06 | 57.67 | 50.41 | 57.86 | 45.94 | 50.1 | | | 50.14 | 52.04 | 51.11 | 54.76 | 56.66 | 61.36 | 78.52 | 90.49 | 56.89 | 59.61 |
| | | Best | 53.94 | 59.03 | 55.42 | 60.58 | 48.58 | 51.84 | | | 51.65 | 53.4 | 51.92 | 54.76 | 57.82 | 64.27 | 83.57 | 90.49 | 57.57 | 60.39 |
| C | 1 | Last | 24.19 | 26.6 | 23.26 | 24.85 | 22.95 | 24.08 | 21.79 | 22.72 | 86.21 | 87.57 | 61.36 | 64.08 | 84.19 | 85.24 | 86.6 | 91.84 | 76.27 | 77.48 |
| | | Best | 26.68 | 28.35 | 25.98 | 26.99 | 26.6 | 27.96 | 24.47 | 26.41 | 86.8 | 88.56 | 61.75 | 64.08 | 84.82 | 86.8 | 88.74 | 91.84 | 77.09 | 79.22 |
| | 2 | Last | 94.99 | **95.34** | 94.49 | **95.92** | 94.37 | **95.15** | 94.37 | 94.76 | 80.39 | 82.33 | 91.84 | 92.62 | 90.72 | 92.82 | 94.56 | **95.34** | 92.15 | 93.01 |
| | | Best | **96.35** | 96.89 | **96.58** | 97.48 | **96.12** | 96.89 | **96.04** | 96.31 | 81.51 | 82.52 | 92.43 | 93.4 | 92.82 | 93.59 | **95.49** | 95.92 | 93.9 | 94.37 |
| | 3 | Last | 22.17 | 25.67 | 25.89 | 28.14 | 27.76 | 31.18 | | | 40.26 | 42.21 | 40.19 | 41.06 | 39.73 | 41.63 | 40.84 | 49.05 | 39.92 | 40.68 |
| | | Best | 28.1 | 32.13 | 28.25 | 29.28 | 30.8 | 32.7 | | | 40.87 | 43.35 | 40.34 | 41.06 | 40.45 | 42.97 | 41.37 | 49.81 | 41.02 | 42.78 |
| | 4 | Last | 89.28 | 89.71 | 88.35 | 89.9 | 88.97 | 91.07 | | | 83.16 | 91.26 | 86.45 | 87.57 | 88.04 | 89.71 | 89.79 | 91.46 | 87.34 | 89.71 |
| | | Best | 89.71 | 90.29 | 89.05 | 90.1 | 90.95 | 92.43 | | | 84.33 | 91.46 | 86.6 | 87.57 | 91.22 | 92.43 | 90.14 | 91.46 | 88.43 | 90.29 |
| | 5 | Last | 54.1 | 57.86 | 53.55 | 60.58 | 50.6 | 53.59 | | | 58.52 | 62.52 | 54.21 | 56.12 | 63.07 | 66.41 | 91.84 | 92.62 | 60.93 | 62.14 |
| | | Best | 57.48 | 60.97 | 56.31 | 62.33 | 53.59 | 56.7 | | | 59.34 | 63.69 | 55.46 | 56.31 | 64 | 69.51 | 92.39 | 92.82 | 61.83 | 63.11 |

MFPT
Results with order split and data augmentation

| | | | AE | | SAE | | DAE | | MLP | | CNN | | LeNet | | AlexNet | | ResNet18 | | LSTM | |
|---|---|---|---|---|---|---|---|---|---|---|---|---|---|---|---|---|---|---|---|---|
| Nor | Input | Loc | Mean | Max | Mean | Max | Mean | Max | Mean | Max | Mean | Max | Mean | Max | Mean | Max | Mean | Max | Mean | Max |
| A | 1 | Last | 35.28 | 40.69 | 36.39 | 41.27 | 35.43 | 41.07 | 31.09 | 37.62 | 86.03 | 90.02 | 69.64 | 73.9 | 80.27 | 83.3 | 92.32 | 93.86 | 79.08 | 81.57 |
| | | Best | 42.65 | 44.72 | 43.8 | 45.68 | 44.3 | 46.64 | 43.72 | 47.22 | 89.83 | 90.4 | 72.78 | 74.66 | 81.42 | 85.03 | 94.74 | **95.59** | 81.84 | 83.3 |
| | 2 | Last | 93.59 | **95.2** | **95.47** | **96.55** | **95.01** | **95.39** | **95.55** | **96.16** | 87.03 | 88.1 | 91.63 | 93.09 | 92.55 | 93.09 | 92.4 | 94.63 | 91.94 | 93.09 |
| | | Best | **96.08** | **96.35** | **96.43** | **97.31** | **96.12** | **96.74** | **96.51** | **96.74** | 88.98 | 89.44 | 93.17 | 94.05 | 93.55 | 94.05 | 94.63 | 94.82 | 93.66 | **95.01** |
| | 3 | Last | 26.07 | 26.82 | 26.52 | 28.12 | 28.34 | 33.33 | | | 39.89 | 41.53 | 39.96 | 41.34 | 34.45 | 43.76 | 44.95 | 45.81 | 39.63 | 42.27 |
| | | Best | 29.31 | 29.98 | 29.79 | 31.28 | 30.43 | 33.33 | | | 43.35 | 44.69 | 44.1 | 45.62 | 35.42 | 46.18 | 48.46 | 49.16 | 42.42 | 44.13 |
| | 4 | Last | 89.25 | 91.94 | 90.06 | 91.55 | 88.79 | 90.6 | | | 77.7 | 80.81 | 78.31 | 81.77 | 90.75 | 91.75 | 90.6 | **96.55** | 80.69 | 81.96 |
| | | Best | 91.71 | 92.51 | 92.28 | 93.86 | 91.52 | 92.71 | | | 78.77 | 81.96 | 80.69 | 82.92 | 92.29 | 92.71 | **97.2** | **97.7** | 85.14 | 86.37 |
| | 5 | Last | 52.55 | 57.2 | 51.32 | 53.74 | 51.28 | 55.66 | | | 50.86 | 55.47 | 53.24 | 55.47 | 65.87 | 69.1 | 74.82 | 89.44 | 60.61 | 63.53 |
| | | Best | 55.47 | 58.16 | 55.01 | 57.58 | 55.36 | 58.35 | | | 57.16 | 57.77 | 56.89 | 57.97 | 68.18 | 73.51 | 92.28 | 92.9 | 63.45 | 66.6 |
| B | 1 | Last | 52.97 | 54.89 | 50.29 | 52.02 | 51.55 | 52.98 | 51.87 | 54.89 | 89.44 | 90.98 | 72.5 | 74.47 | 87.39 | 89.44 | 93.16 | 95.39 | 80.08 | 81.19 |
| | | Best | 54.01 | 55.09 | 51.59 | 53.55 | 54.43 | 56.81 | 53.4 | 56.24 | 90.64 | 91.94 | 73.61 | 75.43 | 88.27 | 89.44 | 94.46 | 96.55 | 81.84 | 83.3 |
| | 2 | Last | 91.82 | 93.86 | 94.74 | **95.2** | 91.79 | 94.05 | 93.38 | **95.39** | 85.58 | 88.48 | 91.99 | 93.86 | 92.35 | 93.47 | 92.44 | 94.05 | 92.09 | 93.28 |
| | | Best | **95.28** | **95.97** | **95.7** | **96.16** | **95.7** | **96.16** | **95.15** | **96.74** | 87.69 | 89.64 | 92.8 | 94.24 | 92.87 | 93.67 | 94.07 | **95.2** | 93.59 | 93.86 |
| | 3 | Last | 27.71 | 30.35 | 27.11 | 30.73 | 27.41 | 29.8 | | | 40.83 | 41.9 | 40.99 | 42.64 | 41.61 | 43.95 | 45.23 | 46.74 | 40.19 | 41.53 |
| | | Best | 31.47 | 32.77 | 30.99 | 33.15 | 30.13 | 32.03 | | | 42.13 | 43.76 | 43.11 | 45.44 | 42.62 | 44.88 | 46.6 | 48.42 | 42.83 | 43.58 |
| | 4 | Last | 88.02 | 90.21 | 86.14 | 89.25 | 88.25 | 90.6 | | | 76.87 | 79.46 | 74.4 | 79.85 | 88.13 | 89.44 | 92.65 | **96.55** | 82.23 | 84.26 |
| | | Best | 90.29 | 92.51 | 89.1 | 91.94 | 90.21 | 91.75 | | | 77.71 | 79.65 | 77.06 | 80.42 | 89.55 | 91.17 | 93.25 | **97.5** | 84.57 | 85.41 |
| | 5 | Last | 51.94 | 55.47 | 51.78 | 55.28 | 51.25 | 56.24 | | | 56.72 | 61.23 | 55.85 | 57.58 | 66.05 | 69.87 | 77.68 | 92.71 | 61.61 | 65.07 |
| | | Best | 57.12 | 62.76 | 54.32 | 56.05 | 53.9 | 57.97 | | | 58.09 | 61.42 | 57.68 | 61.61 | 67.92 | 72.74 | 92.1 | 93.47 | 64.18 | 66.6 |
| C | 1 | Last | 49.75 | 51.06 | 52.67 | 55.28 | 51.63 | 52.98 | 51.06 | 52.78 | 90.71 | 92.13 | 74.51 | 75.24 | 91.63 | 94.24 | 94.05 | **96.16** | 87.75 | 89.44 |
| | | Best | 52.55 | 53.74 | 53.86 | 56.62 | 53.74 | 54.51 | 51.29 | 52.98 | 92.17 | 94.05 | 74.89 | 76.58 | 92.4 | 94.43 | 94.7 | **96.16** | 88.48 | 89.44 |
| | 2 | Last | **95.63** | **96.55** | **95.47** | **96.74** | **95.05** | **95.78** | **95.51** | **96.16** | 87.22 | 88.87 | 92.28 | 93.28 | 92.32 | 93.67 | 91.98 | 93.67 | 92.86 | 93.47 |
| | | Best | **96.74** | **97.12** | **96.51** | **96.74** | **96.36** | **96.74** | **95.85** | **96.35** | 88.02 | 88.87 | 92.78 | 93.47 | 92.63 | 93.67 | 92.86 | 93.67 | 93.13 | 93.47 |
| | 3 | Last | 28.31 | 31.1 | 30.1 | 31.1 | 29.53 | 32.03 | | | 42.16 | 51.29 | 46.97 | 50.95 | 48.8 | 54.19 | 50.98 | 55.59 | 46.47 | 50.78 |
| | | Best | 31.32 | 33.71 | 30.76 | 31.47 | 33.26 | 35.75 | | | 44.53 | 52.1 | 47.38 | 51.65 | 48.98 | 55.06 | 51.82 | 55.59 | 46.86 | 51.37 |
| | 4 | Last | 90.36 | 92.51 | 89.25 | 90.98 | 89.71 | 92.13 | | | 78.23 | 79.46 | 77.7 | 81 | 90.17 | 92.9 | **95.24** | **97.5** | 81.15 | 84.07 |
| | | Best | 92.93 | 93.86 | 92.71 | **95.01** | 92.94 | 94.63 | | | 80.42 | 81 | 78.54 | 81.19 | 90.29 | 92.9 | **95.66** | **97.89** | 83.03 | 84.64 |
| | 5 | Last | 57.62 | 61.42 | 54.51 | 57.2 | 51.32 | 60.08 | | | 60.38 | 62 | 58.35 | 60.08 | 69.29 | 76.01 | 68.91 | **95.59** | 65.91 | 67.56 |
| | | Best | 62.3 | 67.56 | 58.54 | 60.46 | 58.93 | 62.38 | | | 62.42 | 63.34 | 59.04 | 61.61 | 69.75 | 76.01 | 94.63 | **95.59** | 66.68 | 69.48 |

PU
Results with random split and data augmentation

| Nor | Input | Loc | AE Mean | AE Max | DAE Mean | DAE Max | SAE Mean | SAE Max | MLP Mean | MLP Max | CNN Mean | CNN Max | LeNet Mean | LeNet Max | AlexNet Mean | AlexNet Max | ResNet18 Mean | ResNet18 Max | LSTM Mean | LSTM Max |
|---|---|---|---|---|---|---|---|---|---|---|---|---|---|---|---|---|---|---|---|---|
| A | 1 | Last | 47.22 | 53.15 | 44.91 | 53 | 48.79 | 53.3 | 46.18 | 49.16 | 79.02 | 81.87 | 68.76 | 72.66 | 82.46 | 84.33 | 89.74 | 92.17 | 75.02 | 76.04 |
|   |   | Best | 55.08 | 59.14 | 55.21 | 58.06 | 52.69 | 54.53 | 51.68 | 57.16 | 80.68 | 85.87 | 71.43 | 75.12 | 83.1 | 84.33 | 90.85 | 94.01 | 75.6 | 76.34 |
|   | 2 | Last | **97.45** | **97.85** | **97.27** | **97.85** | **97.64** | **98.31** | **98.31** | **99.08** | **95.18** | **96.16** | **94.19** | **96.16** | **96.53** | **98.31** | **97.14** | **97.54** | **94.99** | **95.39** |
|   |   | Best | **98.34** | **98.46** | **98.43** | **98.77** | **98.52** | **98.77** | **98.53** | **99.23** | **95.58** | **96.62** | **94.68** | **97.54** | **96.84** | **98.31** | **97.66** | **98.46** | **95.11** | **95.55** |
|   | 3 | Last | 30.08 | 33.58 | 29.74 | 31.63 | 31.42 | 32.23 |   |   | 36.37 | 39.88 | 33.52 | 38.08 | 33.52 | 35.23 | 41.77 | 47.68 | 37.39 | 38.68 |
|   |   | Best | 33.34 | 36.58 | 33.67 | 34.63 | 34.36 | 36.13 |   |   | 37.24 | 40.63 | 34.27 | 38.83 | 34.69 | 36.73 | 43.93 | 48.28 | 37.9 | 39.28 |
|   | 4 | Last | 94.22 | **95.7** | 95.51 | **96.31** | 93.98 | **96.31** |   |   | 82.52 | 85.71 | 82.3 | 84.18 | 94.26 | **95.55** | 93.03 | **97.7** | 84.55 | 86.94 |
|   |   | Best | **96.59** | **96.93** | **96.96** | **97.39** | **95.79** | **97.08** |   |   | 83.01 | 87.86 | 82.8 | 84.49 | **95.08** | **96.31** | 93.68 | **98.82** | 86.7 | 88.48 |
|   | 5 | Last | 56.99 | 68.36 | 63.96 | 68.05 | 59.36 | 62.67 |   |   | 45.71 | 53.46 | 44.09 | 49.16 | 7.83 | 7.83 | 91.37 | 94.47 | 48.63 | 52.38 |
|   |   | Best | 66.76 | 73.73 | 68.26 | 72.35 | 65.19 | 70.51 |   |   | 48.83 | 56.61 | 47.59 | 51.31 | 7.83 | 7.83 | 94.1 | **95.55** | 53.67 | 59.29 |
| B | 1 | Last | 65.95 | 66.97 | 62.13 | 64.06 | 65.79 | 67.74 | 70.72 | 72.2 | 83.44 | 86.18 | 72.26 | 72.96 | 86.61 | 88.94 | 91.58 | 93.24 | 75.88 | 77.11 |
|   |   | Best | 67.61 | 68.97 | 63.59 | 65.9 | 66.62 | 68.82 | 71.24 | 72.66 | 84.55 | 86.33 | 73.18 | 74.5 | 87.5 | 90.02 | 92.32 | 94.01 | 78.49 | 79.42 |
|   | 2 | Last | **97.06** | **98.31** | **96.26** | **97.24** | **96.49** | **98** | **97.69** | **98.46** | 93.7 | **95.55** | **95.02** | **96.31** | **97.05** | **97.54** | 90.91 | **98** | 94.66 | 95.39 |
|   |   | Best | **98.28** | **98.46** | **98.08** | **98.46** | **97.83** | **98.16** | **97.97** | **98.46** | 95.15 | 95.7 | **95.42** | **96.62** | **97.24** | **97.54** | **97.76** | **98.16** | **96.31** | **96.47** |
|   | 3 | Last | 29.41 | 32.53 | 31.86 | 35.83 | 33.91 | 37.03 |   |   | 37.6 | 40.63 | 38.86 | 40.33 | 33.4 | 35.98 | 45.04 | 47.98 | 39.76 | 41.68 |
|   |   | Best | 32.65 | 34.93 | 34.2 | 38.53 | 37.33 | 38.38 |   |   | 39.52 | 43.18 | 39.46 | 41.53 | 34.06 | 37.63 | 45.43 | 47.98 | 42.82 | 44.53 |
|   | 4 | Last | 91.94 | 94.32 | 92.06 | 93.39 | 93.42 | **95.08** |   |   | 77.17 | 85.71 | 76.19 | 79.42 | 91.06 | 92.78 | **95.14** | **99.23** | 82 | 84.64 |
|   |   | Best | 93.78 | **96.47** | 94.42 | **96.31** | 95.03 | 96.01 |   |   | 77.64 | 85.71 | 78.96 | 81.26 | 91.83 | 93.55 | **99.02** | **99.39** | 87.1 | 89.25 |
|   | 5 | Last | 60.01 | 66.67 | 56.91 | 66.82 | 62.88 | 68.82 |   |   | 33.79 | 55.61 | 50.38 | 53.92 | 49.09 | 61.6 | 90.23 | **95.85** | 57.33 | 59.75 |
|   |   | Best | 66.36 | 72.04 | 63.59 | 72.04 | 65.97 | 70.05 |   |   | 34.87 | 57.3 | 51.58 | 54.85 | 49.09 | 61.6 | 94.87 | **95.85** | 60.09 | 60.83 |
| C | 1 | Last | 69.83 | 71.58 | 64.98 | 68.05 | 70.72 | 71.43 | 71.09 | 72.96 | 85.96 | 88.63 | 78.92 | 80.03 | 90.57 | 92.17 | 93.92 | **95.24** | 80.31 | 82.03 |
|   |   | Best | 70.97 | 74.04 | 66.73 | 69.59 | 71.31 | 72.04 | 72.19 | 72.96 | 86.79 | 88.63 | 78.99 | 80.18 | 90.66 | 92.17 | 94.44 | **96.01** | 81.26 | 83.26 |
|   | 2 | Last | **97.45** | **98** | **97.7** | **98.62** | **97.42** | **97.54** | **97.94** | **98.92** | 95.3 | **96.31** | **95.33** | **95.85** | **97.3** | **98.16** | **97.17** | **98** | **95.02** | **95.55** |
|   |   | Best | **98.25** | **98.62** | **98.5** | **98.62** | **98.34** | **98.62** | **98.22** | **98.92** | **95.55** | **96.47** | **95.85** | **96.93** | **97.73** | **98.16** | **97.64** | **98.77** | **95.36** | **96.16** |
|   | 3 | Last | 25.52 | 33.58 | 29.95 | 32.23 | 33.01 | 34.63 |   |   | 39.1 | 41.68 | 39.61 | 41.53 | 37 | 38.98 | 41.86 | 47.08 | 40.21 | 41.68 |
|   |   | Best | 26.84 | 33.73 | 32.77 | 34.03 | 36.04 | 38.53 |   |   | 40.45 | 41.68 | 39.82 | 42.58 | 38.02 | 41.23 | 45.91 | 47.08 | 41.29 | 41.83 |
|   | 4 | Last | 94.96 | **96.01** | 93.64 | **95.39** | 94.69 | **97.54** |   |   | 82.34 | 85.41 | 79.85 | 82.03 | 91.95 | 93.55 | 92.87 | **98.16** | 85.1 | 86.64 |
|   |   | Best | **96.07** | **96.47** | **95.02** | **96.31** | **96.44** | **97.54** |   |   | 83.41 | 86.64 | 81.04 | 82.64 | 92.47 | 94.62 | 93.18 | **98.92** | 86.33 | 86.94 |
|   | 5 | Last | 59.11 | 65.75 | 56.62 | 66.36 | 62.15 | 69.59 |   |   | 38.55 | 64.36 | 53.09 | 55.61 | 28.45 | 62.37 | 91.55 | **96.62** | 59.51 | 60.83 |
|   |   | Best | 64.27 | 66.05 | 61.26 | 70.51 | 67.77 | 72.96 |   |   | 39.11 | 65.13 | 53.58 | 55.61 | 29.31 | 62.37 | **95.85** | **96.77** | 59.85 | 61.44 |

*A is the max-min normalization; B is the -1-1 normalization; C is the Z-score normalization

1 is the time domain input, 2 is the frequency domain input; 3 is the wavelet domain input; 4 is the time domain sample after STFT; 5 is the time domain sample reshape to a 2D matrix

PU
Results with random split and without data augmentation

| Nor | Input | Loc | AE | | SAE | | DAE | | MLP | | CNN | | LeNet | | AlexNet | | ResNet18 | | LSTM | |
|---|---|---|---|---|---|---|---|---|---|---|---|---|---|---|---|---|---|---|---|---|
| | | | Mean | Max | Mean | Max | Mean | Max | Mean | Max | Mean | Max | Mean | Max | Mean | Max | Mean | Max | Mean | Max |
| A | 1 | Last | 37.7 | 44.19 | 29.59 | 34.33 | 38.31 | 43.13 | 50.39 | 56.84 | 78.96 | 84.95 | 71.49 | 74.96 | 82.74 | 84.49 | 81.85 | 85.41 | 73.24 | 77.11 |
| | | Best | 46.85 | 48.5 | 39.36 | 40.89 | 45.63 | 47.25 | 55.95 | 60.22 | 79.97 | 84.95 | 73.92 | 75.27 | 83.23 | 85.87 | 85.29 | **95.24** | 74.62 | 77.27 |
| | 2 | Last | 74.62 | 93.38 | 91.91 | 93.76 | 92.66 | 93.2 | **97.91** | **98.46** | 90.57 | 91.86 | **95.85** | **97.08** | 95.18 | 96.77 | 98.77 | 99.85 | 93.86 | **95.55** |
| | | Best | 75.48 | 93.88 | 92.86 | 93.95 | 93.5 | 94.01 | **98.43** | **98.92** | 90.72 | 91.86 | **96.13** | **97.54** | 95.57 | 97.85 | 99.48 | 99.85 | 94.29 | **96.01** |
| | 3 | Last | 31.03 | 33.58 | 31.06 | 33.28 | 32.86 | 37.03 | | | 46.3 | 49.36 | 47.1 | 48.59 | 43.36 | 54.62 | 47.94 | 55.64 | 46.65 | 48.56 |
| | | Best | 34.12 | 37.18 | 33.58 | 35.83 | 35.2 | 38.23 | | | 48.04 | 51.26 | 48.26 | 49.37 | 43.5 | 55.07 | 53.56 | 55.8 | 47.34 | 48.56 |
| | 4 | Last | 89.31 | 90.01 | 89.83 | 90.89 | 89.45 | 90.57 | | | 77.63 | 86.94 | 81.38 | 82.03 | 92.35 | **95.7** | **95.48** | **98.46** | 78.06 | 86.64 |
| | | Best | 90.72 | 91.14 | 90.94 | 91.64 | 89.98 | 91.2 | | | 85.74 | 87.71 | 81.81 | 82.49 | 94.01 | **95.7** | 96.99 | 98.62 | 84.73 | 88.17 |
| | 5 | Last | 56.09 | 59.74 | 56.02 | 59.3 | 53.06 | 60.61 | | | 51.21 | 54.22 | 49.65 | 54.84 | 7.83 | 7.83 | 89.59 | 94.47 | 55.36 | 58.37 |
| | | Best | 59.51 | 61.67 | 58.5 | 61.99 | 58.54 | 67.23 | | | 55.21 | 55.91 | 55.24 | 57.76 | 7.83 | 7.83 | 94.99 | **95.39** | 58.37 | 59.91 |
| B | 1 | Last | 55.01 | 55.68 | 31.37 | 38.14 | 53.08 | 55.43 | 70.2 | 72.81 | 84.06 | 85.56 | 72.04 | 73.73 | 87.04 | 87.71 | 89.68 | 93.86 | 72.41 | 76.65 |
| | | Best | 55.98 | 56.55 | 31.85 | 39.26 | 54.79 | 56.18 | 71.27 | 72.81 | 84.67 | 86.48 | 73.49 | 75.73 | 87.65 | 88.79 | 90.87 | **95.39** | 75.67 | 78.34 |
| | 2 | Last | 90.79 | 92.7 | 89.63 | 93.26 | 92.52 | 93.95 | 80.92 | **97.85** | 88.63 | 92.17 | **96.44** | **97.39** | 95.73 | 97.24 | 97.94 | 99.54 | 93.67 | **95.39** |
| | | Best | 94.06 | 94.32 | 93.84 | 94.19 | 94.16 | 94.51 | **95.15** | **98.92** | 88.83 | 92.17 | **96.59** | **97.39** | 96.41 | 97.24 | 99.32 | 99.69 | 94.87 | **95.7** |
| | 3 | Last | 32.47 | 34.78 | 32.71 | 38.68 | 35.35 | 37.93 | | | 50.56 | 52.14 | 50.27 | 51.56 | 53.39 | 53.87 | 53.64 | 55.46 | 48.34 | 48.76 |
| | | Best | 34.54 | 36.13 | 34.75 | 40.18 | 36.7 | 39.13 | | | 51.13 | 52.29 | 50.77 | 51.56 | 53.76 | 55.07 | 54.32 | 56.22 | 48.81 | 49.24 |
| | 4 | Last | 87.88 | 90.32 | 87.45 | 87.83 | 86.22 | 90.14 | | | 83.76 | 88.33 | 76.01 | 81.57 | 91.92 | 92.93 | 81.97 | **98.62** | 81.87 | 86.48 |
| | | Best | 88.71 | 90.32 | 88.4 | 89.58 | 88.26 | 90.45 | | | 84.16 | 88.63 | 77.6 | 83.41 | 92.11 | 93.55 | **98.22** | 99.08 | 82.43 | 86.48 |
| | 5 | Last | 57.7 | 60.61 | 58.94 | 61.17 | 58.69 | 60.86 | | | 51.36 | 51.46 | 51.77 | 52.69 | 51.77 | 68.2 | 94.77 | **95.08** | 55.54 | 57.45 |
| | | Best | 60.43 | 61.92 | 62.66 | 67.79 | 62.26 | 66.42 | | | 53.97 | 54.38 | 55.49 | 56.84 | 54.47 | 72.04 | **95.11** | 95.85 | 59.69 | 61.44 |
| C | 1 | Last | 59.39 | 60.36 | 51.25 | 54.18 | 57.15 | 58.49 | 71.77 | 72.81 | 83.41 | 87.1 | 77.94 | 80.18 | 91.43 | 92.78 | 91.92 | **95.55** | 80.18 | 81.57 |
| | | Best | 60.18 | 62.3 | 52 | 54.43 | 57.81 | 58.93 | 72.53 | 73.12 | 85.53 | 87.1 | 79.02 | 80.18 | 91.98 | 92.78 | 92.75 | **97.08** | 80.83 | 82.64 |
| | 2 | Last | 93.12 | 93.95 | 92.8 | 93.51 | 93.17 | 94.38 | **97.85** | **98.46** | 89.43 | 92.63 | **95.61** | **96.77** | 97.3 | 98.92 | 99.63 | 99.69 | 94.07 | 94.78 |
| | | Best | 94.07 | 94.26 | 94.24 | 94.57 | 94.13 | 94.38 | **98.52** | **98.77** | 89.43 | 92.63 | **95.79** | **96.77** | 97.82 | 98.92 | 99.63 | 99.69 | 94.72 | **96.01** |
| | 3 | Last | 28.04 | 32.23 | 24.71 | 29.39 | 34.45 | 37.03 | | | 51.58 | 52.05 | 50.88 | 51.84 | 54.5 | 55.91 | 55.21 | 56.36 | 48.96 | 49.6 |
| | | Best | 31.12 | 33.88 | 29 | 33.43 | 37.15 | 40.93 | | | 51.99 | 52.91 | 51.36 | 52.33 | 54.86 | 55.91 | 55.4 | 57.24 | 49.4 | 49.93 |
| | 4 | Last | 90.04 | 90.89 | 89.96 | 91.51 | 83.43 | 90.82 | | | 82.8 | 86.33 | 79.17 | 82.03 | 92.17 | **95.39** | 97.79 | **98.62** | 85.38 | 86.94 |
| | | Best | 90.4 | 91.07 | 90.26 | 91.51 | 84.14 | 91.07 | | | 83.07 | 86.64 | 79.45 | 82.03 | 93.24 | **95.39** | 97.94 | 98.62 | 86.54 | 87.71 |
| | 5 | Last | 56.34 | 59.68 | 47.64 | 62.48 | 58.04 | 62.98 | | | 45.16 | 57.3 | 51.43 | 53.76 | 7.83 | 7.83 | 89.28 | **96.62** | 59.63 | 61.14 |
| | | Best | 60.26 | 65.04 | 49.74 | 63.17 | 60.12 | 64.04 | | | 50.31 | 59.45 | 56.84 | 58.68 | 8.35 | 10.45 | 96.44 | **96.93** | 62.18 | 63.13 |

PU
Results with order split and data augmentation

| | | | AE | | SAE | | DAE | | MLP | | CNN | | LeNet | | AlexNet | | ResNet18 | | LSTM | |
|---|---|---|---|---|---|---|---|---|---|---|---|---|---|---|---|---|---|---|---|---|
| Nor | Input | Loc | Mean | Max | Mean | Max | Mean | Max | Mean | Max | Mean | Max | Mean | Max | Mean | Max | Mean | Max | Mean | Max |
| A | 1 | Last | 48.11 | 53.3 | 49.86 | 55.45 | 48.33 | 52.69 | 51.06 | 55.3 | 80.83 | 82.8 | 64.24 | 71.43 | 83.9 | 84.95 | 85.07 | 93.7 | 71.64 | 73.89 |
| | | Best | 54.75 | 56.68 | 54.75 | 57.14 | 55.08 | 57.6 | 54.41 | 62.37 | 82.22 | 85.6 | 67.96 | 71.43 | 84.73 | 86.79 | 90.38 | 93.7 | 72.01 | 74.96 |
| | 2 | Last | **97.63** | **98.46** | **97.63** | **98.92** | **97.67** | **98.46** | 89.54 | 92.31 | 66.15 | 72.31 | 77.54 | 84.62 | 65.54 | 75.38 | 88 | 92.31 | 76.92 | 80 |
| | | Best | **99.14** | **99.23** | **98.89** | **98.92** | **98.77** | **98.92** | 95.69 | 98.46 | 69.23 | 72.31 | 82.77 | 87.69 | 69.23 | 75.38 | 91.08 | 93.85 | 82.77 | 86.15 |
| | 3 | Last | 29.96 | 32.2 | 28.42 | 31.61 | 29.69 | 31.61 | | | 32.59 | 37.52 | 32.94 | 36.48 | 25.61 | 31.02 | 41.54 | 44.46 | 34.5 | 35.89 |
| | | Best | 32.79 | 35.01 | 32.61 | 34.56 | 33.38 | 35.3 | | | 35.3 | 38.4 | 34.53 | 37.37 | 25.61 | 31.02 | 42.63 | 46.23 | 35.63 | 37.52 |
| | 4 | Last | 94.44 | **95.55** | 93.7 | 94.62 | 94.32 | **95.24** | | | 78.74 | 83.56 | 77.42 | 78.65 | 93.24 | **95.24** | **95.45** | **98.92** | 83.96 | 88.33 |
| | | Best | **96.9** | **97.39** | 96.37 | **97.08** | **95.82** | **96.62** | | | 80.18 | 83.56 | 78.83 | 81.72 | 93.52 | **95.55** | **97.97** | **98.92** | 87.07 | 88.33 |
| | 5 | Last | 61.2 | 66.97 | 57.7 | 62.98 | 58 | 65.59 | | | 42.58 | 49.31 | 45.16 | 51.61 | 7.83 | 7.83 | 85.9 | 94.47 | 45.53 | 51.15 |
| | | Best | 64.52 | 68.51 | 62.43 | 69.89 | 62.03 | 67.9 | | | 45.68 | 50.84 | 48.82 | 51.61 | 7.83 | 7.83 | 87.19 | **95.39** | 50.51 | 51.92 |
| B | 1 | Last | 65.5 | 67.9 | 63.1 | 65.28 | 66.82 | 69.59 | 69.65 | 71.58 | 79.66 | 84.64 | 70.2 | 72.04 | 86.95 | 88.33 | 91.58 | 93.39 | 74.38 | 75.88 |
| | | Best | 67.34 | 69.12 | 64.33 | 67.13 | 68.36 | 69.89 | 70.08 | 71.89 | 80.74 | 85.25 | 71.15 | 74.5 | 87.62 | 88.33 | 92.44 | 94.47 | 76.4 | 77.11 |
| | 2 | Last | **97.2** | **98.46** | **98.19** | **98.46** | **96.59** | **98** | 75.69 | 89.23 | 58.15 | 67.69 | 79.08 | 84.62 | 66.77 | 70.77 | 72.31 | 80 | 78.15 | 83.08 |
| | | Best | **98.95** | **99.23** | **99.05** | **99.23** | **98.59** | **98.92** | 92.31 | 93.85 | 68.31 | 73.85 | 82.15 | 86.15 | 68.62 | 72.31 | 88.92 | 90.77 | 82.16 | 84.62 |
| | 3 | Last | 31.82 | 33.68 | 29.87 | 32.35 | 31.28 | 31.76 | | | 37.52 | 39 | 35.66 | 36.93 | 31.34 | 34.86 | 42.51 | 45.2 | 36.1 | 38.85 |
| | | Best | 34.65 | 36.48 | 33.03 | 35.45 | 34.5 | 35.45 | | | 37.84 | 39.44 | 36.87 | 38.7 | 32.26 | 34.86 | 43.4 | 45.2 | 39.44 | 41.36 |
| | 4 | Last | 92.1 | 94.62 | 91.58 | 92.78 | 92.23 | 93.24 | | | 80.15 | 84.02 | 72.72 | 81.72 | 91.52 | 92.47 | 90.38 | **98** | 83.84 | 86.18 |
| | | Best | 94.13 | **97.24** | 94.56 | 95.85 | 94.22 | 95.39 | | | 81.66 | 84.02 | 74.5 | 82.95 | 91.74 | 92.63 | **97.02** | **98.31** | 86.54 | 87.71 |
| | 5 | Last | 61.32 | 66.21 | 60.46 | 63.13 | 56.93 | 66.67 | | | 48.54 | 52.23 | 48.05 | 50.54 | 30.87 | 67.74 | 88.88 | **95.55** | 50.63 | 53.46 |
| | | Best | 64.88 | 67.74 | 64.64 | 66.36 | 67.96 | 71.43 | | | 48.6 | 52.38 | 48.76 | 50.54 | 30.87 | 67.74 | 94.84 | **95.55** | 54.5 | 57.6 |
| C | 1 | Last | 69.4 | 71.43 | 65.83 | 66.97 | 69.31 | 71.58 | 72.93 | 75.73 | 84.73 | 86.64 | 75.7 | 77.57 | 91.15 | 91.71 | 93.67 | **95.39** | 78.43 | 78.96 |
| | | Best | 70.6 | 71.89 | 67.4 | 69.43 | 70.75 | 72.96 | 73.46 | 75.73 | 86.18 | 88.33 | 76.4 | 78.03 | 91.8 | 93.7 | **95.02** | **96.16** | 80.06 | 81.87 |
| | 2 | Last | **98.62** | **99.23** | **99.05** | **99.39** | **98.53** | **99.23** | 88.61 | 90.77 | 64 | 66.15 | 78.77 | 86.15 | 77.23 | 80 | 85.54 | 89.23 | 78.15 | 81.54 |
| | | Best | **99.14** | **99.54** | **99.36** | **99.39** | **99.36** | **99.54** | 93.54 | 93.85 | 74.15 | 76.92 | 82.15 | 86.15 | 81.85 | 84.62 | 92.92 | **95.38** | 84.62 | 86.15 |
| | 3 | Last | 27.5 | 31.76 | 30.07 | 33.68 | 31.7 | 32.79 | | | 37.93 | 40.62 | 36.1 | 37.37 | 34.71 | 37.08 | 44.19 | 46.82 | 38.08 | 38.85 |
| | | Best | 30.46 | 34.27 | 32.41 | 35.01 | 34.92 | 36.63 | | | 38.85 | 40.62 | 37.19 | 38.85 | 35.36 | 37.37 | 45.73 | 46.82 | 38.76 | 40.92 |
| | 4 | Last | 94.35 | **95.7** | 93.95 | **95.85** | 94.75 | **95.55** | | | 82.43 | 85.56 | 81.07 | 82.49 | 92.2 | 94.32 | 94.65 | **98.92** | 84.7 | 85.56 |
| | | Best | **96.44** | **97.08** | **95.79** | **96.62** | **96.31** | **97.24** | | | 82.67 | 85.56 | 81.78 | 82.95 | 93.15 | 94.32 | **98.34** | **98.92** | 85.41 | 88.63 |
| | 5 | Last | 61.97 | 72.81 | 60.89 | 63.29 | 61.78 | 69.12 | | | 44.49 | 58.53 | 50.26 | 52.84 | 35.51 | 63.29 | **96.5** | **97.24** | 54.5 | 55.61 |
| | | Best | 67.1 | 75.88 | 64.45 | 69.89 | 66.21 | 73.43 | | | 46.48 | 58.53 | 50.78 | 52.84 | 37.26 | 63.29 | **96.65** | **97.24** | 55.85 | 56.37 |

SEU
Results with random split and data augmentation

| Nor | Input | Loc | AE Mean | AE Max | DAE Mean | DAE Max | SAE Mean | SAE Max | MLP Mean | MLP Max | CNN Mean | CNN Max | LeNet Mean | LeNet Max | AlexNet Mean | AlexNet Max | ResNet18 Mean | ResNet18 Max | LSTM Mean | LSTM Max |
|---|---|---|---|---|---|---|---|---|---|---|---|---|---|---|---|---|---|---|---|---|
| A | 1 | Last | 50.25 | 53.19 | 45.1 | 52.7 | 46.91 | 52.7 | 60.64 | 65.69 | 88.24 | 94.36 | 83.19 | 86.27 | 53.43 | 75.74 | **95.73** | **99.75** | 87.38 | 91.67 |
|   |   | Best | 61.72 | 66.18 | 62.55 | 64.22 | 65.1 | 68.14 | 63.87 | 66.91 | 90.98 | **95.59** | 85.25 | 87.75 | 55.15 | 78.43 | **97.15** | **99.75** | 87.93 | 92.65 |
|   | 2 | Last | **97.16** | **97.79** | **97.45** | **98.28** | **97.16** | **98.04** | **97.35** | **97.55** | 97.3 | 98.04 | 96.81 | 98.53 | 96.03 | 96.81 | 97.25 | 97.55 | 96.02 | 96.81 |
|   |   | Best | **98.14** | **98.53** | **98.38** | **98.77** | **98.48** | **98.77** | 97.79 | 98.28 | 97.55 | 98.53 | 97.3 | 98.53 | 97.4 | 98.04 | 97.7 | 98.53 | 96.5 | 97.55 |
|   | 3 | Last | 50.39 | 54.57 | 53.94 | 57.69 | 55.24 | 63.46 |   |   | 52.94 | 59.62 | 45.34 | 54.81 | 64.95 | 66.59 | 59.52 | 65.14 | 54.69 | 55.77 |
|   |   | Best | 52.4 | 57.45 | 58.99 | 62.02 | 60.39 | 66.35 |   |   | 53.7 | 59.62 | 45.48 | 54.81 | 66.02 | 67.55 | 61.88 | 66.59 | 56.07 | 57.69 |
|   | 4 | Last | **97.5** | **99.02** | **97.99** | **99.02** | **98.09** | **98.77** |   |   | 84.41 | 88.73 | 84.31 | 89.46 | 78.04 | 98.04 | **96.22** | **100** | 88.48 | 90.44 |
|   |   | Best | **98.77** | **99.51** | **99.07** | **99.51** | **98.72** | **99.26** |   |   | 86.81 | 90.2 | 85 | 90.2 | 79.12 | 98.77 | **99.66** | **100** | 89.95 | 91.67 |
|   | 5 | Last | 52.89 | 86.76 | 66.32 | 83.58 | 56.37 | 79.9 |   |   | 20.24 | 43.38 | 34.61 | 43.14 | 4.9 | 4.9 | 67.11 | **100** | 43.38 | 46.08 |
|   |   | Best | 54.22 | 88.48 | 69.27 | 88.73 | 61.42 | 81.13 |   |   | 21.92 | 50.52 | 42.7 | 49.75 | 4.9 | 4.9 | 82.94 | **100** | 44.98 | 47.06 |
| B | 1 | Last | 77.55 | 79.9 | 79.85 | 82.84 | 78.24 | 80.39 | 82.6 | 83.82 | 92.84 | **95.83** | 86.08 | 89.71 | 89.26 | 94.36 | **99.46** | **99.75** | 90.35 | 91.91 |
|   |   | Best | 79.61 | 82.6 | 81.08 | 83.09 | 78.97 | 80.39 | 83.48 | 85.78 | 93.14 | **96.81** | 87.79 | 91.91 | 90.78 | 94.36 | **99.66** | **100** | 91.03 | 93.63 |
|   | 2 | Last | **95.59** | **97.3** | **95.74** | **97.06** | 94.9 | **97.06** | **96.47** | **97.79** | 97.84 | 98.28 | 97.25 | 97.55 | 96.96 | 97.55 | 96.17 | 97.79 | 96.86 | 97.55 |
|   |   | Best | **97.7** | **98.04** | **97.89** | **98.28** | **97.84** | **98.04** | 96.96 | 97.79 | 98.09 | 98.77 | 97.44 | 97.99 | 97.25 | 97.79 | 96.76 | 98.28 | 97.3 | 98.28 |
|   | 3 | Last | 54.85 | 56.97 | 54.9 | 60.58 | 56.2 | 61.06 |   |   | 52.74 | 59.62 | 50.34 | 54.09 | 62.31 | 65.38 | 60.67 | 66.59 | 56.63 | 60.1 |
|   |   | Best | 58.89 | 62.26 | 57.5 | 61.78 | 59.57 | 64.18 |   |   | 54.47 | 60.1 | 51.68 | 56.01 | 63.85 | 66.35 | 62.65 | 71.88 | 57.98 | 62.5 |
|   | 4 | Last | **97.4** | **98.28** | **95.93** | **98.53** | **97.45** | **98.04** |   |   | 87.25 | 89.71 | 86.32 | 89.71 | **96.32** | **98.53** | **99.26** | **100** | 88.38 | 91.42 |
|   |   | Best | **98.92** | **99.51** | **97.45** | **98.77** | **98.87** | **99.51** |   |   | 88.68 | 90.69 | 87.06 | 91.42 | 96.86 | 98.53 | **99.8** | **100** | 89.26 | 91.42 |
|   | 5 | Last | 86.08 | 88.48 | 81.23 | 86.52 | 73.43 | 87.75 |   |   | 14.02 | 50.49 | 43.48 | 47.55 | 59.41 | 85.05 | 97.16 | 99.51 | 46.77 | 49.02 |
|   |   | Best | 88.87 | 92.16 | 87.3 | 91.91 | 78.48 | 90.44 |   |   | 14.12 | 50.49 | 45.34 | 47.55 | 61.42 | 90.44 | 97.4 | **100** | 47.6 | 50.49 |
| C | 1 | Last | 84.31 | 87.25 | 83.72 | 86.03 | 83.33 | 90.44 | 87.79 | 91.18 | **95.83** | **97.3** | 93.33 | 94.85 | **99.02** | **99.51** | **100** | **100** | 96.86 | 97.55 |
|   |   | Best | 85.78 | 87.25 | 85.15 | 87.25 | 85.1 | 90.44 | 88.38 | 91.18 | **96.27** | **97.3** | 93.97 | 96.32 | **99.26** | **100** | **100** | **100** | 97.3 | 98.28 |
|   | 2 | Last | **97.74** | **98.04** | **97.79** | **98.28** | **97.84** | **98.53** | **97.84** | **98.28** | 98.82 | 99.26 | 97.6 | 98.04 | 98.18 | 99.51 | 97.84 | 98.53 | 97.5 | 98.28 |
|   |   | Best | **98.58** | **98.77** | **99.02** | **99.02** | **98.67** | **98.77** | **98.15** | **99.02** | 99.12 | 99.51 | 97.99 | 99.26 | 98.67 | **100** | 98.09 | 99.02 | 97.7 | 98.53 |
|   | 3 | Last | 51.68 | 54.09 | 49.42 | 53.85 | 56.4 | 59.86 |   |   | 57.93 | 63.94 | 56.97 | 59.86 | 63.12 | 65.62 | 71.88 | 74.76 | 60.29 | 63.94 |
|   |   | Best | 54.9 | 58.65 | 52.02 | 55.53 | 59.71 | 61.54 |   |   | 61.59 | 63.94 | 57.74 | 61.54 | 64.09 | 66.59 | 72.89 | 74.76 | 61.06 | 63.94 |
|   | 4 | Last | **96.96** | **99.26** | **98.63** | **98.77** | **97.55** | **99.02** |   |   | 90 | 92.16 | 87.4 | 90.2 | **96.81** | **99.02** | **99.7** | **100** | 90.05 | 91.18 |
|   |   | Best | **98.48** | **99.75** | **99.46** | **99.75** | **99.26** | **100** |   |   | 90.34 | 92.65 | 88.38 | 90.2 | **97.85** | **99.02** | **99.8** | **100** | 90.98 | 91.91 |
|   | 5 | Last | 89.95 | 92.16 | 88.04 | 89.95 | 86.47 | 92.16 |   |   | 31.72 | 53.92 | 51.23 | 56.13 | 63.04 | 80.64 | **99.26** | **100** | 52.84 | 56.13 |
|   |   | Best | 93.19 | **96.57** | 91.22 | 94.36 | 92.26 | 94.12 |   |   | 32.5 | 54.66 | 52.3 | 56.13 | 63.63 | 81.13 | **99.41** | **100** | 55.07 | 58.18 |

*A is the max-min normalization; B is the -1-1 normalization; C is the Z-score normalization

1 is the time domain input, 2 is the frequency domain input; 3 is the wavelet domain input; 4 is the time domain sample after STFT; 5 is the time domain sample reshape to a 2D matrix

## SEU
### Results with random split and without data augmentation

| | | | AE | | SAE | | DAE | | MLP | | CNN | | LeNet | | AlexNet | | ResNet18 | | LSTM | |
|---|---|---|---|---|---|---|---|---|---|---|---|---|---|---|---|---|---|---|---|---|
| Nor | Input | Loc | Mean | Max | Mean | Max | Mean | Max | Mean | Max | Mean | Max | Mean | Max | Mean | Max | Mean | Max | Mean | Max |
| A | 1 | Last | 40.13 | 44.85 | 31.52 | 34.56 | 36.42 | 42.16 | 37.45 | 44.61 | 88.09 | 91.42 | 83.24 | 87.01 | 69.93 | 72.55 | **98.04** | **99.02** | 79.22 | 81.37 |
| | | Best | 44.67 | 48.53 | 35.54 | 37.99 | 43.78 | 45.59 | 42.94 | 48.28 | 88.38 | 92.16 | 83.58 | 87.5 | 72.57 | 77.77 | **98.28** | **99.51** | 80.78 | 82.35 |
| | 2 | Last | **96.71** | **97.55** | **97.35** | **98.04** | **97.11** | **97.55** | **97.6** | **98.28** | 96.86 | 99.26 | 98.09 | 98.53 | 97.2 | 97.55 | **99.85** | **100** | 97.35 | 97.79 |
| | | Best | **98.38** | **98.53** | **98.58** | **98.77** | **98.43** | **98.77** | 97.99 | 98.28 | 97.4 | 99.75 | 98.43 | 99.26 | 97.5 | 98.28 | 99.9 | 100 | 97.94 | 98.77 |
| | 3 | Last | 57.6 | 63.46 | 57.45 | 60.1 | 54.18 | 60.34 | | | 48.51 | 56.97 | 50.53 | 54.57 | 62.62 | 65.87 | 52.16 | 62.98 | 54.28 | 56.73 |
| | | Best | 61.54 | 65.38 | 60.53 | 62.74 | 60.14 | 62.5 | | | 53.89 | 57.69 | 52.39 | 54.57 | 63.16 | 66.35 | 56.3 | 72.6 | 55.91 | 57.93 |
| | 4 | Last | **98.58** | **99.02** | **98.18** | **98.77** | **98.09** | **98.77** | | | 85.69 | 90.69 | 88.24 | 90.2 | **96.42** | **97.06** | **99.85** | **100** | 88.78 | 90.2 |
| | | Best | **98.97** | **99.51** | **99.21** | **99.75** | **98.77** | **99.51** | | | 86.52 | 90.69 | 89 | 91.42 | **97.35** | **98.77** | **99.9** | **100** | 89.81 | 90.69 |
| | 5 | Last | 64.12 | 86.03 | 69.8 | 86.03 | 80.59 | 89.46 | | | 14.31 | 42.65 | 33.97 | 39.46 | 4.9 | 4.9 | **98.72** | **100** | 41.42 | 48.04 |
| | | Best | 66.76 | 87.5 | 75.05 | 89.95 | 83.14 | 89.71 | | | 14.46 | 42.65 | 36.52 | 44.85 | 5.05 | 5.15 | **98.77** | **100** | 46.08 | 50.49 |
| B | 1 | Last | 44.97 | 48.28 | 36.13 | 39.71 | 43.22 | 44.61 | 43.87 | 47.06 | 87.35 | 89.22 | 81.62 | 84.31 | 85.25 | **97.06** | 97.89 | 99.26 | 79.66 | 82.35 |
| | | Best | 49.02 | 52.7 | 37.94 | 41.42 | 47.96 | 50.98 | 44.27 | 47.55 | 87.99 | 89.95 | 82.65 | 84.31 | 85.78 | **97.55** | **98.77** | 99.26 | 80.64 | 82.35 |
| | 2 | Last | 92.61 | 97.55 | 78.58 | 91.91 | 79.7 | **97.79** | 93.87 | **97.3** | 98.73 | 99.51 | 97.89 | 98.28 | 97.5 | 98.28 | 99.66 | 100 | **97.5** | **98.04** |
| | | Best | **98.16** | **98.28** | **98.08** | **98.77** | **98.28** | **98.53** | 94.85 | **98.04** | 98.92 | 99.51 | 98.23 | 99.02 | 97.6 | 98.28 | 99.66 | 100 | 97.89 | 98.28 |
| | 3 | Last | 55.89 | 61.06 | 57.11 | 58.89 | 59.05 | 65.62 | | | 56.39 | 59.62 | 48.75 | 51.2 | 62.36 | 65.87 | 59.71 | 66.59 | 55.14 | 58.41 |
| | | Best | 58.97 | 64.42 | 58.75 | 60.34 | 60.38 | 65.87 | | | 58.46 | 59.62 | 51.64 | 55.77 | 63.99 | 66.35 | 64.81 | 72.6 | 57.07 | 61.06 |
| | 4 | Last | 94.08 | **99.02** | **97.01** | **98.28** | **97.74** | **99.26** | | | 89.22 | 91.67 | 79.22 | 90.69 | **96.13** | **96.57** | **99.85** | **100** | 86.37 | 88.24 |
| | | Best | **98.49** | **99.02** | **98.23** | **99.26** | **98.43** | **99.26** | | | 89.66 | 91.67 | 88.68 | 91.91 | **96.71** | **98.28** | **99.85** | **100** | 87.4 | 90.93 |
| | 5 | Last | 85.7 | 92.4 | 86.52 | 90.44 | 83.68 | 90.93 | | | 13.38 | 47.3 | 45.64 | 49.51 | 14.8 | 54.41 | **99.46** | **99.75** | 45.1 | 46.08 |
| | | Best | 86.97 | 92.4 | 88.33 | 90.93 | 85.15 | 92.16 | | | 13.38 | 47.3 | 46.82 | 51.23 | 14.85 | 54.41 | **99.61** | **100** | 46.81 | 50.25 |
| C | 1 | Last | 46.03 | 48.77 | 34.9 | 39.46 | 48.23 | 49.75 | 43.53 | 48.04 | 88.14 | 92.16 | 87.7 | 91.67 | **96.91** | **98.53** | 99.56 | 100 | 91.81 | 93.38 |
| | | Best | 50.88 | 53.43 | 40.25 | 42.4 | 50.34 | 52.21 | 45.44 | 50.74 | 91.62 | 92.89 | 88.43 | 93.14 | **97.74** | **100** | 99.56 | 100 | 92.74 | 93.38 |
| | 2 | Last | **98.14** | **99.02** | **97.89** | **99.02** | **97.79** | **98.28** | **97.84** | **98.28** | 98.38 | 99.51 | 99.16 | 99.75 | 99.65 | 100 | **100** | **100** | 98.33 | 99.26 |
| | | Best | **98.87** | **99.02** | **99.36** | **99.51** | **98.87** | **99.26** | 97.99 | 98.53 | 98.58 | 99.51 | 99.31 | 99.75 | 99.75 | 100 | **100** | **100** | 98.63 | 99.51 |
| | 3 | Last | 43.37 | 61.3 | 40.67 | 50.72 | 54.71 | 57.69 | | | 59.61 | 61.54 | 55.86 | 58.17 | 61.54 | 63.46 | 69.33 | 72.12 | 59.23 | 63.7 |
| | | Best | 45.05 | 62.5 | 42.26 | 52.16 | 57.16 | 59.13 | | | 63.3 | 64.42 | 57.45 | 58.65 | 62.69 | 64.9 | 70.92 | 75.96 | 60.63 | 63.7 |
| | 4 | Last | **98.87** | **99.51** | **98.53** | **99.26** | **98.58** | **99.51** | | | 89.56 | 91.18 | 87.55 | 88.73 | 95.39 | 96.81 | **100** | **100** | 90.15 | 92.65 |
| | | Best | **99.02** | **99.51** | **98.68** | **99.51** | **99.12** | **99.51** | | | 90.88 | 92.65 | 88.09 | 90.2 | 95.83 | 97.79 | **100** | **100** | 91.96 | 94.36 |
| | 5 | Last | 91.32 | 93.63 | 91.96 | 94.85 | 90.83 | 92.65 | | | 4.9 | 4.9 | 47.01 | 53.43 | 51.23 | 87.5 | **99.9** | **100** | 53.04 | 55.15 |
| | | Best | 92.4 | 94.61 | 93.04 | **96.32** | 91.77 | 93.14 | | | 4.95 | 5.15 | 48.97 | 55.64 | 51.28 | 87.5 | **99.95** | **100** | 53.83 | 57.11 |

SEU

Results with order split and data augmentation

| Nor | Input | Loc | AE Mean | AE Max | SAE Mean | SAE Max | DAE Mean | DAE Max | MLP Mean | MLP Max | CNN Mean | CNN Max | LeNet Mean | LeNet Max | AlexNet Mean | AlexNet Max | ResNet18 Mean | ResNet18 Max | LSTM Mean | LSTM Max |
|---|---|---|---|---|---|---|---|---|---|---|---|---|---|---|---|---|---|---|---|---|
| A | 1 | Last | 38.81 | 45.95 | 31.81 | 34.76 | 39.9 | 49.52 | 38.05 | 46.67 | 83.24 | 90.71 | 67.05 | 74.76 | 67.62 | 73.1 | **95.53** | **98.33** | 77.76 | 79.76 |
| | | Best | 44.48 | 47.86 | 48.43 | 52.14 | 46.33 | 49.52 | 41.24 | 46.67 | 84.05 | 90.71 | 67.38 | 74.76 | 69.48 | 75.1 | **96.05** | **99.05** | 79.24 | 81.9 |
| | 2 | Last | **99.24** | **99.76** | 99 | 100 | 98.71 | 99.52 | **99.38** | **99.76** | 98.48 | 99.76 | 98.24 | 99.52 | 98.62 | 99.05 | 98.38 | 99.52 | 99.48 | 99.76 |
| | | Best | **100** | **100** | 100 | 100 | 100 | 100 | 99.57 | 99.76 | 98.86 | 99.76 | 98.33 | 99.52 | 98.9 | 99.52 | 98.81 | 99.76 | 99.52 | 100 |
| | 3 | Last | 55.57 | 59.76 | 54.38 | 55.71 | 55.43 | 64.05 | | | 51.48 | 58.33 | 50.28 | 61.9 | 57.29 | 60 | 53 | 64.76 | 53.62 | 55 |
| | | Best | 59.47 | 62.14 | 60.9 | 62.14 | 60.1 | 64.05 | | | 55.1 | 59.05 | 55 | 61.9 | 59.52 | 63.57 | 54.52 | 68.57 | 55.72 | 60.24 |
| | 4 | Last | 88 | 90.48 | 87.62 | 92.38 | 89.62 | 95 | | | 80.62 | 87.38 | 84.05 | 86.19 | **97.08** | **97.86** | 98.81 | 99.29 | 88.62 | 90.24 |
| | | Best | 93.48 | **96.43** | 94.48 | **96.67** | 94.09 | **96.67** | | | 82.71 | 88.33 | 84.95 | 87.62 | **97.86** | **98.57** | 98.91 | 99.29 | 90.62 | 92.38 |
| | 5 | Last | 47.24 | 63.33 | 60.05 | 73.33 | 50.29 | 53.57 | | | 13.62 | 27.62 | 34.67 | 41.43 | 5 | 5 | **94.86** | **98.1** | 30.62 | 38.57 |
| | | Best | 51.19 | 65.95 | 66.33 | 76.19 | 55.53 | 65 | | | 13.76 | 27.62 | 38.91 | 41.67 | 5 | 5 | **97.1** | **98.57** | 32.57 | 41.43 |
| B | 1 | Last | **99** | **99.52** | 95.33 | 99.29 | 96.57 | 99.05 | 54.52 | 58.33 | 88.95 | 92.38 | 64.81 | 69.05 | 89.43 | 94.76 | **99.19** | **99.76** | 85.59 | 86.67 |
| | | Best | **99.86** | **100** | 99.9 | 100 | 100 | 100 | 56.24 | 62.62 | 91.14 | 93.33 | 67.05 | 70 | 91.19 | **96.9** | **99.43** | **99.76** | 86.61 | 88.1 |
| | 2 | Last | 50.14 | 57.14 | 52.29 | 59.52 | 53.81 | 54.29 | **98.33** | **99.76** | 97.67 | 98.81 | 99.28 | 99.76 | 98.28 | 99.52 | 95.67 | 99.29 | 98.93 | 99.76 |
| | | Best | 53.95 | 57.14 | 57.38 | 60.95 | 59.48 | 61.43 | **99.48** | **99.76** | 98.29 | 99.29 | 99.52 | 99.76 | 99 | 99.52 | 96.57 | 99.76 | 99.52 | 99.76 |
| | 3 | Last | 91.95 | 92.86 | 89.43 | 93.81 | 88.76 | 92.86 | | | 50.43 | 53.81 | 54.29 | 56.43 | 60.81 | 63.81 | 64.62 | 67.86 | 52.98 | 56.19 |
| | | Best | **95.43** | **96.43** | 94 | **97.62** | 93.38 | 95.71 | | | 53.19 | 56.9 | 55 | 59.05 | 62.29 | 65.95 | 65.86 | 69.29 | 56.38 | 58.62 |
| | 4 | Last | 54.76 | 57.86 | 53.81 | 58.33 | 41.14 | 55.24 | | | 83.87 | 87.38 | 84.57 | 86.67 | **95.48** | **96.67** | 93.76 | **99.05** | 85.36 | 86.43 |
| | | Best | 59.9 | 64.52 | 60 | 63.57 | 46.91 | 61.67 | | | 85.42 | 87.38 | 85.33 | 86.9 | **96.33** | **98.57** | 94.09 | 99.52 | 86.49 | 88.33 |
| | 5 | Last | **99** | **99.52** | 95.33 | 99.29 | 96.57 | 99.05 | | | 44.29 | 44.29 | 37.86 | 44.52 | 64.52 | 68.81 | **97.33** | **98.57** | 39.94 | 42.38 |
| | | Best | **99.86** | **100** | 99.9 | 100 | 100 | 100 | | | 46.67 | 46.67 | 39.27 | 44.52 | 68.65 | 76.9 | **97.62** | **98.57** | 40.48 | 42.38 |
| C | 1 | Last | 56.52 | 59.29 | 63.24 | 68.1 | 59.19 | 60.95 | 56.29 | 58.57 | 93.52 | **95.24** | 82.95 | 85.48 | **97.24** | **99.29** | 97.29 | 99.29 | 93.29 | 94.76 |
| | | Best | 59.91 | 64.05 | 66 | 68.33 | 62.19 | 62.86 | 58.81 | 63.57 | 94.71 | **97.14** | 82.95 | 85.48 | **98.1** | **99.29** | 98.05 | 99.29 | 93.52 | **95** |
| | 2 | Last | **99.52** | **100** | 99.52 | 99.76 | 99.1 | 99.29 | 99.24 | 99.52 | 99.14 | 99.52 | 99 | 99.52 | 99.38 | 99.76 | 99.09 | 99.76 | 99 | 99.29 |
| | | Best | **99.95** | **100** | 99.95 | 100 | 99.9 | 100 | 99.33 | 99.76 | 99.48 | 100 | 99.28 | 99.52 | 99.48 | 100 | 99.81 | 100 | 99.29 | 99.76 |
| | 3 | Last | 46.48 | 48.81 | 51.14 | 56.43 | 53.24 | 55.95 | | | 56.24 | 58.33 | 54.24 | 55.24 | 62.38 | 63.57 | 62.53 | 69.05 | 57.71 | 62.14 |
| | | Best | 50.38 | 55.24 | 53.29 | 60.24 | 56.81 | 58.57 | | | 59 | 62.62 | 56.14 | 58.81 | 62.95 | 64.76 | 65.48 | 73.57 | 58.76 | 62.14 |
| | 4 | Last | 86.14 | 91.19 | 91.67 | 94.29 | 88.62 | 95.95 | | | 83.95 | 91.43 | 84.67 | 86.67 | **95.67** | **97.86** | 99.09 | 99.52 | 88.86 | 91.67 |
| | | Best | 92.71 | **97.38** | 94.33 | 95.48 | 94.05 | 98.1 | | | 86.33 | 91.43 | 86.05 | 88.57 | **97.48** | **98.57** | 99.19 | 99.76 | 91.81 | 93.81 |
| | 5 | Last | 55.33 | 58.81 | 55 | 59.05 | 55.28 | 62.62 | | | 22.48 | 53.57 | 42.91 | 48.1 | 31 | 77.14 | **97.95** | **99.52** | 45.28 | 48.1 |
| | | Best | 58.09 | 59.52 | 60.71 | 65.48 | 62.47 | 65.24 | | | 23.29 | 53.57 | 44.29 | 48.81 | 33.09 | 82.14 | **98.52** | **99.76** | 45.76 | 48.1 |

UoC
Results with random split and data augmentation

| Nor | Input | Loc | AE | | DAE | | SAE | | MLP | | CNN | | LeNet | | AlexNet | | ResNet18 | | LSTM | |
|---|---|---|---|---|---|---|---|---|---|---|---|---|---|---|---|---|---|---|---|---|
| | | | Mean | Max | Mean | Max | Mean | Max | Mean | Max | Mean | Max | Mean | Max | Mean | Max | Mean | Max | Mean | Max |
| A | 1 | Last | 27.09 | 29.22 | 27.49 | 29.83 | 25.57 | 28.01 | 26.88 | 28.92 | 41.98 | 48.25 | 31.93 | 33.49 | 11.11 | 11.11 | 66.18 | 76.26 | 33.12 | 34.09 |
| | | Best | 29.92 | 31.35 | 31.66 | 32.57 | 28.46 | 29.38 | 30.2 | 31.2 | 47.55 | 51.45 | 33.94 | 35.46 | 11.45 | 12.79 | 76.8 | 78.39 | 36.89 | 37.9 |
| | 2 | Last | 92.94 | 94.06 | 90.78 | 91.63 | 92.76 | 93.46 | 92.66 | 93.46 | 68.49 | 71.08 | 79.97 | 81.58 | 70.62 | 75.49 | 88.16 | 89.65 | 79.21 | 80.67 |
| | | Best | 94.7 | **95.13** | 93.3 | 93.76 | 94.55 | **95.13** | 94.91 | **95.13** | 69.53 | 71.08 | 84.78 | 86.45 | 73.3 | 78.69 | 90.59 | 91.93 | 80.58 | 81.74 |
| | 3 | Last | 15.25 | 19.44 | 19.11 | 24.04 | 21.39 | 24.78 | | | 24.57 | 31.6 | 34.21 | 38.13 | 11.13 | 11.13 | 62.34 | 63.8 | 37.48 | 40.36 |
| | | Best | 17.57 | 23.15 | 20.65 | 24.04 | 23.95 | 26.71 | | | 26.44 | 32.05 | 36.26 | 40.36 | 11.16 | 11.28 | 67.3 | 69.14 | 38.9 | 42.43 |
| | 4 | Last | 52.42 | 57.69 | 50.71 | 55.1 | 51.14 | 53.88 | | | 31.32 | 32.42 | 34.19 | 36.99 | 45.91 | 47.34 | 73.52 | 74.43 | 34.4 | 35.77 |
| | | Best | 55.98 | 60.27 | 53 | 56.47 | 54.43 | 56.16 | | | 34.43 | 35.01 | 37.57 | 39.12 | 48.04 | 50.53 | 79.18 | 80.67 | 37.26 | 37.75 |
| | 5 | Last | 35.37 | 49.01 | 34.03 | 45.97 | 37.26 | 46.27 | | | 21.92 | 23.59 | 24.35 | 26.03 | 11.11 | 11.11 | 78.51 | 83.56 | 20.12 | 22.53 |
| | | Best | 37.56 | 49.01 | 36.83 | 49.01 | 39.39 | 48.1 | | | 25.69 | 27.85 | 26.88 | 29.07 | 11.11 | 11.11 | 85.93 | 87.82 | 24.99 | 26.64 |
| B | 1 | Last | 26.61 | 27.55 | 29.98 | 31.81 | 28.01 | 29.68 | 27.67 | 28.92 | 46.18 | 47.79 | 32.17 | 34.09 | 30.11 | 40.79 | 69.16 | 72.15 | 35.13 | 37.29 |
| | | Best | 29.47 | 31.2 | 31.81 | 32.72 | 30.14 | 32.12 | 30.44 | 31.96 | 47.34 | 48.86 | 34.19 | 35.01 | 31.29 | 42.01 | 76.84 | 77.78 | 36.71 | 38.81 |
| | 2 | Last | 79.87 | 84.32 | 80.67 | 85.54 | 81.4 | 87.37 | 81.19 | 90.11 | 62.89 | 67.28 | 78.72 | 82.34 | 68.77 | 74.89 | 83.87 | 88.43 | 77.62 | 78.84 |
| | | Best | 88.05 | 89.19 | 90.02 | 91.63 | 89.53 | 91.63 | 89.5 | 90.87 | 67.91 | 68.65 | 81.19 | 82.34 | 72.33 | 76.56 | 89.8 | 90.87 | 80.7 | 82.19 |
| | 3 | Last | 12.85 | 19.73 | 17.86 | 21.81 | 21.19 | 23.15 | | | 27.18 | 33.09 | 36.29 | 38.72 | 13.21 | 21.51 | 65.19 | 67.66 | 37.62 | 40.65 |
| | | Best | 13.5 | 22.4 | 19.97 | 24.18 | 24.6 | 25.67 | | | 28.66 | 34.12 | 38.9 | 40.5 | 14.22 | 24.48 | 67.89 | 68.99 | 39.91 | 42.73 |
| | 4 | Last | 47.76 | 54.03 | 44.87 | 47.03 | 48.01 | 55.4 | | | 30.93 | 32.42 | 34.4 | 36.38 | 33 | 48.25 | 75.04 | 79.15 | 34 | 36.07 |
| | | Best | 52.48 | 56.16 | 48.07 | 52.36 | 52.06 | 56.47 | | | 34.58 | 35.01 | 38.39 | 39.73 | 35.13 | 52.51 | 79.63 | 80.97 | 37.99 | 39.88 |
| | 5 | Last | 39.7 | 51.45 | 41.95 | 52.97 | 35.59 | 44.9 | | | 16.96 | 22.68 | 27 | 29.22 | 11.11 | 11.11 | 71.38 | 87.21 | 20.7 | 22.68 |
| | | Best | 40.82 | 52.82 | 44.2 | 53.58 | 39.36 | 47.03 | | | 19.15 | 28.46 | 28.31 | 29.22 | 11.17 | 11.42 | 86.03 | 88.13 | 26.12 | 27.7 |
| C | 1 | Last | 26.27 | 30.14 | 28.1 | 28.46 | 25.15 | 28.16 | 27.67 | 32.12 | 42.98 | 46.42 | 29.86 | 32.57 | 34.95 | 46.58 | 67.09 | 73.52 | 37.63 | 39.73 |
| | | Best | 29.13 | 30.59 | 30.26 | 32.88 | 27.64 | 30.29 | 29.77 | 32.12 | 45.9 | 50.68 | 32.69 | 35.31 | 35.92 | 46.58 | 76.16 | 77.32 | 40.67 | 42.92 |
| | 2 | Last | 93.12 | 94.22 | 94.09 | 94.82 | 94.49 | **95.59** | 94.12 | 94.98 | 70.23 | 71.84 | 85.87 | 86.45 | 83.68 | 85.69 | 86.91 | 89.35 | 82.41 | 84.47 |
| | | Best | 94.95 | **95.28** | **95.22** | **95.89** | **95.22** | **95.74** | **95.68** | **96.19** | 73.55 | 74.73 | 87.73 | 88.74 | 86.27 | 87.21 | 89.2 | 90.26 | 84.05 | 85.84 |
| | 3 | Last | 12.88 | 19.88 | 18.1 | 21.66 | 20.03 | 21.81 | | | 34.69 | 38.87 | 38.6 | 39.17 | 11.13 | 11.13 | 61.45 | 65.58 | 36.94 | 40.5 |
| | | Best | 13.47 | 22.85 | 19.38 | 23.15 | 23.26 | 24.93 | | | 35.99 | 40.5 | 41.25 | 42.43 | 11.22 | 11.42 | 66.08 | 67.95 | 39.97 | 42.58 |
| | 4 | Last | 52.05 | 53.42 | 53.45 | 57.99 | 52.91 | 59.06 | | | 30.59 | 31.2 | 36.1 | 40.03 | 45.75 | 51.75 | 75.95 | 77.78 | 37.05 | 38.05 |
| | | Best | 55.8 | 57.08 | 55.56 | 59.06 | 57.57 | 59.97 | | | 34.8 | 35.46 | 39.36 | 41.7 | 47.73 | 52.82 | 80.31 | 81.28 | 39.63 | 40.18 |
| | 5 | Last | 44.57 | 49.01 | 44.9 | 48.25 | 49.25 | 53.58 | | | 16.29 | 26.64 | 26 | 27.09 | 11.11 | 11.11 | 71.29 | 88.74 | 22.59 | 23.29 |
| | | Best | 47.03 | 49.47 | 47.64 | 51.29 | 51.9 | 56.16 | | | 18.06 | 28.77 | 27.52 | 29.22 | 11.11 | 11.11 | 87.34 | 88.74 | 27.95 | 30.29 |

*A is the max-min normalization; B is the -1-1 normalization; C is the Z-score normalization

1 is the time domain input, 2 is the frequency domain input; 3 is the wavelet domain input; 4 is the time domain sample after STFT; 5 is the time domain sample reshape to a 2D matrix

UoC
Results with random split and without data augmentation

| | | | AE | | SAE | | DAE | | MLP | | CNN | | LeNet | | AlexNet | | ResNet18 | | LSTM | |
|---|---|---|---|---|---|---|---|---|---|---|---|---|---|---|---|---|---|---|---|---|
| Nor | Input | Loc | Mean | Max | Mean | Max | Mean | Max | Mean | Max | Mean | Max | Mean | Max | Mean | Max | Mean | Max | Mean | Max |
| A | 1 | Last | 21.55 | 25.27 | 22.13 | 25.57 | 21.89 | 25.11 | 24.41 | 27.55 | 36.23 | 42.92 | 28.07 | 31.05 | 11.11 | 11.11 | 72.12 | 88.28 | 29.71 | 32.42 |
| | | Best | 25.39 | 27.7 | 26.7 | 27.7 | 24.47 | 25.11 | 27.37 | 28.46 | 44.08 | 48.55 | 31.96 | 34.4 | 11.26 | 11.87 | 88.04 | 89.5 | 34.43 | 36.38 |
| | 2 | Last | 92.53 | 94.67 | 91.45 | 93.15 | 92.15 | 94.06 | 91.69 | 92.69 | 65.36 | 69.25 | 83.38 | 84.93 | 75.04 | 78.54 | 88.13 | 90.72 | 80.82 | 82.34 |
| | | Best | 93.94 | 94.67 | 92.95 | 93.61 | 93.59 | 94.37 | 94.07 | 94.52 | 67.03 | 69.25 | 85.33 | 86.76 | 76.29 | 78.69 | 91.9 | 92.39 | 82.5 | 84.02 |
| | 3 | Last | 18.22 | 21.51 | 18.22 | 21.36 | 18.37 | 19.44 | | | 27.15 | 29.38 | 31.54 | 33.68 | 11.13 | 11.13 | 63.02 | 66.17 | 35.7 | 42.14 |
| | | Best | 21.72 | 23.29 | 18.66 | 21.81 | 21.87 | 22.85 | | | 29.26 | 31.9 | 35.34 | 38.72 | 11.13 | 11.13 | 67.36 | 69.44 | 39.17 | 44.66 |
| | 4 | Last | 55.4 | 57.99 | 51.78 | 54.49 | 53.52 | 58.75 | | | 31.6 | 32.12 | 34.15 | 37.29 | 44.23 | 52.21 | 75.89 | 78.69 | 35.68 | 36.99 |
| | | Best | 58.3 | 59.67 | 54.98 | 56.62 | 55.65 | 59.36 | | | 34.1 | 35.01 | 37.93 | 39.57 | 46.91 | 54.19 | 79.15 | 80.06 | 38.33 | 39.57 |
| | 5 | Last | 35.43 | 50.23 | 30.23 | 32.57 | 36.38 | 42.92 | | | 18.2 | 25.57 | 22.16 | 24.05 | 11.11 | 11.11 | 71.14 | 85.39 | 19.39 | 21.31 |
| | | Best | 37.75 | 50.84 | 32.12 | 33.33 | 38.69 | 45.97 | | | 19.27 | 27.7 | 27.06 | 28.01 | 11.17 | 11.42 | 86.24 | 87.98 | 23.87 | 25.88 |
| B | 1 | Last | 22.98 | 23.59 | 24.2 | 28.16 | 23.68 | 24.96 | 24.41 | 26.64 | 39.73 | 43.07 | 27.64 | 31.05 | 36.04 | 42.31 | 81.74 | 87.37 | 28.52 | 30.75 |
| | | Best | 26.39 | 27.4 | 26.94 | 28.61 | 26.06 | 28.77 | 27.76 | 29.07 | 43.74 | 45.97 | 32.21 | 33.79 | 38.93 | 42.31 | 87.06 | 87.98 | 33.3 | 34.09 |
| | 2 | Last | 78.17 | 84.78 | 68.16 | 78.08 | 77.11 | 82.34 | 73.85 | 78.84 | 64.32 | 67.58 | 81.43 | 83.71 | 78.26 | 81.58 | 89.93 | 91.63 | 80.37 | 82.8 |
| | | Best | 85.75 | 88.13 | 86.03 | 88.43 | 85.91 | 86.91 | 86.48 | 88.28 | 65.72 | 68.8 | 84.05 | 85.69 | 80.12 | 83.56 | 92.33 | 93.3 | 82.71 | 84.17 |
| | 3 | Last | 13 | 20.47 | 20.18 | 21.96 | 20.77 | 22.85 | | | 29.94 | 31.01 | 32.17 | 36.2 | 11.13 | 11.13 | 64.01 | 67.51 | 36.86 | 39.76 |
| | | Best | 13.68 | 22.7 | 22.46 | 24.63 | 23.65 | 26.11 | | | 31.75 | 32.94 | 34.95 | 39.91 | 11.51 | 12.31 | 68.25 | 69.14 | 39.35 | 41.99 |
| | 4 | Last | 49.07 | 51.75 | 53.21 | 55.56 | 50.5 | 54.79 | | | 30.69 | 31.51 | 32.6 | 34.09 | 39.63 | 48.4 | 67.52 | 76.41 | 35.95 | 37.29 |
| | | Best | 51.14 | 53.73 | 55.8 | 59.06 | 53.52 | 56.01 | | | 34.49 | 35.62 | 37.35 | 39.57 | 42.31 | 49.62 | 79.6 | 81.74 | 39.21 | 40.64 |
| | 5 | Last | 42.59 | 53.42 | 47.36 | 50.68 | 40.25 | 46.73 | | | 19.66 | 23.74 | 24.99 | 26.79 | 15.19 | 22.98 | 71.63 | 81.28 | 22.04 | 23.59 |
| | | Best | 44.14 | 54.03 | 49.1 | 51.6 | 41.43 | 47.95 | | | 22.04 | 27.09 | 27.55 | 28.31 | 16.47 | 24.96 | 87.12 | 88.13 | 27.37 | 28.01 |
| C | 1 | Last | 22.68 | 24.66 | 26.24 | 29.07 | 22.46 | 25.11 | 26.12 | 27.85 | 37.66 | 42.92 | 28.22 | 32.88 | 28.89 | 45.21 | 83.23 | 87.82 | 34.21 | 37.44 |
| | | Best | 26.15 | 27.4 | 29.22 | 31.35 | 26.4 | 27.25 | 28.53 | 29.38 | 41.34 | 44.44 | 31.78 | 35.62 | 30.59 | 45.81 | 88.34 | 89.35 | 38.26 | 40.33 |
| | 2 | Last | 93.27 | 93.91 | 93.45 | 94.52 | 93.58 | 94.52 | 93.97 | **95.74** | 70.47 | 76.41 | 85.54 | 86.61 | 82.89 | 84.02 | 89.92 | 92.54 | 85.36 | 86.45 |
| | | Best | 94.58 | **95.13** | **95.58** | **96.04** | 94.89 | **95.74** | **95.4** | **95.74** | 72.51 | 77.93 | 86.64 | 87.82 | 84.29 | 85.69 | 92.91 | 93.46 | 87.15 | 87.67 |
| | 3 | Last | 12.74 | 20.92 | 13.55 | 18.55 | 22.28 | 25.96 | | | 30.45 | 36.2 | 37.42 | 38.87 | 17.24 | 41.84 | 58.22 | 63.35 | 40.09 | 41.1 |
| | | Best | 13.11 | 23 | 14.64 | 21.96 | 24.76 | 27.6 | | | 31.78 | 36.2 | 40.12 | 41.99 | 17.72 | 43.92 | 65.67 | 67.21 | 42.67 | 44.36 |
| | 4 | Last | 52.54 | 54.03 | 49.96 | 57.99 | 55.59 | 58.45 | | | 31.2 | 32.57 | 34.06 | 35.46 | 44.02 | 45.97 | 75.19 | 78.69 | 35.65 | 36.99 |
| | | Best | 54.1 | 55.86 | 52.39 | 58.14 | 57.63 | 59.51 | | | 35.53 | 36.83 | 37.41 | 38.51 | 47.76 | 50.99 | 79.54 | 80.97 | 39.51 | 40.18 |
| | 5 | Last | 46.45 | 49.32 | 42.83 | 48.25 | 49.22 | 60.88 | | | 13.36 | 22.37 | 25.54 | 28.16 | 11.11 | 11.11 | 71.72 | 77.78 | 22.4 | 24.35 |
| | | Best | 47.12 | 49.77 | 44.05 | 48.25 | 53.79 | 60.88 | | | 14.37 | 25.27 | 27.86 | 30.14 | 11.11 | 11.11 | 86.67 | 87.98 | 26.64 | 27.55 |

UoC
Results with order split and data augmentation

| | | | AE | | SAE | | DAE | | MLP | | CNN | | LeNet | | AlexNet | | ResNet18 | | LSTM | |
|---|---|---|---|---|---|---|---|---|---|---|---|---|---|---|---|---|---|---|---|---|
| Nor | Input | Loc | Mean | Max | Mean | Max | Mean | Max | Mean | Max | Mean | Max | Mean | Max | Mean | Max | Mean | Max | Mean | Max |
| A | 1 | Last | 25.75 | 27.09 | 26.3 | 29.53 | 24.72 | 26.03 | 25.72 | 27.7 | 28.83 | 32.72 | 29.95 | 31.05 | 11.11 | 11.11 | 38.96 | 43.99 | 34.79 | 35.46 |
| | | Best | 28.55 | 31.05 | 29.53 | 31.81 | 28 | 28.46 | 29.65 | 31.05 | 33.18 | 34.09 | 32.11 | 32.88 | 11.41 | 12.63 | 44.23 | 44.75 | 36.99 | 38.51 |
| | 2 | Last | 62.74 | 64.54 | 63.23 | 67.88 | 64.02 | 65.3 | 65.36 | 70.02 | 52.24 | 53.27 | 53.06 | 54.79 | 53.36 | 55.86 | 53.82 | 56.32 | 52.48 | 55.56 |
| | | Best | 67.88 | 70.47 | 68.31 | 68.95 | 68.52 | 71.08 | 70.04 | 71.99 | 55.86 | 57.99 | 58.02 | 59.82 | 56.99 | 57.84 | 61.67 | 63.32 | 55.98 | 57.53 |
| | 3 | Last | 17.9 | 22.37 | 15.79 | 19.56 | 21.72 | 23.26 | | | 24.89 | 27.7 | 26.43 | 29.48 | 13.04 | 20.74 | 43.41 | 45.63 | 27.23 | 28.3 |
| | | Best | 20.27 | 25.19 | 16.62 | 20.74 | 23.59 | 25.48 | | | 26.81 | 28 | 29.07 | 30.96 | 13.18 | 21.48 | 49.04 | 50.67 | 29.54 | 31.7 |
| | 4 | Last | 35.89 | 37.75 | 37.9 | 40.18 | 36.28 | 37.44 | | | 30.65 | 34.55 | 30.93 | 33.18 | 36.47 | 38.81 | 43.8 | 45.66 | 31.38 | 33.33 |
| | | Best | 39.82 | 42.16 | 40.43 | 41.55 | 39.42 | 42.01 | | | 36.32 | 37.6 | 36.29 | 38.2 | 39.12 | 39.73 | 48.46 | 50.23 | 35.74 | 37.44 |
| | 5 | Last | 21.52 | 32.57 | 22.8 | 28.31 | 24.05 | 27.85 | | | 18.6 | 26.33 | 24.32 | 26.79 | 11.11 | 11.11 | 34.61 | 39.88 | 21.58 | 23.14 |
| | | Best | 24.63 | 35.31 | 24.78 | 32.57 | 26.91 | 31.51 | | | 20.64 | 27.09 | 29.01 | 30.59 | 11.11 | 11.11 | 41.89 | 45.36 | 25.69 | 26.64 |
| B | 1 | Last | 24.87 | 26.64 | 24.9 | 28.16 | 26.33 | 28.31 | 25.42 | 27.7 | 29.65 | 32.88 | 30.07 | 31.35 | 24.48 | 32.12 | 37.87 | 41.7 | 34.25 | 35.01 |
| | | Best | 27.89 | 28.92 | 27.52 | 28.31 | 28.92 | 30.44 | 28.71 | 29.68 | 33.76 | 37.6 | 32.94 | 35.01 | 26.76 | 33.18 | 43.38 | 44.6 | 36.74 | 37.29 |
| | 2 | Last | 52.27 | 59.21 | 55.95 | 60.43 | 46.76 | 53.58 | 57.26 | 63.47 | 48.8 | 50.99 | 54.7 | 56.16 | 49.77 | 54.03 | 51.81 | 55.1 | 50.99 | 52.51 |
| | | Best | 62.41 | 64.08 | 65.36 | 67.88 | 62.31 | 64.99 | 65.9 | 68.49 | 53.42 | 56.01 | 57.2 | 58.9 | 53.36 | 57.23 | 59.97 | 61.64 | 54.7 | 55.1 |
| | 3 | Last | 13.95 | 20.59 | 14.72 | 20.59 | 19.23 | 21.48 | | | 22.55 | 24.3 | 24.03 | 25.33 | 12.56 | 17.78 | 46.19 | 48.44 | 20.89 | 21.61 |
| | | Best | 14.73 | 23.7 | 16.12 | 23.7 | 22.04 | 23.7 | | | 23.76 | 24.3 | 27.73 | 29.33 | 13.21 | 20.89 | 49.07 | 50.07 | 23.75 | 23.99 |
| | 4 | Last | 32.94 | 35.77 | 36.01 | 39.73 | 32.88 | 34.55 | | | 31.54 | 33.03 | 29.01 | 30.14 | 37.02 | 38.96 | 40.21 | 49.16 | 31.35 | 33.49 |
| | | Best | 36.71 | 40.33 | 39.18 | 41.7 | 38.23 | 39.57 | | | 37.2 | 38.05 | 35.4 | 36.99 | 41.25 | 43.23 | 49.22 | 49.92 | 35.62 | 36.07 |
| | 5 | Last | 26.79 | 28.77 | 19.72 | 27.85 | 26.36 | 29.22 | | | 18.32 | 24.81 | 25.48 | 28.46 | 17.47 | 21.61 | 30.75 | 40.03 | 23.01 | 24.96 |
| | | Best | 31.17 | 35.01 | 23.01 | 31.51 | 30.05 | 32.57 | | | 20.97 | 28.31 | 29.41 | 31.2 | 20.4 | 24.51 | 41.61 | 43.53 | 27.27 | 28.92 |
| C | 1 | Last | 23.62 | 27.25 | 24.47 | 27.09 | 23.93 | 26.48 | 23.93 | 25.11 | 31.2 | 32.88 | 29.68 | 32.72 | 29.96 | 31.81 | 38.63 | 45.66 | 36.44 | 38.2 |
| | | Best | 27.08 | 29.38 | 27.2 | 28.46 | 27.37 | 29.22 | 28.13 | 28.77 | 35.68 | 37.9 | 31.48 | 34.4 | 32.33 | 34.25 | 44.63 | 45.66 | 38.39 | 39.12 |
| | 2 | Last | 63.18 | 67.28 | 62.97 | 66.97 | 62.59 | 66.21 | 62.59 | 64.84 | 51.29 | 54.03 | 57.9 | 59.21 | 54.46 | 57.69 | 53.36 | 56.16 | 50.9 | 51.75 |
| | | Best | 68.25 | 70.62 | 67.78 | 70.47 | 68.51 | 72.15 | 69.89 | 70.78 | 55.68 | 56.62 | 61.8 | 62.71 | 58.66 | 60.12 | 59.21 | 61.04 | 56.89 | 58.14 |
| | 3 | Last | 12.8 | 19.56 | 15.82 | 20.89 | 20.89 | 24.15 | | | 23.53 | 28.44 | 26.49 | 28.89 | 23.11 | 34.52 | 39.73 | 48.59 | 25.96 | 27.26 |
| | | Best | 13.39 | 21.63 | 16.68 | 21.48 | 22.67 | 25.33 | | | 26.16 | 30.67 | 30.28 | 31.11 | 24.62 | 35.11 | 47.32 | 48.89 | 29.63 | 30.81 |
| | 4 | Last | 37.17 | 40.33 | 37.09 | 39.27 | 37.28 | 38.51 | | | 31.05 | 33.03 | 30.78 | 32.42 | 28.34 | 35.92 | 43.35 | 49.62 | 30.93 | 32.72 |
| | | Best | 40.62 | 43.68 | 40.76 | 42.62 | 41.48 | 43.84 | | | 35.83 | 36.68 | 36.13 | 37.14 | 31.29 | 37.29 | 49.86 | 52.51 | 35.59 | 36.68 |
| | 5 | Last | 28.17 | 32.88 | 28.1 | 32.27 | 28.11 | 35.77 | | | 11.11 | 11.11 | 24.29 | 27.55 | 11.11 | 11.11 | 31.84 | 37.6 | 23.65 | 26.03 |
| | | Best | 32.04 | 34.86 | 30.59 | 34.86 | 31.93 | 36.99 | | | 11.96 | 13.09 | 28.71 | 30.44 | 11.14 | 11.26 | 39.64 | 41.25 | 28.31 | 29.38 |

XJTU-SY
Results with random split and data augmentation

| Nor | Input | Loc | AE Mean | AE Max | DAE Mean | DAE Max | SAE Mean | SAE Max | MLP Mean | MLP Max | CNN Mean | CNN Max | LeNet Mean | LeNet Max | AlexNet Mean | AlexNet Max | ResNet18 Mean | ResNet18 Max | LSTM Mean | LSTM Max |
|---|---|---|---|---|---|---|---|---|---|---|---|---|---|---|---|---|---|---|---|---|
| A | 1 | Last | 68.59 | 74.22 | 71.98 | 80.99 | 70.62 | 78.12 | 66.72 | 73.7 | 94.64 | **99.74** | 96.09 | 99.22 | 84.12 | **98.18** | 88.39 | **99.74** | 96.82 | 97.92 |
|   |   | Best | 74.17 | 76.3 | 79.84 | 83.07 | 75.62 | 80.21 | 77.24 | 80.21 | **99.95** | 100 | **98.6** | 99.22 | 86.36 | **98.18** | **99.79** | 100 | **98.91** | 99.48 |
|   | 2 | Last | **100** | 100 | **100** | 100 | **100** | 100 | **100** | 100 | 96.82 | 100 | 90.89 | 100 | 6.51 | 6.51 | **100** | 100 | 95.83 | 100 |
|   |   | Best | **100** | 100 | **100** | 100 | **100** | 100 | **100** | 100 | **100** | 100 | **100** | 100 | 6.72 | 6.77 | **100** | 100 | **100** | 100 |
|   | 3 | Last | 37.81 | 48.7 | 30.78 | 40.89 | 38.85 | 47.14 |   |   | 71.2 | 73.44 | 65.05 | 71.88 | 6.51 | 6.51 | 74.38 | 92.19 | 67.45 | 70.57 |
|   |   | Best | 39.01 | 49.48 | 34.9 | 47.66 | 41.15 | 48.18 |   |   | 75.47 | 76.56 | 75.52 | 76.3 | 7.08 | 7.81 | 91.83 | 92.71 | 72.55 | 74.22 |
|   | 4 | Last | **99.95** | 100 | **99.84** | 100 | **99.69** | 100 |   |   | 98.65 | 99.22 | 98.33 | 99.48 | 99.58 | 100 | **100** | 100 | 99.17 | **99.74** |
|   |   | Best | **100** | 100 | **100** | 100 | **100** | 100 |   |   | 98.75 | 99.22 | 99.48 | 99.74 | **100** | 100 | **100** | 100 | **99.79** | 100 |
|   | 5 | Last | 42.97 | 90.62 | 68.02 | 90.89 | 59.17 | 89.84 |   |   | 6.51 | 6.51 | 60.16 | 74.48 | 6.51 | 6.51 | **100** | 100 | 71.41 | 76.56 |
|   |   | Best | 56.04 | 92.71 | 72.71 | 93.75 | 65.68 | 91.93 |   |   | 6.72 | 6.77 | 74.58 | 76.82 | 6.72 | 6.77 | **100** | 100 | 77.97 | 80.99 |
| B | 1 | Last | 86.67 | 88.54 | 88.9 | 90.36 | 85.57 | 88.54 | 87.08 | 89.06 | **99.53** | 100 | 97.55 | 98.7 | 97.03 | 98.44 | 98.96 | **99.74** | 96.35 | 98.18 |
|   |   | Best | 89.17 | 90.89 | 90.52 | 91.93 | 88.23 | 89.58 | 89.27 | 91.41 | **100** | 100 | 98.75 | 98.96 | 98.86 | **99.74** | **100** | 100 | 98.23 | 99.22 |
|   | 2 | Last | 11.93 | 21.88 | 46.25 | 57.03 | 12.39 | 16.93 | 13.18 | 17.45 | 61.56 | **99.22** | 93.54 | 100 | 6.51 | 6.51 | 68.75 | 100 | **100** | 100 |
|   |   | Best | 41.3 | 68.49 | 55.42 | 63.8 | 42.13 | 67.97 | 31.56 | 40.62 | **99.48** | 99.74 | **100** | 100 | 6.72 | 6.77 | **100** | 100 | **100** | 100 |
|   | 3 | Last | 28.91 | 45.31 | 37.03 | 47.66 | 44.43 | 49.74 |   |   | 74.22 | 77.34 | 72.87 | 77.34 | 79.63 | 83.85 | 89.79 | 91.15 | 70.89 | 73.7 |
|   |   | Best | 31.3 | 46.35 | 39.69 | 51.82 | 47.45 | 53.65 |   |   | 77.19 | 79.69 | 76.92 | 77.86 | 82.34 | 84.11 | 91.93 | 92.19 | 74.32 | 76.3 |
|   | 4 | Last | **99.9** | 100 | **100** | 100 | **99.84** | 100 |   |   | 98.39 | 99.22 | 86.62 | **99.74** | 99.32 | 100 | **99.9** | 100 | 98.65 | 100 |
|   |   | Best | **100** | 100 | **100** | 100 | **100** | 100 |   |   | 98.65 | 99.48 | 99.32 | 99.74 | **100** | 100 | **100** | 100 | **99.9** | 100 |
|   | 5 | Last | 87.08 | 90.89 | 80.78 | 90.89 | 72.4 | 86.72 |   |   | 38.65 | 67.97 | 72.5 | 75.52 | 40.16 | 90.89 | **100** | 100 | 76.09 | 80.21 |
|   |   | Best | 88.39 | 93.23 | 84.89 | 94.27 | 82.13 | 87.24 |   |   | 42.13 | 76.56 | 76.56 | 78.91 | 40.52 | 91.41 | **100** | 100 | 80.21 | 83.33 |
| C | 1 | Last | 88.07 | 90.36 | 90.36 | 91.93 | 86.41 | 88.54 | 88.07 | 89.58 | **97.6** | 100 | 97.61 | 99.22 | 98.86 | 99.48 | 99.22 | 100 | 98.23 | 98.7 |
|   |   | Best | 89.01 | 91.15 | 91.3 | 92.19 | 89.27 | 90.89 | 90.36 | 91.41 | **99.9** | 100 | 98.7 | 100 | 99.79 | 100 | **100** | 100 | 99.01 | 99.22 |
|   | 2 | Last | **100** | 100 | **100** | 100 | **100** | 100 | 99.95 | 100 | 99.84 | 100 | 99.84 | 100 | 99.95 | 100 | **100** | 100 | **100** | 100 |
|   |   | Best | **100** | 100 | **100** | 100 | **100** | 100 | **100** | 100 | **100** | 100 | **100** | 100 | **100** | 100 | **100** | 100 | **100** | 100 |
|   | 3 | Last | 10.89 | 28.39 | 16.04 | 40.89 | 39.74 | 51.82 |   |   | 74.17 | 78.39 | 75.05 | 76.82 | 83.44 | 84.64 | 91.93 | 92.71 | 70.89 | 72.4 |
|   |   | Best | 11.98 | 33.07 | 16.2 | 40.89 | 42.34 | 53.39 |   |   | 77.4 | 79.95 | 77.03 | 78.39 | 85.89 | 86.46 | 92.76 | 93.23 | 75.1 | 76.82 |
|   | 4 | Last | **100** | 100 | **99.9** | 100 | **100** | 100 |   |   | 96.88 | 98.96 | 98.6 | 98.96 | 99.79 | 100 | **100** | 100 | 98.96 | **99.74** |
|   |   | Best | **100** | 100 | **100** | 100 | **100** | 100 |   |   | 98.8 | 99.22 | 99.32 | 99.74 | **100** | 100 | **100** | 100 | **100** | 100 |
|   | 5 | Last | 90 | 93.75 | 88.23 | 94.79 | 80.57 | 91.15 |   |   | 34.64 | 78.39 | 69.79 | 75.26 | 88.85 | 94.79 | 92.14 | 100 | 80.94 | 85.42 |
|   |   | Best | 94.11 | 96.61 | 89.64 | 97.14 | 88.8 | 95.31 |   |   | 36.25 | 80.47 | 79.53 | 82.03 | 91.25 | 95.83 | **100** | 100 | 85.36 | 87.24 |

*A is the max-min normalization; B is the -1-1 normalization; C is the Z-score normalization

1 is the time domain input, 2 is the frequency domain input; 3 is the wavelet domain input; 4 is the time domain sample after STFT; 5 is the time domain sample reshape to a 2D matrix

XJTU-SY
Results with random split and without data augmentation

| | | | AE | | SAE | | DAE | | MLP | | CNN | | LeNet | | AlexNet | | ResNet18 | | LSTM | |
|---|---|---|---|---|---|---|---|---|---|---|---|---|---|---|---|---|---|---|---|---|
| Nor | Input | Loc | Mean | Max | Mean | Max | Mean | Max | Mean | Max | Mean | Max | Mean | Max | Mean | Max | Mean | Max | Mean | Max |
| A | 1 | Last | 36.04 | 48.18 | 55.21 | 57.29 | 48.85 | 52.6 | 60 | 62.24 | **99.58** | **100** | 93.8 | 96.35 | 73.44 | 93.23 | 94.69 | **100** | 94.95 | **97.14** |
| | | Best | 56.98 | 58.85 | 57.55 | 59.64 | 54.95 | 57.03 | 61.72 | 63.02 | **99.79** | **100** | 97.4 | 98.7 | 75.16 | **95.83** | 99.95 | **100** | **97.87** | **98.18** |
| | 2 | Last | **100** | **100** | **100** | **100** | **100** | **100** | **100** | **100** | 99.64 | 99.74 | **100** | **100** | 6.51 | 6.51 | **100** | **100** | **100** | **100** |
| | | Best | **100** | **100** | **100** | **100** | **100** | **100** | **100** | **100** | 99.79 | **100** | **100** | **100** | 6.72 | 6.77 | **100** | **100** | **100** | **100** |
| | 3 | Last | 39.64 | 48.96 | 36.2 | 43.75 | 37.92 | 41.41 | | | 70.36 | 76.56 | 71.46 | 75.52 | 43.85 | 72.4 | 82.08 | 90.62 | 69.53 | 72.14 |
| | | Best | 42.14 | 50.52 | 39.43 | 46.88 | 41.51 | 42.97 | | | 77.18 | 77.6 | 76.09 | 79.43 | 46.67 | 76.56 | 91.72 | 92.45 | 73.02 | 75 |
| | 4 | Last | **100** | **100** | **100** | **100** | 99.9 | **100** | | | 98.34 | 98.7 | 99.27 | **100** | 99.74 | **100** | 99.06 | **100** | 98.91 | 99.74 |
| | | Best | **100** | **100** | **100** | **100** | 99.9 | **100** | | | 98.49 | 98.96 | 99.53 | **100** | **100** | **100** | **100** | **100** | 99.84 | **100** |
| | 5 | Last | 54.43 | 92.19 | 36.93 | 86.72 | 63.28 | 92.97 | | | 6.51 | 6.51 | 52.97 | 66.15 | 6.51 | 6.51 | 78.54 | **100** | 59.9 | 75.78 |
| | | Best | 54.95 | 92.71 | 38.65 | 87.24 | 63.8 | 93.49 | | | 7.66 | 10.94 | 74.53 | 79.69 | 6.82 | 7.03 | **100** | **100** | 77.61 | 79.69 |
| B | 1 | Last | 57.87 | 60.16 | 57.24 | 59.11 | 57.97 | 59.9 | 60 | 61.2 | 95.68 | **100** | 93.91 | 96.61 | 96.88 | 98.18 | 99.79 | **100** | 95.83 | 98.44 |
| | | Best | 60.37 | 62.5 | 58.33 | 60.16 | 59.85 | 60.68 | 61.82 | 63.54 | 99.84 | **100** | 97.97 | 98.96 | 98.6 | 99.22 | **100** | **100** | 98.02 | 98.44 |
| | 2 | Last | 68.18 | **100** | 70.89 | 84.64 | 62.66 | **100** | 62.66 | **100** | **99.84** | **100** | **100** | **100** | 6.51 | 6.51 | **100** | **100** | **99.95** | **100** |
| | | Best | **100** | **100** | 85.68 | 89.06 | **100** | **100** | **100** | **100** | 99.95 | **100** | **100** | **100** | 6.77 | 6.77 | **100** | **100** | **100** | **100** |
| | 3 | Last | 20.47 | 42.71 | 24.58 | 43.23 | 44.95 | 50.78 | | | 75.78 | 77.6 | 72.97 | 75.26 | 79.89 | 83.85 | 90.52 | 91.93 | 70.26 | 72.14 |
| | | Best | 21.98 | 45.31 | 26.98 | 46.35 | 47.34 | 52.6 | | | 78.28 | 80.21 | 75.21 | 76.56 | 81.93 | 84.38 | 92.14 | 92.71 | 74.38 | 75.78 |
| | 4 | Last | 99.64 | **100** | **100** | **100** | **100** | **100** | | | 98.39 | 98.96 | 98.44 | 98.96 | 99.48 | **100** | **100** | **100** | 96.15 | **100** |
| | | Best | **99.79** | **100** | **100** | **100** | **100** | **100** | | | 98.6 | 99.22 | 99.01 | 99.22 | **100** | **100** | **100** | **100** | **100** | **100** |
| | 5 | Last | 86.77 | 89.84 | 50.78 | 89.06 | 68.13 | 87.24 | | | 31.93 | 76.56 | 72.19 | 75.78 | 37.5 | 91.15 | 98.91 | **100** | 77.5 | 80.99 |
| | | Best | 88.75 | 93.23 | 51.46 | 90.36 | 85.47 | 93.49 | | | 35.05 | 78.65 | 76.25 | 80.21 | 38.91 | 92.97 | **100** | **100** | 80.42 | 82.55 |
| C | 1 | Last | 62.29 | 63.02 | 62.19 | 64.84 | 61.67 | 64.84 | 61.1 | 64.84 | **99.22** | **99.74** | 96.93 | 98.18 | 98.54 | 99.48 | 99.95 | **100** | 97.03 | 98.44 |
| | | Best | 64.16 | 65.62 | 63.75 | 65.62 | 65.42 | 66.15 | 62.66 | 65.1 | **99.48** | **100** | 98.7 | 99.22 | 99.12 | 99.74 | **100** | **100** | 98.86 | 99.22 |
| | 2 | Last | 99.9 | **100** | **100** | **100** | **100** | **100** | **100** | **100** | 99.74 | **100** | **100** | **100** | **100** | **100** | **100** | **100** | **100** | **100** |
| | | Best | **100** | **100** | **100** | **100** | **100** | **100** | **100** | **100** | 99.9 | **100** | **100** | **100** | **100** | **100** | **100** | **100** | **100** | **100** |
| | 3 | Last | 17.76 | 36.46 | 17.97 | 40.1 | 35.78 | 41.15 | | | 77.13 | 81.77 | 74.27 | 76.04 | 82.76 | 84.9 | 91.88 | 92.97 | 72.08 | 76.3 |
| | | Best | 19.58 | 40.62 | 19.01 | 43.49 | 38.23 | 42.71 | | | 78.65 | 81.77 | 77.5 | 78.39 | 85.37 | 87.24 | 93.39 | 94.53 | 75 | 76.3 |
| | 4 | Last | **100** | **100** | **100** | **100** | **100** | **100** | | | 98.34 | 99.22 | 98.6 | 99.22 | 99.74 | **100** | 99.84 | **100** | 99.48 | **100** |
| | | Best | **100** | **100** | **100** | **100** | **100** | **100** | | | 98.75 | 99.48 | 98.86 | 99.48 | **100** | **100** | **100** | **100** | **100** | **100** |
| | 5 | Last | 93.96 | **97.66** | 85.52 | **96.35** | 90.26 | 93.75 | | | 34.84 | 77.86 | 76.77 | 79.43 | 89.79 | 94.79 | **98.75** | **99.74** | 82.61 | 85.16 |
| | | Best | 94.9 | **98.7** | 86.3 | **98.18** | 92.34 | 94.79 | | | 35.62 | 79.95 | 80.16 | 81.51 | 91.77 | **95.31** | **100** | **100** | 85.62 | 87.76 |

XJTU-SY
Results with order split and data augmentation

| | | | AE | | SAE | | DAE | | MLP | | CNN | | LeNet | | AlexNet | | ResNet18 | | LSTM | |
|---|---|---|---|---|---|---|---|---|---|---|---|---|---|---|---|---|---|---|---|---|
| Nor | Input | Loc | Mean | Max | Mean | Max | Mean | Max | Mean | Max | Mean | Max | Mean | Max | Mean | Max | Mean | Max | Mean | Max |
| A | 1 | Last | 62.97 | 73.59 | 68.36 | 76.41 | 60.46 | 71.54 | 68.41 | 75.38 | **98.97** | **99.74** | **96.31** | **96.67** | 91.9 | **95.13** | 89.54 | 96.41 | **96.26** | **97.95** |
| | | Best | 72.67 | 76.92 | 81.08 | 83.33 | 70.92 | 72.05 | 77.64 | 80.26 | **99.84** | **100** | **98.1** | **98.97** | 94.2 | **96.15** | 99.69 | 100 | 98 | 98.97 |
| | 2 | Last | **100** | **100** | **100** | **100** | **100** | **100** | 73.33 | 73.33 | 73.02 | 73.33 | 72.2 | 73.33 | 6.67 | 6.67 | 73.33 | 73.33 | 71.38 | 73.33 |
| | | Best | **100** | **100** | **100** | **100** | **100** | **100** | 73.33 | 73.33 | 73.33 | 73.33 | 73.33 | 73.33 | 6.67 | 6.67 | 73.33 | 73.33 | 73.33 | 73.33 |
| | 3 | Last | 26.05 | 45.64 | 29.95 | 40.77 | 41.95 | 46.67 | | | 72.3 | 76.15 | 75.28 | 78.21 | 14.72 | 46.92 | 73.8 | 88.46 | 72.77 | 74.62 |
| | | Best | 27.64 | 47.44 | 33.13 | 43.08 | 46.72 | 51.28 | | | 76.77 | 78.21 | 78.26 | 79.74 | 15.03 | 46.92 | 90.51 | 91.79 | 77.38 | 80 |
| | 4 | Last | **99.95** | **100** | **99.84** | **100** | **99.9** | **100** | | | 97.38 | 98.97 | 97.49 | 99.49 | 99.64 | **100** | **100** | **100** | 98.15 | 99.23 |
| | | Best | **100** | **100** | **100** | **100** | **100** | **100** | | | 98.26 | 98.97 | 99.08 | 99.74 | **100** | **100** | **100** | **100** | 99.59 | 99.74 |
| | 5 | Last | 58.36 | 92.56 | 54.05 | 88.46 | 78.1 | 89.74 | | | 19.44 | 70.51 | 53.28 | 75.9 | 6.67 | 6.67 | 95.18 | 99.74 | 71.79 | 78.97 |
| | | Best | 60.56 | 93.59 | 55.39 | 89.23 | 82.31 | 95.13 | | | 21.33 | 79.74 | 74 | 76.92 | 6.67 | 6.67 | 100 | 100 | 78.05 | 79.49 |
| B | 1 | Last | 82.67 | 85.9 | 84.26 | 86.92 | 82.05 | 83.59 | 85.13 | 87.44 | **98.87** | **99.74** | 97.59 | 97.95 | 97.54 | 98.46 | 97.44 | 99.49 | 96.51 | 97.69 |
| | | Best | 84.41 | 85.9 | 87.03 | 87.44 | 85.03 | 85.9 | 87.54 | 88.21 | **99.79** | **100** | 98.46 | 99.23 | 98.51 | 99.49 | 99.84 | 100 | 97.9 | 98.46 |
| | 2 | Last | 14 | 36.15 | 29.23 | 32.82 | 20.82 | 47.44 | 11.39 | 18.46 | 54.2 | 72.05 | 73.28 | 73.33 | 6.67 | 6.67 | 52.77 | 73.33 | 73.33 | 73.33 |
| | | Best | 42.92 | 58.21 | 52.31 | 63.59 | 32.51 | 47.44 | 33.95 | 38.97 | 73.28 | 73.33 | 73.33 | 73.33 | 6.67 | 6.67 | 73.33 | 73.33 | 73.33 | 73.33 |
| | 3 | Last | 42.87 | 47.44 | 33.54 | 47.18 | 45.59 | 47.69 | | | 75.74 | 78.46 | 76.41 | 79.23 | 81.64 | 82.56 | 87.49 | 88.97 | 75.13 | 76.67 |
| | | Best | 45.9 | 50.77 | 36.05 | 48.72 | 49.85 | 52.31 | | | 78.62 | 79.23 | 78.67 | 79.74 | 85.28 | 85.9 | 90.62 | 91.28 | 77.79 | 79.23 |
| | 4 | Last | **100** | **100** | 98.51 | **100** | 99.59 | **100** | | | 96.87 | 98.46 | 96.1 | 98.72 | 81.23 | **100** | **100** | **100** | 98.62 | 99.49 |
| | | Best | **100** | **100** | 98.82 | **100** | **100** | **100** | | | 98.41 | 98.97 | 98.51 | 99.23 | 81.33 | **100** | **100** | **100** | 99.44 | 99.74 |
| | 5 | Last | 89.33 | 92.56 | 55.28 | 91.79 | 66.67 | 94.1 | | | 26.72 | 76.67 | 70.72 | 74.36 | 35.39 | 84.62 | **99.33** | **100** | 72.31 | 75.64 |
| | | Best | 91.18 | 93.59 | 56.46 | 93.85 | 72.36 | 94.87 | | | 27.44 | 76.67 | 76.72 | 79.23 | 38.05 | 89.74 | **100** | **100** | 78.15 | 82.05 |
| C | 1 | Last | 81.44 | 83.85 | 85.38 | 87.44 | 81.08 | 83.59 | 83.9 | 85.13 | **96.67** | **99.74** | 98.05 | 98.97 | 98.82 | 98.97 | 98.92 | 100 | 97.29 | 98.21 |
| | | Best | 83.59 | 86.15 | 87.44 | 89.23 | 84.26 | 85.64 | 85.9 | 87.18 | **99.95** | **100** | 99.33 | 99.74 | 99.64 | 99.74 | **100** | **100** | 99.18 | 99.49 |
| | 2 | Last | **99.95** | **100** | **100** | **100** | **100** | **100** | 73.33 | 73.33 | 73.33 | 73.33 | 73.23 | 73.33 | 73.33 | 73.33 | 73.33 | 73.33 | 73.33 | 73.33 |
| | | Best | **100** | **100** | **100** | **100** | **100** | **100** | 73.33 | 73.33 | 73.33 | 73.33 | 73.33 | 73.33 | 73.33 | 73.33 | 73.33 | 73.33 | 73.33 | 73.33 |
| | 3 | Last | 6.67 | 6.67 | 6.67 | 6.67 | 49.69 | 57.44 | | | 75.64 | 76.92 | 76.97 | 81.54 | 83.69 | 85.9 | 90.46 | 91.54 | 75.49 | 76.92 |
| | | Best | 6.67 | 6.67 | 6.67 | 6.67 | 51.85 | 61.54 | | | 78.05 | 80.26 | 78.82 | 81.54 | 85.95 | 88.46 | 93.18 | 93.59 | 78.87 | 80 |
| | 4 | Last | **99.95** | **100** | 99.79 | **100** | 99.95 | **100** | | | 98.26 | 98.97 | 97.59 | 98.46 | 99.95 | **100** | **100** | **100** | 98.87 | 99.74 |
| | | Best | **100** | **100** | **100** | **100** | **100** | **100** | | | 98.77 | 98.97 | 98.15 | 98.72 | **100** | **100** | **100** | **100** | 99.44 | 99.74 |
| | 5 | Last | 74.05 | 93.08 | 69.95 | 92.56 | 90.1 | 93.85 | | | 57.95 | 73.85 | 77.38 | 81.28 | 90.51 | 94.62 | 90.67 | **100** | 79.23 | 82.82 |
| | | Best | 76.77 | 95.13 | 73.33 | **96.92** | 94.56 | 98.21 | | | 64.46 | 82.31 | 81.28 | 82.31 | 93.69 | 97.69 | **100** | **100** | 85.79 | 88.72 |


\end{landscape}

\end{document}